\documentclass[aps,prd,twocolumn,showpacs,superscriptaddress,
longbibliography]{revtex4-1}

\usepackage{amsmath,amssymb}
\usepackage{graphicx}\usepackage{capt-of}
\usepackage{booktabs}
\usepackage{multirow}
\usepackage{hyperref}\hypersetup{hidelinks}
\usepackage{siunitx}
\usepackage[utf8]{inputenc}
\usepackage[T1]{fontenc}

\graphicspath{{figures/}}

\begin{document}

\title{Inter-detector correlations in LIGO--Virgo strain: an exploratory study and possible astrophysical implications}

\author{S. C. Ulhoa}\email[]{sc.ulhoa@gmail.com}
\affiliation{Instituto de F\'isica, Universidade de Bras\'ilia, 70910-900, Bras\'ilia, DF, Brazil} \affiliation{Canadian Quantum Research Center,\\ 
204-3002 32 Ave Vernon, BC V1T 2L7  Canada} 

\author{B. C. C. Carneiro}\email[]{bcccarneiro@gmail.com}
\affiliation{Instituto Federal do Tocantins,  77760-000, Colinas do Tocantins, TO, Brazil} 

\author{F. L. Carneiro}\email[]{fernandolessa45@gmail.com}
\affiliation{Universidade Federal do Norte do Tocantins,
Centro de Ci\^encias Integradas,
77824-838 Aragua\'ina, TO, Brazil}

\author{L. V. A. Cunha}\email[]{l.v.a.cunha100@gmail.com}
\affiliation{Instituto de F\'isica, Universidade de Bras\'ilia, 70910-900, Bras\'ilia, DF, Brazil}

\date{November 16, 2026}

\begin{abstract}
We investigate inter-detector correlations in public LIGO--Virgo strain
around nine O2 and O3 events, using a band-pass-only statistic without
waveform-template fitting. The primary sample comprises 27 event--pair
measurements, with three complementary measurements for GW191127.
Event-window maxima over $|\tau|\leq35$ ms
are compared with 200 surrounding off-source maxima. With verified
fractional catalog epochs, seven of the 27 primary event--pair
measurements, all
H1$\times$L1, reach the local finite-background floor $p=1/201$.
No individual comparison passes a Holm correction at 5\% for this
family; the available background limits the attainable tail resolution.
Complementary comparisons of simultaneous and time-displaced off-source
data show processing-dependent structure, including a local excess near
GW200219 that does not recur at comparable size in the additional
segments examined. We discuss the possibility that part of the strain
classified operationally as noise contains an astrophysical contribution,
including an accumulated gravitational-radiation background associated
with Bondi radiative processes.
\end{abstract}
\maketitle

\section{Introduction}\label{sec:intro}
The direct detection of gravitational waves by the LIGO
Scientific Collaboration in September 2015, reported in
February 2016~\cite{GW150914}, stands as one of the most
consequential experimental results in modern physics.
The signal, designated GW150914 and interpreted as the
merger of two stellar-mass black holes, provided the first
direct confirmation of a prediction of Einstein's general
relativity that had remained unverified for a century.
The achievement was recognized with the Nobel Prize in
Physics in 2017.

The road to that detection, however, was neither short nor
uncontroversial. The theoretical status of gravitational
waves was itself disputed for decades after Einstein first
derived them in 1916~\cite{Einstein1916}. A central question
was whether the waves carried physical energy or were merely
coordinate artefacts of the linearized theory. At the Chapel
Hill conference in 1957, Richard Feynman introduced what
became known as the ``sticky bead argument'': a bead free to
slide along a rod would be heated by a passing gravitational
wave---an argument that gave the debate a
particularly vivid and intuitive resolution~\cite{Feynman1957,Kennefick2007}.
The same period saw the question of whether gravitational
radiation transports energy addressed in formal terms,
notably by Trautman, who developed a treatment of
the energy of gravitational radiation and showed it to be
non-negative under appropriate asymptotic conditions~\cite{Trautman1958}. The modern mathematical
formulation of the observable effects of gravitational waves
in the linearized theory---including the geodesic deviation
equation as the basis for detector response---was laid out by
Pirani in 1956~\cite{Pirani1956}.

The first serious proposal for a gravitational-wave detector
came from Joseph Weber in the 1960s, who constructed
resonant-bar detectors and reported candidate
detections~\cite{Weber1960,Weber1969}.
Weber's claims were ultimately not reproduced, but his work
established the experimental program that would eventually
lead to the kilometer-scale laser interferometers of
today~\cite{Cervantes2016}.
The LIGO detectors --- two Michelson interferometers with
4\,km arms located in Hanford, Washington, and Livingston,
Louisiana --- and the Virgo detector near Cascina, Italy,
were designed precisely to measure the fractional length
change $h = \Delta L / L \sim 10^{-21}$ induced by a
passing gravitational wave.
GW150914 appeared to vindicate this design, and was widely
regarded as closing the century-long controversy over the
physical reality of gravitational radiation.

The interpretation of strain beyond identified transient events remains
an interesting physical question. In this work, ``off-source background''
denotes the measured data used as an empirical reference; a
``gravitational-wave background'' denotes a possible physical contribution
to those data. Classifying a component operationally as noise does not
by itself determine its physical origin.

Inter-detector correlations provide a way to investigate the data
without fitting a particular compact-binary waveform. Creswell
et al.~\cite{Creswell2017} studied time-lag correlations in LIGO strain
and residuals. Subsequent studies emphasized the effects of signal
subtraction, spectral leakage, and statistical calibration, finding no
significant residual correlation in the analyses they
performed~\cite{LIGOresponse,Nielsen2019}. These issues motivate explicit
control comparisons and a precise statement of the statistic used.

Gravitational-wave searches include both matched-filter and unmodelled
coherent analyses, together with data-quality assessment and empirical
background estimation~\cite{GW150914,LIGOresponse}. Our short-window
correlation statistic is an exploratory observable, rather than a
reproduction of the full detection decision. Its sensitivity can differ
between pairs because of detector noise, antenna response and the
portion of a signal contained in the window.

We investigate two questions: whether an event-window maximum is
exceptional relative to nearby off-source maxima, and how simultaneous
off-source correlations compare with time-displaced controls. The first
comparison alone does not test independence of the detector noise.
We retain a band-pass-only primary analysis, permitting inspection of
structure without waveform-template fitting. Alternative processing is
used to assess sensitivity of the conclusions to spectral weighting.

Our physical motivation includes the possibility that radiation from
astrophysical systems contributes to the strain background. In
particular, the treatment of radiative energy in the teleparallel
equivalent of general relativity (TEGR) on Bondi--Sachs space-time
motivates discussion of an accumulated radiative
component~\cite{Maluf2024BackgroundEnergy}. We present this as a possible
interpretation to investigate. Its observational identification requires
additional predictions and controls beyond showing sample correlations.

\section{Data and methodology}\label{sec:method}
\subsection{Sample, epochs and data provenance}
The illustrative primary sample comprises the nine events in Table~\ref{tab:results},
with H1$\times$L1, H1$\times$V1 and L1$\times$V1 for each event.
The complementary event GW191127\_050227 contributes three further
measurements, one for each detector pair, presented in the Supplemental
Material. Thus the analysis comprises 27 primary event--pair
measurements plus three complementary measurements. The primary
statistical family contains 27 comparisons; combining the primary and
complementary sets defines an expanded family of 30.
The sample is exploratory and is not a complete or randomly selected
catalog sample; no event-population prevalence is inferred.

We use 256 s of public strain centered on each catalog epoch, at
4096 Hz, accessed through GWOSC and GWpy~\cite{GWOSC,gwpy}.
Table~\ref{tab:gps} records the entries explicitly selected for the
revised event analysis. Strain URLs, retained samples, quality and
injection flags, package versions and hashes are archived with the
analysis products. Public strain is calibrated and may already have
undergone cleaning; ``band-pass-only'' describes our additional processing.

The catalog epoch is a specified centering convention, not an optimized
window position or an exact posterior geocentric merger time.

\begin{table*}[t]
\caption{Catalog epochs used in the revised event-window analysis.
The strain-file release is separately recorded in the source manifests.}
\label{tab:gps}\centering
\begin{ruledtabular}\begin{tabular}{lrl}
Event & GPS epoch (s) & Catalog entry\\
GW170814 & 1186741861.5 & GWTC-2.1-confident, v4\\
GW170818 & 1187058327.1 & GWTC-2.1-confident, v2\\
GW190412 & 1239082262.1 & GWTC-2.1-confident, v4\\
GW190512\_180714 & 1241719652.4 & GWTC-2.1-confident, v2\\
GW190521 & 1242442967.4 & GWTC-2.1-confident, v4\\
GW190828\_063405 & 1251009263.7 & GWTC-2.1-confident, v2\\
GW200129\_065458 & 1264316116.4 & GWTC-3-confident, v1\\
GW200219\_094415 & 1266140673.1 & GWTC-3-confident, v1\\
GW200224\_222234 & 1266618172.4 & GWTC-3-confident, v1\\
\end{tabular}\end{ruledtabular}
\end{table*}

\subsection{Band-pass filter}
\begin{figure}[t]\centering
\includegraphics[width=\linewidth]{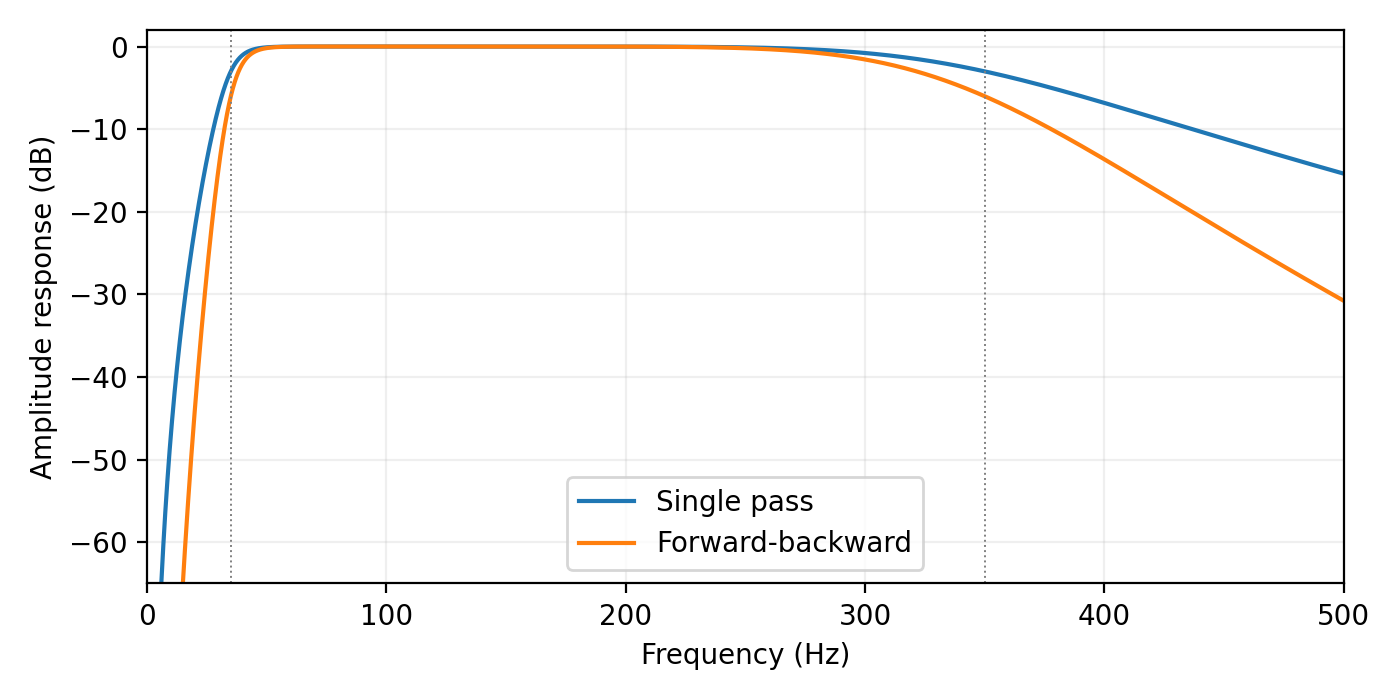}
\caption{Calculated amplitude response of the actual Butterworth design
at 4096 Hz. The forward--backward curve describes the interior response;
finite-segment edge transients require separate treatment.}
\label{fig:filter}\end{figure}

Each full segment is filtered before its short windows are extracted.
The implementation calls \texttt{butter} with prototype order 4,
normalized critical frequencies $35/(f_s/2)$ and $350/(f_s/2)$,
and \texttt{btype="band"}; its coefficients are then supplied to
\texttt{filtfilt}~\cite{scipy,SciPyButter}. The band-pass transformation
has order 8 in a single pass. Forward--backward application yields
zero phase away from finite-record boundary effects and a response
\begin{equation}
H_{\rm fb}(f)=H(f)H^*(f)=|H(f)|^2.
\end{equation}
Thus the critical frequencies of the single-pass design, approximately
$-3$ dB in amplitude, become approximately $-6$ dB after the two passes
(Fig.~\ref{fig:filter}). This is not a sharp frequency cut, and
zero phase does not mean that waveform amplitudes are unchanged.
Nonfinite input or output samples are rejected. The original coefficient
representation and default edge treatment are retained for reproducibility.
As a numerical check, the same design implemented with second-order
sections differs by at most $4.35\times10^{-9}$ in relative RMS over
the 30 detector segments. This check does not assess edge transients.
No whitening or notch filter is applied in the primary analysis.

\subsection{Statistic and reference windows}
For complete windows $x,y$ of $n$ samples, the implemented statistic is
\begin{equation}
C(k)=\frac{\displaystyle{\sum_{i\in I_k}}(x_i-\bar x)(y_{i-k}-\bar y)}
{n\sigma_x\sigma_y},\qquad |I_k|=n-|k|,
\label{eq:cc}
\end{equation}
where means and population standard deviations are computed on the
complete windows and held fixed for all lags. This equals Pearson
correlation at zero lag, but is not Pearson recomputed on each overlap
or between complete shifted windows. Positive lag in this convention
means that the first detector is delayed relative to the second.

An event window has width 0.2 s, centered at the specified GPS epoch;
inclusive time selection can produce 819 or 820 samples. We report
the signed value $C^*$ at the maximum of $|C(k)|$, with
$\tau^*=k/f_s$, over $|\tau|\leq35$ ms. At 4096 Hz this retains
lags through 143 samples. Anticorrelation is included in the search.
The common lag interval is an analysis choice, applied identically to
events and references. Different polarization projections and noise can
displace a sample maximum; the maximizing lag is not assumed to be a
geometric propagation-delay estimate.

For each pair, 200 centers are sampled without replacement from a
one-second grid in the surrounding segment, excluding centers within
$\pm2$ s of the event. The grid endpoints reserve the half-window and
0.1 s from the segment boundaries. Seeds are fixed and archived.
This separation prevents window overlap but does not guarantee
independent realizations in the presence of long-lived lines or
nonstationarity. The statistic is evaluated on all reference windows
with the same lag search. Its empirical local p-value is
\begin{equation}
p_{\rm emp}=\frac{k+1}{N+1},\quad
k=\#\{j:|C_j^*|\geq|C_{\rm ev}^*|\},\quad N=200.
\label{eq:pval}
\end{equation}
The floor $1/201$ denotes zero observed exceedances, rather than a
resolved tail probability below that scale. Interpretation requires
comparability of event and reference windows. Searching over lags is
included in this reference construction; multiplicity across events
and pairs is addressed separately with Holm's procedure.

The event-table reproduction records quality flags without adding new
vetoes to the supplied pipeline. In particular, L1 at GW200129 has
CBC and burst category-2/3 flags unset. Its correlations require a
specific glitch assessment. Continuous-wave hardware-injection flags
are also present in several LIGO records; they and spectral lines
must be addressed before assigning an astrophysical origin to
persistent structure. The stricter off-source controls below have a
separately stated selection and are not interchangeable with this
reference sample.

\section{Event-window results}\label{sec:results}
Table~\ref{tab:results} presents all 27 primary measurements. Seven
H1$\times$L1 comparisons have $k=0$ and $p=1/201$: GW170814,
GW190412, GW190521, GW190828\_063405, GW200129\_065458,
GW200219\_094415 and GW200224\_222234. Two further comparisons
have local $p\leq0.05$: GW170814 L1$\times$V1 and
GW190512\_180714 H1$\times$L1. No Virgo pair reaches the local
1\% level. A large p-value is a lack of exceptionalness under this
comparison, not a demonstration of equivalence or absence of a signal.

For each of the seven floor values, the primary Holm-adjusted p is
$27/201\simeq0.1343$. No individual comparison passes Holm at 5\%.
The floor already exceeds the first threshold $0.05/27$, so this
background cannot resolve the required tail. This limits discovery
claims without implying that the local excesses are absent. No
combined test of the number of excesses is reported. For the expanded
30-test family the corresponding adjusted value is $30/201\simeq0.1493$.

\begin{table*}[t]\centering
\caption{Corrected event-window measurements with the supplied estimator.
Bold local p-values denote zero exceedances among 200 references.
Holm correction uses the 27 primary tests. Results involving L1 for
GW200129 require the quality qualification in the text.}
\label{tab:results}
\begin{ruledtabular}\begin{tabular}{llrrrr}
Event & Pair & $C^*_{\rm ev}$ & $\tau^*$ (ms) & $p_{\rm emp}$ & $p_{\rm Holm}$\\
GW170814 & H1$\times$L1 & -0.557 & +8.06 & \textbf{0.00498} & 0.13433\\
GW170814 & H1$\times$V1 & +0.317 & +14.89 & 0.23383 & 1.00000\\
GW170814 & L1$\times$V1 & +0.336 & +17.58 & 0.02985 & 0.59701\\
GW170818 & H1$\times$L1 & -0.197 & +5.37 & 0.56716 & 1.00000\\
GW170818 & H1$\times$V1 & +0.318 & +2.20 & 0.23881 & 1.00000\\
GW170818 & L1$\times$V1 & +0.234 & +10.99 & 0.21891 & 1.00000\\
GW190412 & H1$\times$L1 & -0.535 & +3.17 & \textbf{0.00498} & 0.13433\\
GW190412 & H1$\times$V1 & -0.056 & -10.50 & 0.37313 & 1.00000\\
GW190412 & L1$\times$V1 & +0.049 & -14.65 & 0.50746 & 1.00000\\
GW190512\_180714 & H1$\times$L1 & -0.293 & -5.62 & 0.03483 & 0.66169\\
GW190512\_180714 & H1$\times$V1 & +0.158 & +28.81 & 0.12438 & 1.00000\\
GW190512\_180714 & L1$\times$V1 & +0.080 & -30.03 & 0.30348 & 1.00000\\
GW190521 & H1$\times$L1 & -0.641 & +1.95 & \textbf{0.00498} & 0.13433\\
GW190521 & H1$\times$V1 & +0.059 & -12.21 & 0.92537 & 1.00000\\
GW190521 & L1$\times$V1 & +0.128 & -5.37 & 0.47761 & 1.00000\\
GW190828\_063405 & H1$\times$L1 & -0.547 & -0.49 & \textbf{0.00498} & 0.13433\\
GW190828\_063405 & H1$\times$V1 & -0.111 & -23.68 & 0.16418 & 1.00000\\
GW190828\_063405 & L1$\times$V1 & -0.088 & -13.43 & 0.46766 & 1.00000\\
GW200129\_065458 & H1$\times$L1 & -0.786 & -5.13 & \textbf{0.00498} & 0.13433\\
GW200129\_065458 & H1$\times$V1 & +0.067 & +15.62 & 0.43781 & 1.00000\\
GW200129\_065458 & L1$\times$V1 & -0.060 & +21.00 & 0.52239 & 1.00000\\
GW200219\_094415 & H1$\times$L1 & -0.397 & -2.44 & \textbf{0.00498} & 0.13433\\
GW200219\_094415 & H1$\times$V1 & +0.093 & +1.22 & 0.66169 & 1.00000\\
GW200219\_094415 & L1$\times$V1 & +0.083 & -6.84 & 0.70149 & 1.00000\\
GW200224\_222234 & H1$\times$L1 & -0.497 & +2.20 & \textbf{0.00498} & 0.13433\\
GW200224\_222234 & H1$\times$V1 & +0.167 & +10.01 & 0.12438 & 1.00000\\
GW200224\_222234 & L1$\times$V1 & -0.116 & -10.74 & 0.51244 & 1.00000\\
\end{tabular}\end{ruledtabular}\end{table*}

\begin{figure*}[t]\centering
\includegraphics[width=.49\textwidth]{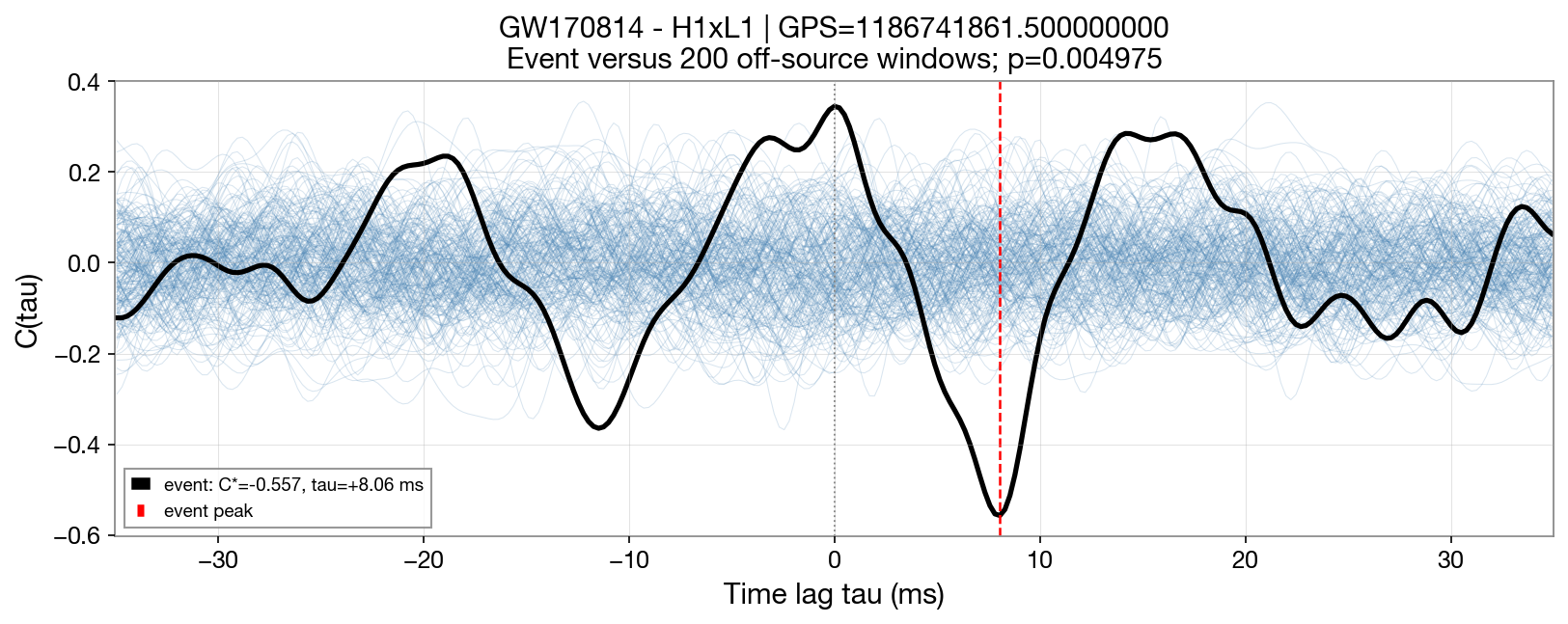}
\includegraphics[width=.49\textwidth]{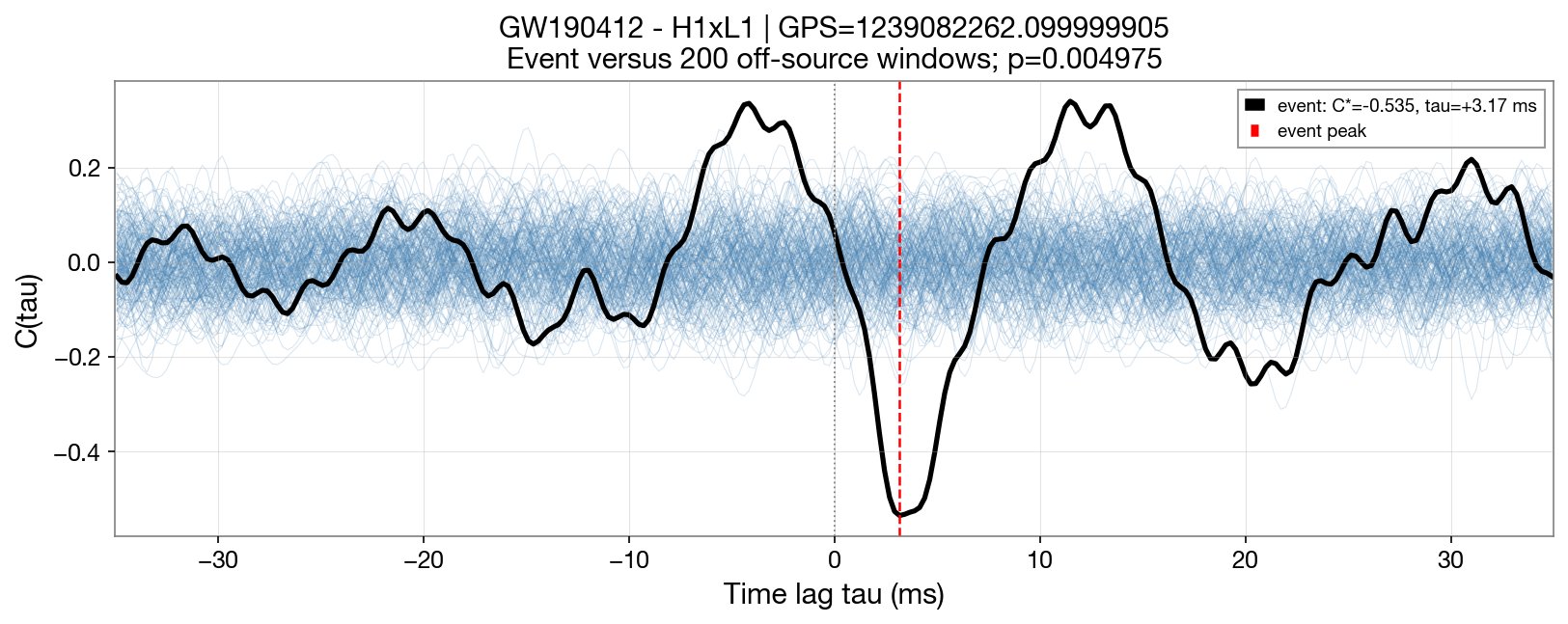}\\
\includegraphics[width=.49\textwidth]{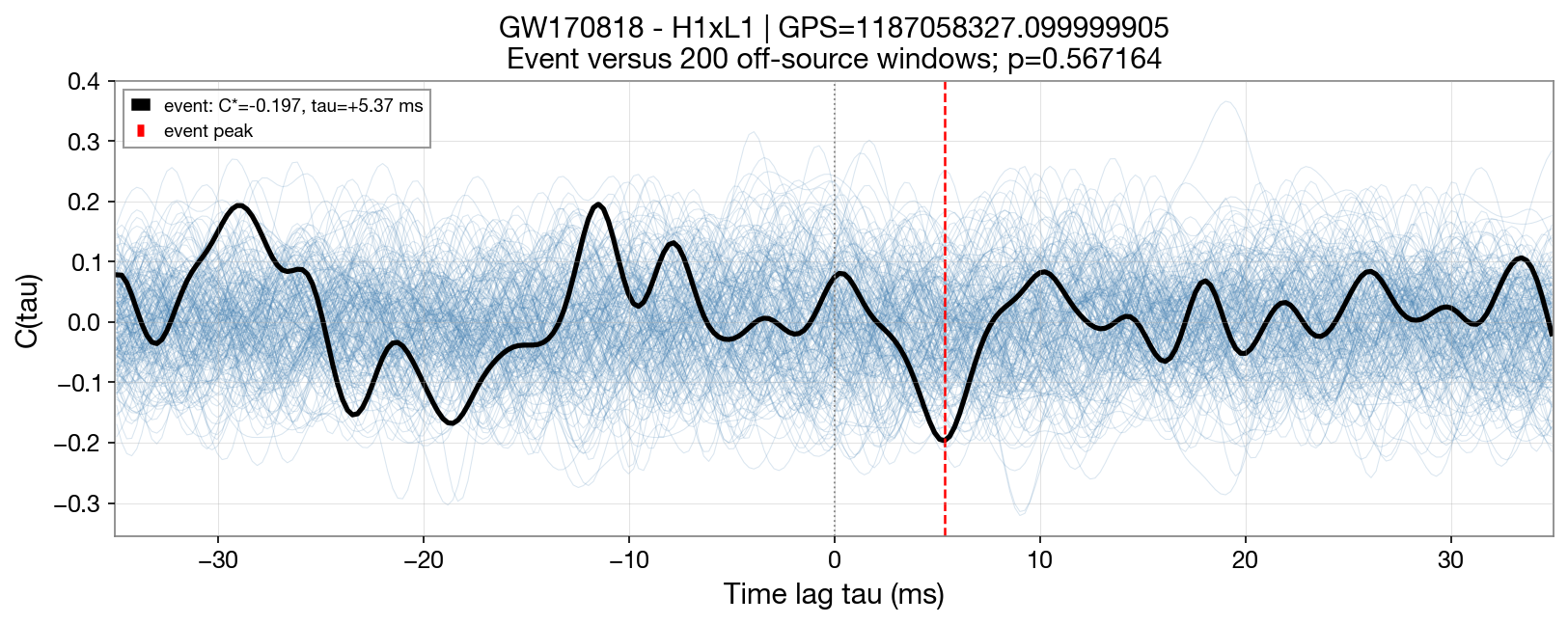}
\includegraphics[width=.49\textwidth]{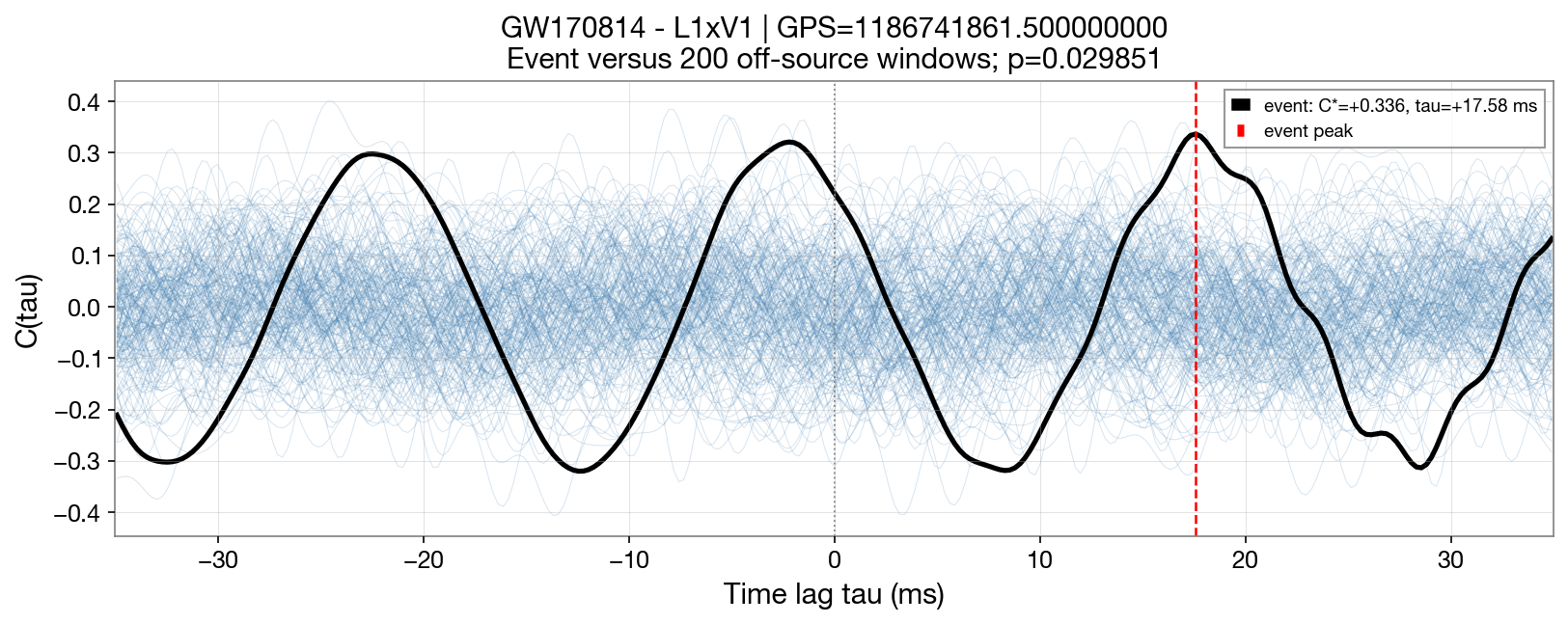}
\caption{Representative event curves (black) and 200 reference curves
(blue). Top: local H1$\times$L1 excesses for GW170814 and GW190412.
Bottom: GW170818 H1$\times$L1 and GW170814 L1$\times$V1, showing
other degrees of separation. All displayed p-values are local.}
\label{fig:events}\end{figure*}

Figure~\ref{fig:events} illustrates both local excesses and a case
without an H1$\times$L1 excess. The three complementary measurements for
GW191127 have local p-values 0.05473, 0.92040 and 0.75124 in
H1$\times$L1, H1$\times$V1 and L1$\times$V1, respectively.
The Supplemental Material~\cite{SupplementalMaterial}, Figs. S1--S20,
contains every event curve and the reference distributions of peak
magnitude and maximizing lag, including GW191127. Its Table S1 lists
the complete timing, quality-restricted and spectral-sensitivity results.

\subsection{Window centering and quality-restricted comparison}
\label{sec:timing}
We assessed timing sensitivity on a fixed grid of 21 window-center
offsets from $-100$ to $+100$ ms in 10-ms increments, rounded to samples.
This diagnostic uses 819-sample windows, while the inclusive masks in
Table~\ref{tab:results} can contain 819 or 820 samples. Reference centers
are 200 distinct integer seconds selected from $[-123,123]$ s with
$|t|>3$ s, using the same archived seed prescription. These references
are common to all offsets and reserve more than 4 s at the segment
edges. We retain the specified zero-offset result rather than selecting
a favorable center for the primary table.

For a scan-calibrated comparison, the largest $\max|C|$ over all 21
centers is compared with reference maxima over exactly the same 21
centers, using Eq.~\ref{eq:pval}. This incorporates the additional timing
search in the reference statistic. The adjacent windows are strongly overlapping.

\begin{figure*}[t]\centering
\includegraphics[width=.98\textwidth]{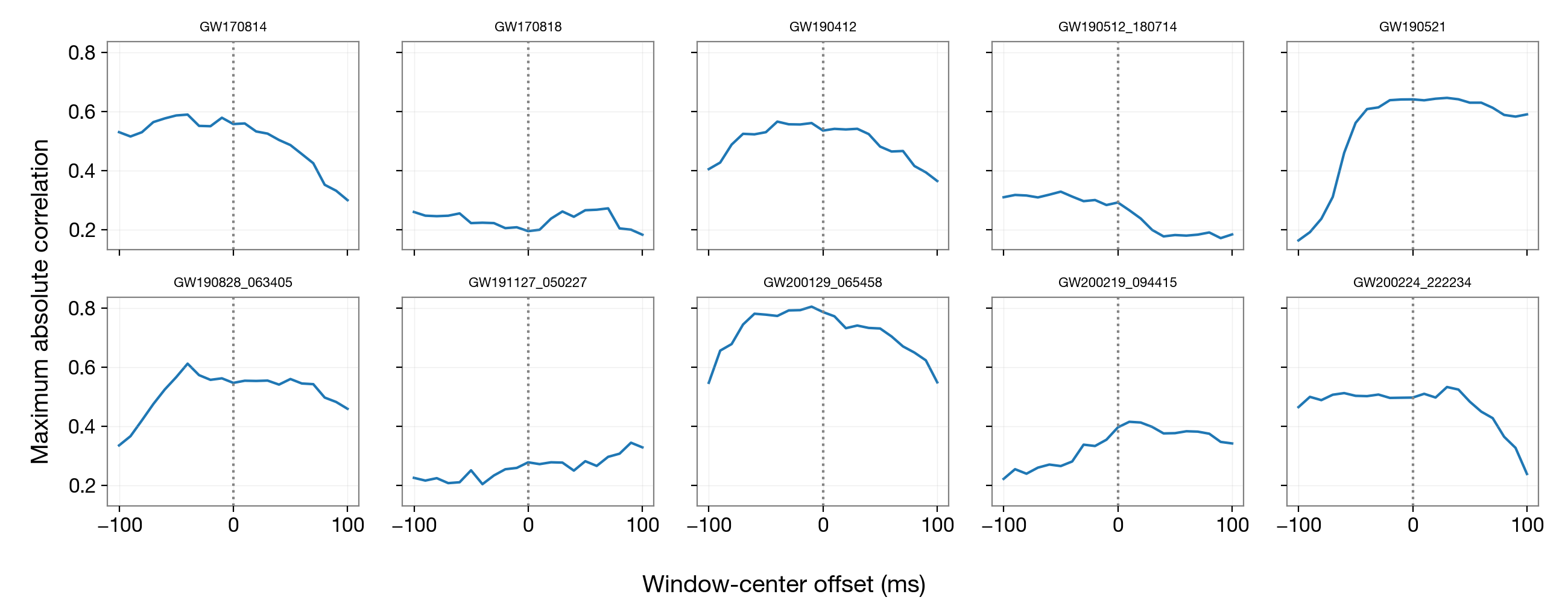}
\caption{H1$\times$L1 window-centering sensitivity at fixed 819-sample
length. All ten events are shown, including the supplementary GW191127.
The plot shows peak magnitude.
GW200129 involving L1 fails the quality-restricted event selection.}
\label{fig:timing}\end{figure*}

Figure~\ref{fig:timing} shows that peak amplitude can depend appreciably
on centering. In H1$\times$L1, the number of offsets attaining local
$p\leq0.01$ is 19/21 for GW170814, 21/21 for GW190412, 17/21 for
GW190521, 21/21 for GW190828, 21/21 for GW200129, 11/21 for GW200219
and 19/21 for GW200224, with the unvetoed diagnostic references.
These fractions describe overlapping offsets and are not binomial
recovery probabilities. All seven have scan-calibrated local $p=1/201$.
Some originally less prominent cases also change rank, emphasizing
why a center selected from the data needs its own calibration.

We repeat the comparison requiring DATA, CBC\_CAT1/2 and
BURST\_CAT1/2, and all non-CW hardware-injection absence flags, across
the entire scan support and an additional 2-s margin. Selection uses
flags rather than measured correlation. Of the 27 primary event--pair
windows, 25 are eligible: GW200129 H1$\times$L1 and L1$\times$V1
are excluded. Livingston's CBC and burst category-2 flags are unset
from GPS 1264316116 to 1264316124, containing this event epoch.
We therefore do not use its event-window correlations as evidence
robust to the quality restriction. Such exclusion does not assess a
glitch-subtraction model or establish the true astrophysical waveform.

The eligible primary comparisons retain 171--200 references per pair.
The six remaining H1$\times$L1 excesses have zero-offset and
scan-calibrated local $p\leq0.01$. None of the 25 comparisons passes
Holm at 5\% for either diagnostic. These sensitivity families are
reported separately and are not pooled with the main family to select
a stronger significance.

Requiring the CW absence flag as well leaves no eligible primary event
pair. Hence these data do not provide a simultaneous, entirely
hardware-injection-free event comparison. The flags identify a
potential contribution, not its amplitude in this statistic. GWOSC
documents CW hardware injections during O2 and O3, including Virgo
subtraction and possible residuals in part of O3b~\cite{GWOSCO2Inj,GWOSCO3Inj}.
We assess spectral sensitivity below without equating notching with
verified removal of all injected power.

\section{Off-source controls with observed data}\label{sec:controls}
To investigate off-source structure directly, we filter the real strain
again and recompute the per-window scores for the same nine event
neighborhoods using fixed time displacements. This produces 108
summary entries: 27 pairs and four processing variants.

The positive displacements are $5.137+0.731j$ s for $j=0,\ldots,20$,
rounded to the sampling grid, together with their negatives. Candidate
centers lie on the grid $-102+1.173j$ s through approximately +102 s.
Only centers whose supports are valid for every displacement, including
zero, are retained. A $\pm10$ s event exclusion and segment-edge
margins are used. The selection requires DATA, BURST\_CAT1 and
BURST\_CAT2, with a two-second quality margin, and the absence of
selected transient hardware-injection flags. Continuous-wave injection
flags remain a limitation. Windows have 819 samples, and the internal
lag maximization extends to 143 samples. Between 28 and 123 windows
remain per pair. The most restrictive GW200129 H1$\times$L1 selection
contains only pre-event windows.

The summary is the mean of $T=\max|C|$. We compare its simultaneous
value with 42 displaced values, using identical selected centers within
each pair. The displacement range is descriptive: it is neither a
confidence interval nor a calibrated probability distribution. Reusing
the data under shifts does not create independent observing time.
The original statistic is the principal comparison; complete-window
Pearson correlation, whitening and 50/60 Hz harmonic notches are
additional diagnostics. The complete-window Pearson variant shifts
one window within an extended record, maintaining 819 samples at every
lag. Off-source whitening uses mean periodograms of 4-s Hann windows
centered every 4 s from $-122$ to $+122$ s, with $|t|>14$ s and the
same quality selection. The ASD supplies a 2-s GWpy whitening FIR with
a 20-Hz high-pass corner, followed by the band-pass. The harmonic-notch
variant applies forward--backward IIR notches at positive multiples of
50 and 60 Hz strictly below 350 Hz, with $Q=f/(1\,\mathrm{Hz})$.
The protocol was specified after earlier
exploration and is not an external preregistration.

\begin{figure*}[t]\centering
\includegraphics[width=.78\textwidth]{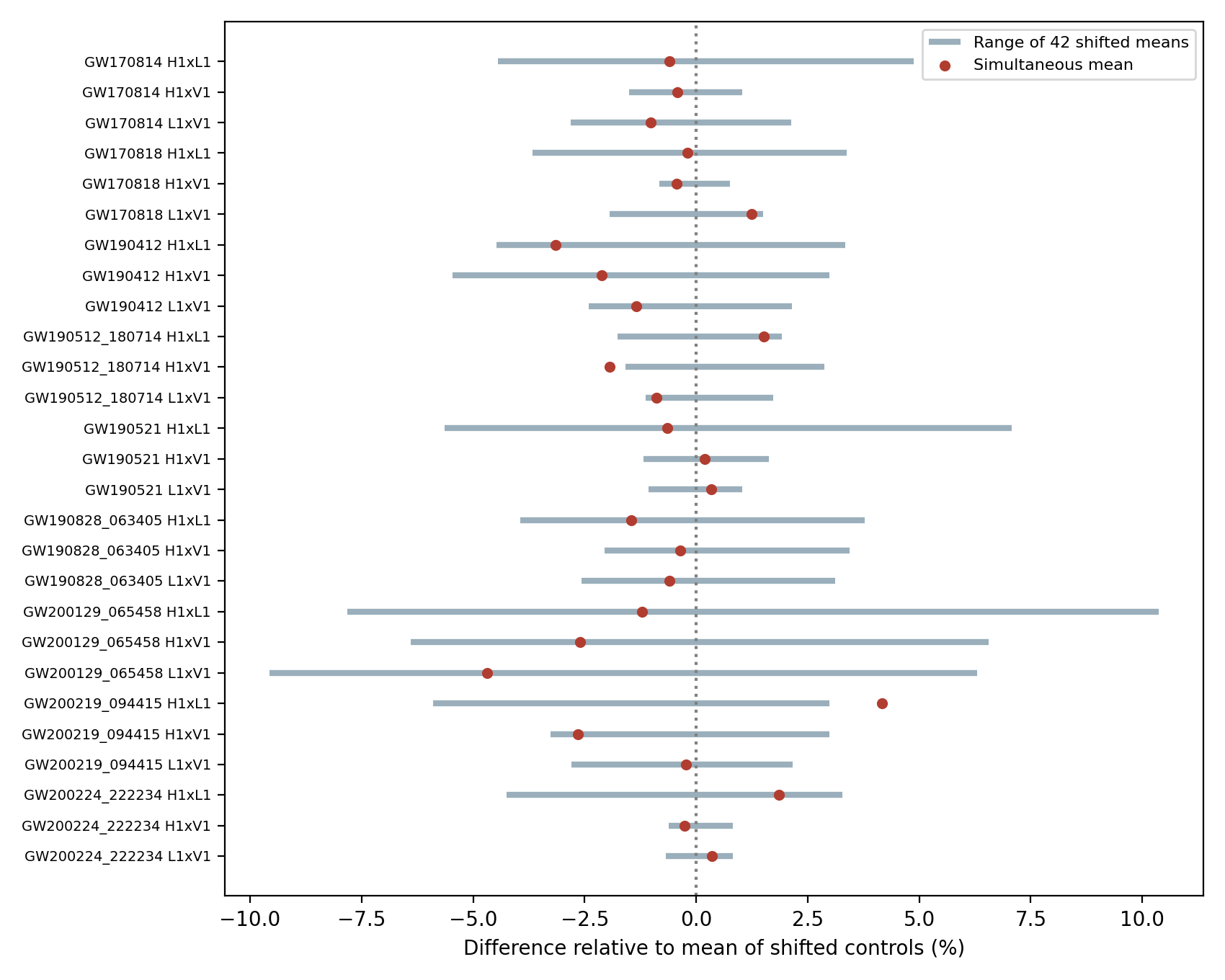}
\caption{Off-source means of the original maximum-correlation statistic.
Points show simultaneous values relative to the mean of 42 displaced
controls; bars span those displaced means. Bars are not confidence
intervals and their ranks are not calibrated p-values.}
\label{fig:controls}\end{figure*}

In the original-statistic comparison, seven of 27 simultaneous means
exceed the mean of the displaced controls, and twenty are smaller.
Relative differences range from $-4.69\%$ to $+4.16\%$.
Only GW200219 H1$\times$L1 exceeds every displaced mean
(Fig.~\ref{fig:controls}). Its simultaneous mean is 0.243903,
versus 0.234159 averaged over displacements. The displaced means
range from 0.220355 to 0.241162. The relative excess is $+3.93\%$
before and $+4.40\%$ after the event. However, the simultaneous
95th percentile is 0.307538, compared with 0.310236 for the average
of the displaced 95th percentiles; the distribution is not uniformly
elevated in its tail.

For GW200219, the relative difference becomes $+2.74\%$ with
complete-window Pearson correlation, $-1.89\%$ after whitening,
and $+3.08\%$ with the specified harmonic notches. This is a
processing-dependent local observation. Notches do not exhaust the
instrumental-line inventory, and a change after whitening does not
identify the physical origin of the removed or reweighted power.

\begin{figure}[t]\centering
\includegraphics[width=\linewidth]{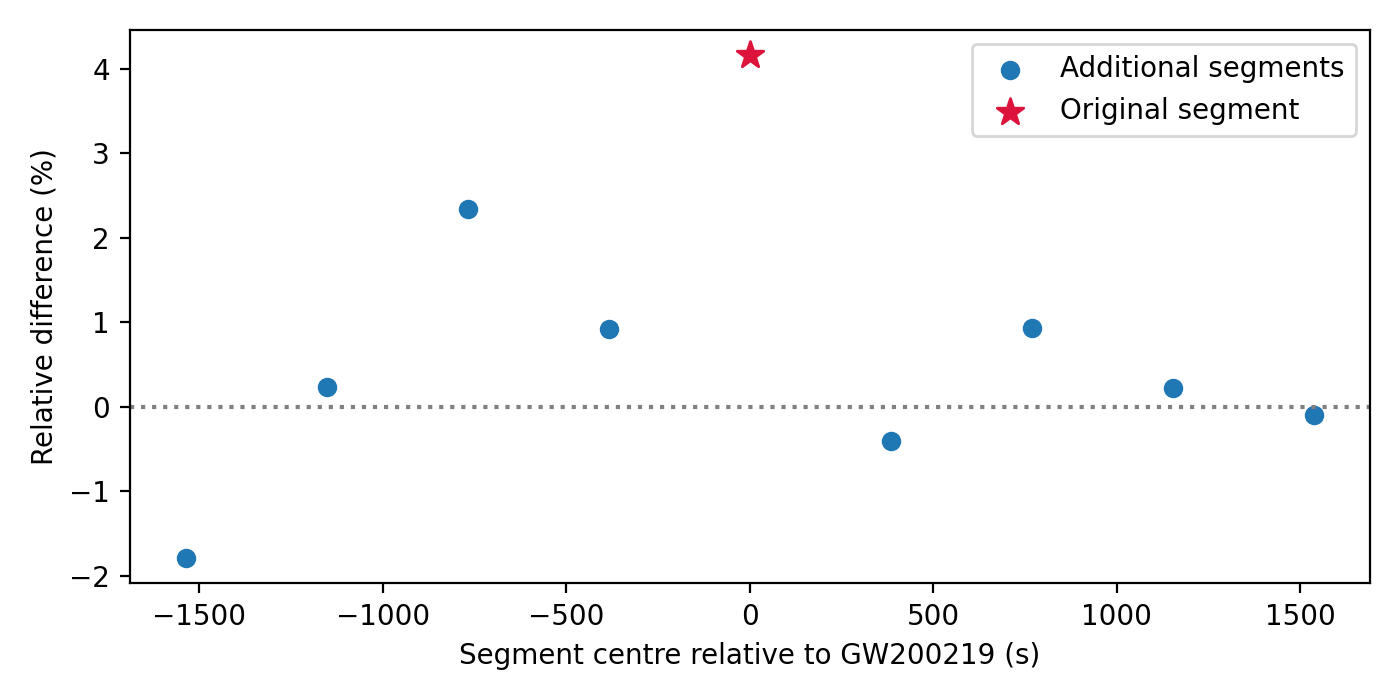}
\caption{GW200219 H1$\times$L1: original neighborhood and eight
additional segments, with the original statistic and fixed controls.
Values describe selected windows; no uncertainty or discovery
significance is inferred from the displacement ranks.}
\label{fig:followup}\end{figure}

Eight further 256-s segments centered at $\pm384$, $\pm768$,
$\pm1152$ and $\pm1536$ s from GW200219 were examined with the
same controls and selection. They do not overlap the original segment,
but belong to the same instrumental records and are not assumed to be
independent. The follow-up retains 909 windows. Its window-weighted
relative difference is $+0.27\%$ for the original statistic, and none
of the eight segments exceeds all displaced controls in any of the four
variants (Fig.~\ref{fig:followup}). The initial excess thus does not
recur as a persistent elevation of comparable size in this follow-up.
The case was selected after observing its excess. This result does not
exclude a transient contribution or provide a sensitivity limit on a
weak background.

\subsection{Sensitivity to documented injection frequencies}
We apply an additional diagnostic with forward--backward IIR notches
of 1-Hz single-pass bandwidth at the documented LIGO CW injection
frequencies inside 35--350 Hz, in addition to the harmonic notches.
For each event the published reference frequencies and frequency
derivatives are propagated to the catalog epoch~\cite{GWOSCCWParamsO2,GWOSCCWParamsO3}.
The same filter is applied to both detectors, preserving the control
centers and displacements. The corresponding event diagnostic uses
the same inclusive windows and reference centers as the main table.
This spectral veto is not coherent injection subtraction, a complete
instrumental-line inventory, or proof of negligible residual injections;
it can also alter sensitivity to physical signals. Virgo's separate
subtracted injection residuals are not modeled by this LIGO list.

All seven H1$\times$L1 event excesses retain local $p\leq0.01$ after
this combined notching. The GW170814 and GW200224 values are $2/201$;
the other five remain at $1/201$. GW170814 L1$\times$V1, however,
changes from $p=0.02985$ to $p=0.96020$, and GW190512 H1$\times$L1
changes from $p=0.03483$ to $p=0.27363$. The secondary local cases
therefore depend strongly on spectral weighting.

In the off-source comparison, GW200219 H1$\times$L1 changes to
$+1.41\%$, with seven of 42 displaced means at least as large as the
simultaneous mean. GW190512 H1$\times$L1 becomes the only case
above every displaced mean in this processing variant ($+2.92\%$).
These ranks remain descriptive and reflect exploration across
processing choices; the latter case was not subjected to the additional
segment follow-up used for GW200219. Neither case establishes a
persistent common component or its physical origin.

\section{Dependence on preprocessing}\label{white}
Band-pass-only processing is retained as a deliberate exploratory
choice. PSD whitening reweights frequencies using a measured spectral
estimate and does not require a binary template. A common physical
component could contribute to that estimate, but this observation alone
does not establish that whitening removes such a component. For ideal,
nonzero deterministic linear filters, normalized spectral coherence is
unchanged because transfer factors cancel. Finite estimates and
integrated time-domain statistics need their own sensitivity assessment.

An explicitly specified event diagnostic was performed for GW170814
and GW190412. Welch PSDs use Hann segments of 4 s at 2-s steps,
excluding the central $\pm2$ s and keeping within $\pm126$ s.
Mean periodograms define the ASD for GWpy whitening with a 2-s FIR
and a 20-Hz high-pass corner, followed by the original band-pass.
Both the whitened and band-pass-only branches discard 2 s at each
edge and use the same reference centers. This changes some references
relative to Table~\ref{tab:results}.

\begin{figure*}[t]\centering
\includegraphics[width=.85\textwidth]{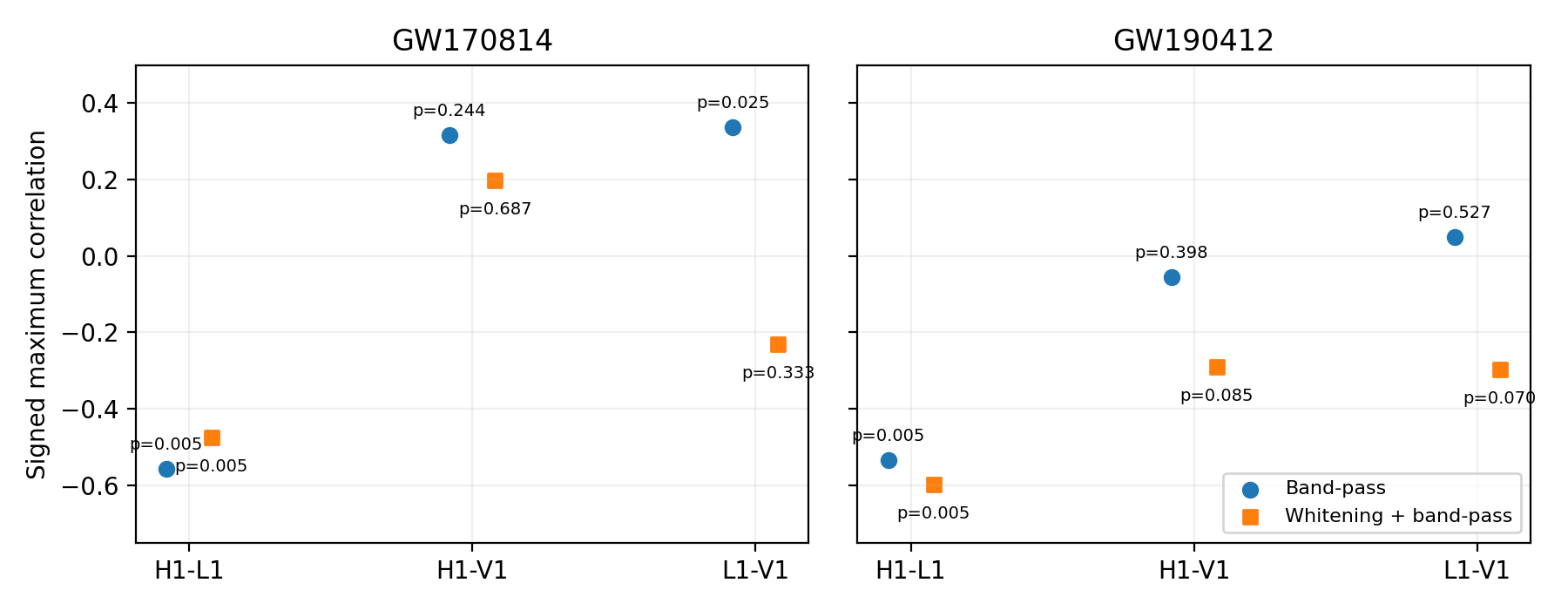}
\caption{Signed peaks and local p-values for the independent event
whitening diagnostic. Branches share reference-window centers.
Both H1$\times$L1 excesses remain at the local floor. Changes in
other pairs illustrate sensitivity to spectral weighting.}
\label{fig:white}\end{figure*}

Figure~\ref{fig:white} compares the signed peaks and local p-values
of the two processing branches. The H1$\times$L1 values change from
$-0.557$ to $-0.475$ for
GW170814 and from $-0.535$ to $-0.599$ for GW190412. All four
values have local $p=1/201$. Thus this diagnostic retains both local
excesses. It is a sensitivity
comparison and is not used to select a more favorable discovery
statistic. The off-source whitening comparison in
Sec.~\ref{sec:controls} concerns a different observable and selection.

\section{Physical interpretation}\label{sec:discuss}
\subsection{An astrophysical possibility}
The empirical analysis motivates continued investigation of the physical
content of off-source strain. We consider the possibility that some
part of the component operationally classified as noise has an
astrophysical origin. An accumulated field of gravitational radiation
from radiating systems is one possibility considered here, alongside
instrumental and environmental contributions.

The theoretical motivation is the TEGR analysis of radiative energy
in Bondi--Sachs space-time~\cite{Maluf2024BackgroundEnergy}. In that
treatment, the energy expression depends on the functions generating
the news functions, and its analysis suggests that radiative energy
may remain in the background space-time after a burst. Extending this
motivation to radiation from many systems across cosmic history leads
to the hypothesis discussed here. The theoretical energy expression
does not yet specify a strain cross-spectrum for this detector network.

We therefore present the accumulated component as a possibility for
further physical investigation. We do not infer its existence from
large off-source sample maxima, or its absence from a null control.
Independent colored records also produce nonzero maxima in finite
windows. A hypothesis compatible with the observations is not thereby
preferred over alternatives. Quantitative discrimination would require
a model connecting source populations and radiative evolution to
polarizations, spectra and detector responses, and comparison with
appropriate instrumental and environmental controls.

A template-free investigation can address structure outside a chosen
binary waveform family. Conversely, successful identification of
compact-binary signals does not by itself classify all remaining strain
by physical origin. These observations motivate the exploratory scope
without making the particular statistic used here a universal detector
of unknown gravitational signals.

\subsection{Response, sensitivity and future tests}
A common incident field produces responses
\begin{equation}
d_i(t)=F_i^+h_+(t-\tau_i)+F_i^\times h_\times(t-\tau_i)+n_i(t).
\end{equation}
Different projections and noise levels can reduce the prominence of a
cross-correlation maximum. Under the illustrative assumptions that
noise is uncorrelated between detectors and with the signals,
\begin{equation}
\rho(d_1,d_2)=\rho(s_1,s_2)
\frac{q_1q_2}{\sqrt{(1+q_1^2)(1+q_2^2)}},
\end{equation}
where $q_i^2=\mathrm{Var}(s_i)/\mathrm{Var}(n_i)$ within the chosen
band. This population identity explains dilution of correlation; it is
not the sampling distribution of the maximized estimator, and $q_i$
is not the catalog matched-filter SNR. The complementary binary
recovery exercise in Appendix~\ref{app:calibration} illustrates the
restricted sensitivity of this statistic in some detector pairs.

Virgo's distinct design and geographical location broaden the comparison,
but do not guarantee absence of environmental correlations across the
network. Extending the analysis to KAGRA~\cite{KAGRA2021} would add
geometry and instrumental diversity. Its evidential value would depend
on sensitivity, observing conditions and the predicted network response;
a null result would not automatically locate the cause in the other
three detectors.

Existing gravitational-background constraints must be compared with
quantitative predictions under their stated spectral, statistical and
response assumptions. Including a putative signal in an estimated PSD
does not by itself invalidate those constraints. Deriving observables
for the proposed accumulated component and studying its compatibility
with applicable bounds remain tasks for a more discriminating test.

\section{Conclusions}\label{sec:conclusions}
With verified catalog epochs and the supplied band-pass-only estimator,
seven of 27 primary event--pair comparisons show local excesses at the
finite-background floor, all in H1$\times$L1. The reference size does
not permit individual Holm rejection at 5\%. A documented whitening diagnostic preserves the local
H1$\times$L1 excesses for GW170814 and GW190412.

The off-source controls show a processing-dependent local deviation
near GW200219, without a comparable persistent enhancement in the
additional segments examined. These observations characterize the
chosen statistic and sample. The quality-restricted diagnostic retains six local H1$\times$L1
excesses while excluding both GW200129 pairs involving L1. Timing
scans and injection-frequency notches quantify additional sensitivity,
but hardware injections, other lines, nonstationarity and limited
sensitivity continue to constrain physical attribution.

We retain the possibility that part of the measured strain background
has an astrophysical origin, including accumulated gravitational
radiation associated with Bondi radiative processes. This hypothesis
provides a motivation for further analysis of correlations beyond
specific waveform-template searches. The present results neither
identify that component nor establish preference for its origin;
connecting the physical proposal to measurable network predictions
is the next step toward testing it.

\begin{acknowledgments}
F. L. C. acknowledges financial support from the Brazilian National
Council for Scientific and Technological Development (CNPq), Brazil,
through a Productivity Fellowship (PQ), Grant No.~303574/2026-7.
We thank Antoine Olivier (Kepler Circle) for helpful correspondence.
\end{acknowledgments}

\section*{Data and code availability}
Public strain is obtained from GWOSC and posterior samples from the
public parameter-estimation releases identified in the source manifests.
Machine-readable event, quality, timing and control tables, selected
posterior samples, injection-trial products, fixed seeds and analysis
scripts accompany this work. The manifests specify input URLs, versions
and hashes. These products support recomputation of the reported
statistics; reproducing waveform generation also requires the documented
LALSimulation environment and the public strain inputs.
\appendix
\section{Complementary recovery calibration}
\label{app:calibration}
We calibrated recovery by regenerating the specified waveform ensemble
and adding it to real off-source strain. The calibration is conditional
on the waveform family, posterior samples and reference windows; it
does not identify the origin of the off-source strain.

This complementary test addresses the objection that
an event correlation can be indistinguishable from the
off-source distribution even when a binary signal is present.
It leaves the experimental event--background analysis and
Table~\ref{tab:results} as the main investigation. For each of
the nine events, 128 joint parameter samples were drawn without
replacement from the public \texttt{C01:IMRPhenomXPHM} posterior
group. Masses, spins, distance, orientation, sky position, phase
and time were retained jointly. Polarizations were generated
with IMRPhenomXPHM through LALSimulation and projected into H1,
L1 and V1 with antenna responses and fractional arrival-time
delays. These are signals generated with an existing waveform
model, not new numerical-relativity simulations. The ensemble
is conditioned on the event data and that model; native amplitude
is the amplitude associated with each posterior sample.

Each signal was added to real off-source data at one of
128 distinct epochs per event, on a 4\,s grid outside
$\pm160\,\mathrm{s}$ of the event. These epochs differ from
the centres used to define the reference threshold; the 256\,s
filtering blocks can overlap. Event geometry was preserved
when transplanting the waveform. The 1,152 source-sample/epoch
combinations were reused for the amplitude and timing
diagnostics and are not independent new realizations at
each diagnostic setting. The main injections use nominal
instrumental calibration.

Recovery uses the same band-pass-only analysis choices:
the Butterworth design with prototype order 4 in 35--350\,Hz, forward--backward
application, 0.2\,s windows and maximization over
$|\tau|\leq35\,\mathrm{ms}$. The calibration implementation uses
a normalization fixed for each window, with decreasing overlap
as the lag changes, rather than recomputing Pearson normalization
over each overlap. The 200 off-source reference maxima fix the
threshold before evaluating recovery. An injection is counted
as recovered when Eq.~(\ref{eq:pval}), applied to its maximum,
gives $p\leq0.01$. No template bank is searched and no waveform
is fitted to the recovery window. Simulated waveforms specify
the added signal; they do not enter the correlation statistic
as templates.

\begin{table*}[htbp]
\centering
\caption{Complementary calibration at native posterior
amplitude. Each event uses 128 source-sample/real-data-epoch
combinations and the nominal local $p\leq0.01$ criterion.
H1$\times$L1 intervals are conditional 95\% Wilson intervals.
The three pairs share the same realizations.}
\label{tab:calibration}
\begin{ruledtabular}
\begin{tabular}{lrrrr}
Event & H1$\times$L1 &
95\% interval & H1$\times$V1 &
L1$\times$V1 \\
GW170814 & 89.1\% (114/128) & 82.5--93.4\% & 1.6\% & 4.7\% \\
GW170818 & 20.3\% (26/128) & 14.3--28.1\% & 0.0\% & 4.7\% \\
GW190412 & 97.7\% (125/128) & 93.3--99.2\% & 0.8\% & 0.0\% \\
GW190512\_180714 & 20.3\% (26/128) & 14.3--28.1\% & 0.8\% & 2.3\% \\
GW190521 & 99.2\% (127/128) & 95.7--99.9\% & 0.8\% & 0.0\% \\
GW190828\_063405 & 98.4\% (126/128) & 94.5--99.6\% & 0.0\% & 0.0\% \\
GW200129\_065458 & 100.0\% (128/128) & 97.1--100.0\% & 0.0\% & 0.0\% \\
GW200219\_094415 & 37.5\% (48/128) & 29.6--46.1\% & 0.0\% & 0.0\% \\
GW200224\_222234 & 100.0\% (128/128) & 97.1--100.0\% & 1.6\% & 0.8\% \\
\end{tabular}
\end{ruledtabular}
\end{table*}

\begin{figure*}[t]\centering
\includegraphics[width=.75\textwidth]{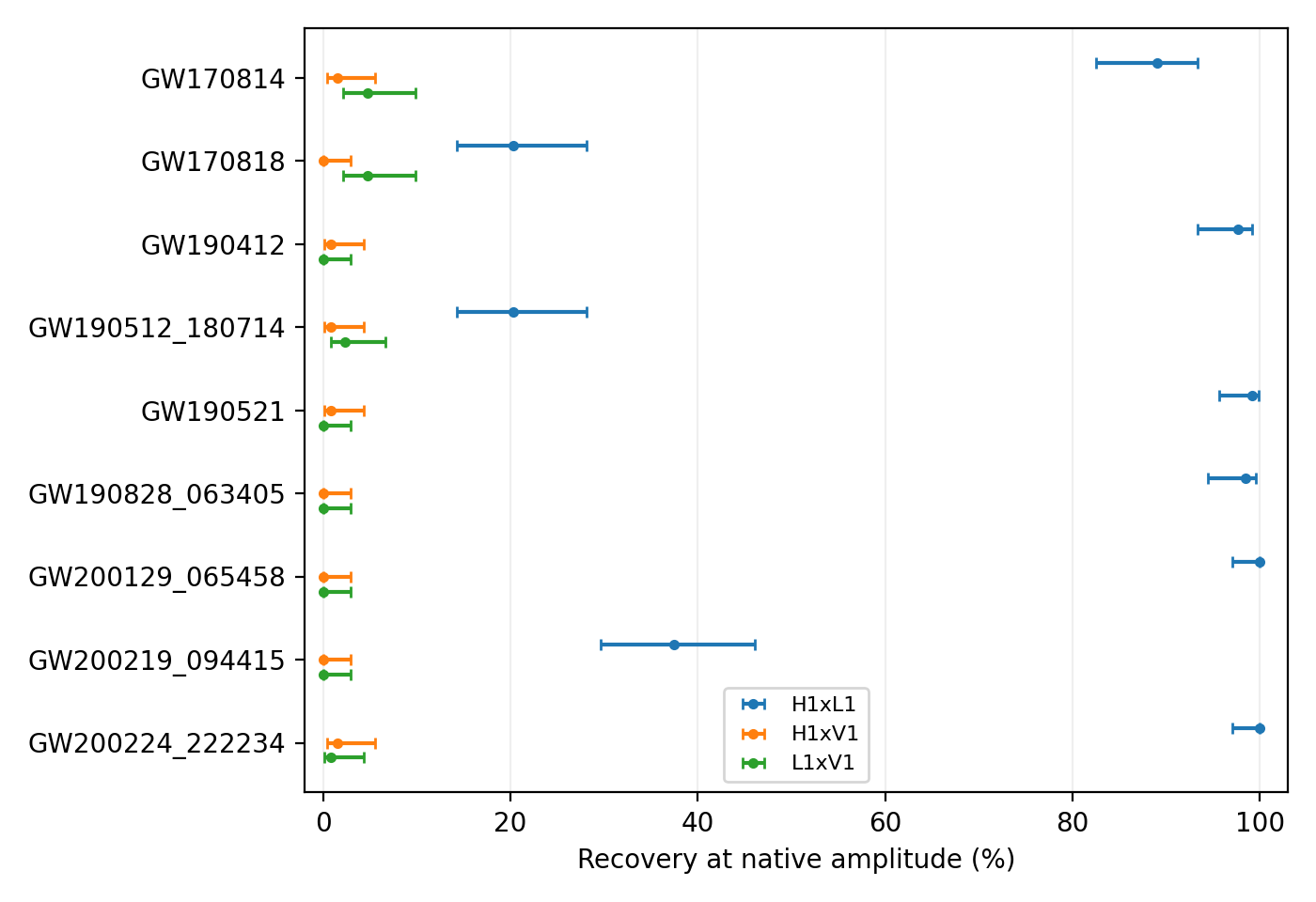}
\caption{Regenerated native-amplitude recovery fractions with conditional
95\% Wilson intervals. Each pair uses 128 trials per event; pairs share
realizations. These intervals omit dependencies and other uncertainties
discussed in the text.}\label{fig:recovery}\end{figure*}

Table~\ref{tab:calibration} and Fig.~\ref{fig:recovery} show substantial
event-dependent recovery in H1$\times$L1 and low recovery in
every Virgo pair. Without adding a signal, exceedance of the
same threshold ranges from 0\% to 3.9\%; the 1\% criterion
is therefore a nominal local level, not a certified false-alarm
frequency or a global significance for the 27 comparisons.
In Virgo, the median ratio of filtered native-signal standard
deviation to background standard deviation is 0.7\%--3.3\%.
The measured asymmetry concerns recovery by this statistic,
not a demonstration that the waveform is physically compatible
only with the LIGO instruments.

The principal prescription retains the offset between
the catalogue GPS epoch and each posterior geocentric time.
For GW190412, a diagnostic centred on the geocentric time
changes H1$\times$L1 recovery from 97.7\% to 60.9\%; the
median offset is approximately 82\,ms. This sensitivity makes
identical timing and implementation essential when comparing
event measurements with injection efficiencies. No significance
for an observed-versus-injected discrepancy is inferred here.
The conditional Wilson intervals do not include all effects
of overlapping data blocks, nonstationarity, threshold uncertainty
or waveform and instrumental-calibration uncertainty. The
calibration measures recovery of the specified binary ensemble,
not the recovery or physical origin of a stochastic background.

\onecolumngrid\clearpage\twocolumngrid\bibliography{refs}

\clearpage\onecolumngrid
\setcounter{figure}{0}\setcounter{table}{0}
\renewcommand{\thefigure}{S\arabic{figure}}
\renewcommand{\thetable}{S\arabic{table}}
\providecommand{\theHfigure}{}\renewcommand{\theHfigure}{supp.\arabic{figure}}
\providecommand{\theHtable}{}\renewcommand{\theHtable}{supp.\arabic{table}}

\begin{center}
{\Large\bfseries Supplemental Material\par}\medskip
{\large Inter-detector correlations in LIGO--Virgo strain:\\
an exploratory study and possible astrophysical implications\par}\medskip
S. C. Ulhoa, B. C. C. Carneiro, F. L. Carneiro, and L. V. A. Cunha
\end{center}
\section*{Contents and conventions}
Figures~\ref{fig:S1}--\ref{fig:S20} present the complete event curves
and reference distributions. Each figure groups the H1$\times$L1,
H1$\times$V1 and L1$\times$V1 pairs, from top to bottom.
Table~\ref{tab:S1} collects the additional timing, quality and spectral
sensitivity comparisons for all 27 primary pairs. The binary-injection
calibration is in Appendix A of the main article.

Band-pass-only analysis, corrected catalog epochs, 200 reference windows and a common $\pm35$ ms lag search. Values in plot titles are local p-values. The nine-event primary family contains 27 event--pair comparisons. GW191127 contributes three complementary event--pair comparisons, one for each detector pair, and is not included in the primary family. The combined primary and complementary sets contain 30 comparisons. Neither family yields an individual Holm rejection at 5\% with this background resolution. Positive lag denotes the first detector delayed relative to the second. Correlation maxima are not guaranteed propagation-delay estimates.
\clearpage
\section*{GW170814 -- Correlation curves}
Figure~\ref{fig:S1} shows the correlation curves.
\par\noindent\begin{minipage}{\linewidth}\centering
\includegraphics[width=.94\linewidth,height=.235\textheight,keepaspectratio]{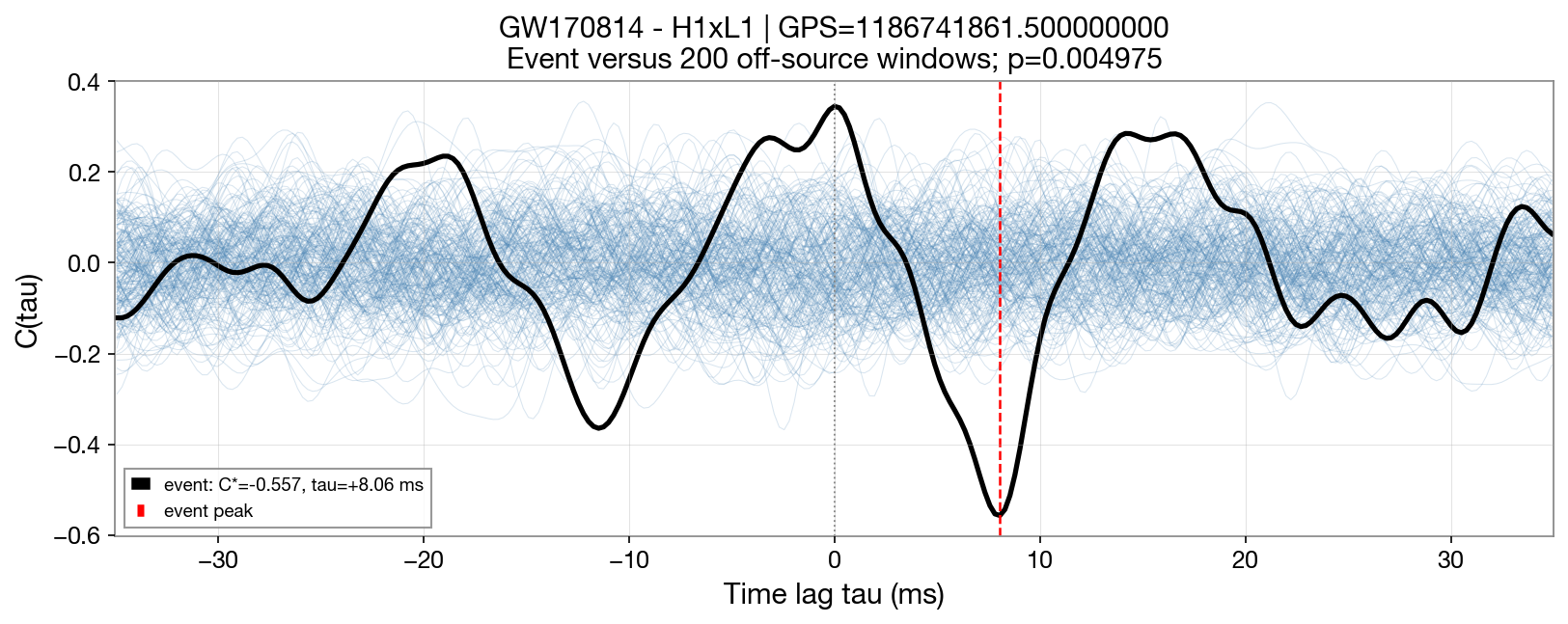}

\includegraphics[width=.94\linewidth,height=.235\textheight,keepaspectratio]{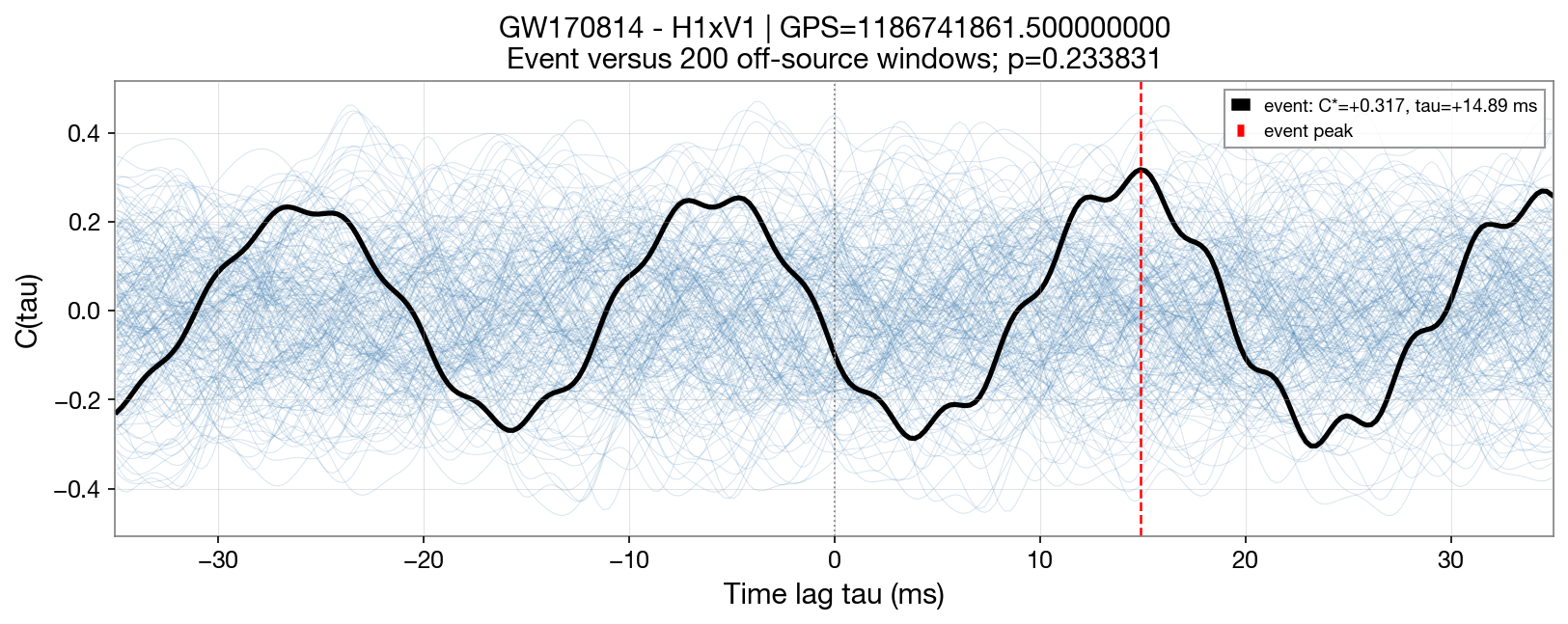}

\includegraphics[width=.94\linewidth,height=.235\textheight,keepaspectratio]{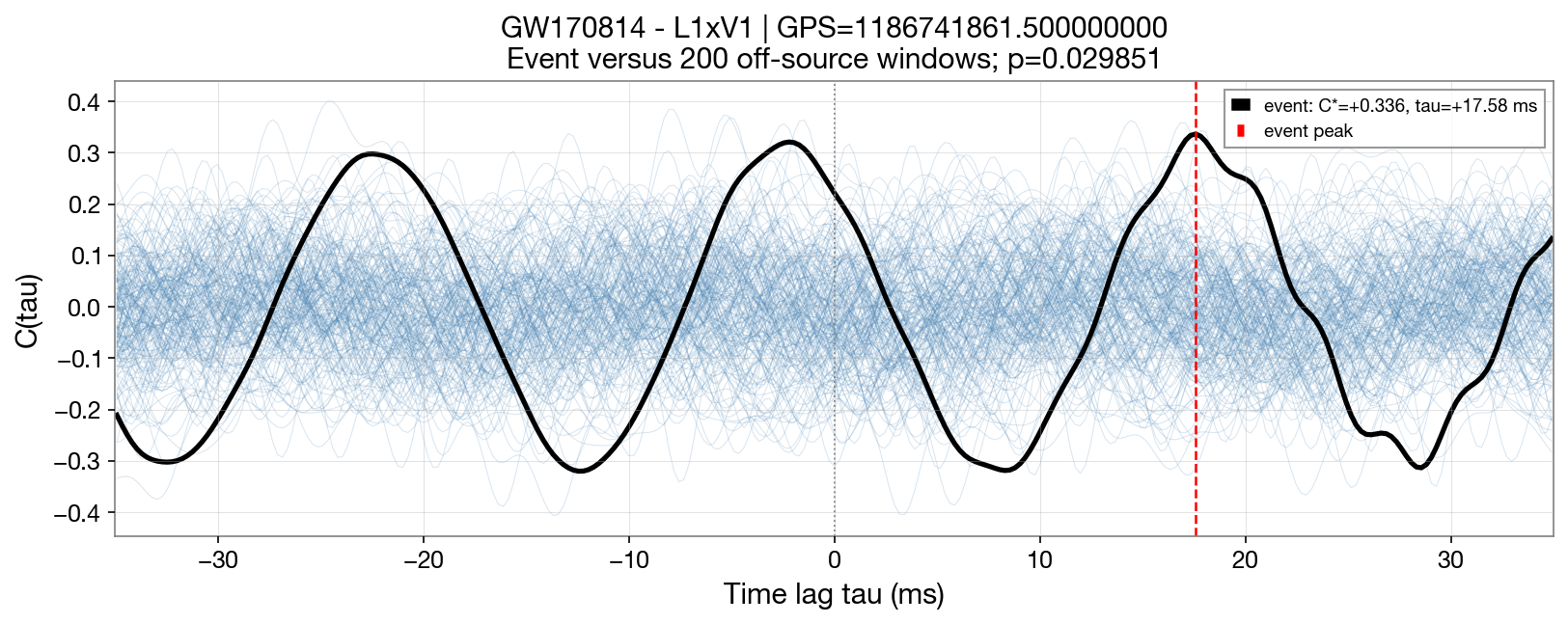}

\captionof{figure}{GW170814: event correlations (black) and 200 reference curves (blue). Positive lag means the first detector is delayed relative to the second. All p-values are local.}\label{fig:S1}
\end{minipage}\par
\clearpage
\section*{GW170814 -- Peak magnitudes and lags}
Figure~\ref{fig:S2} shows the reference distributions.
\par\noindent\begin{minipage}{\linewidth}\centering
\includegraphics[width=.94\linewidth,height=.235\textheight,keepaspectratio]{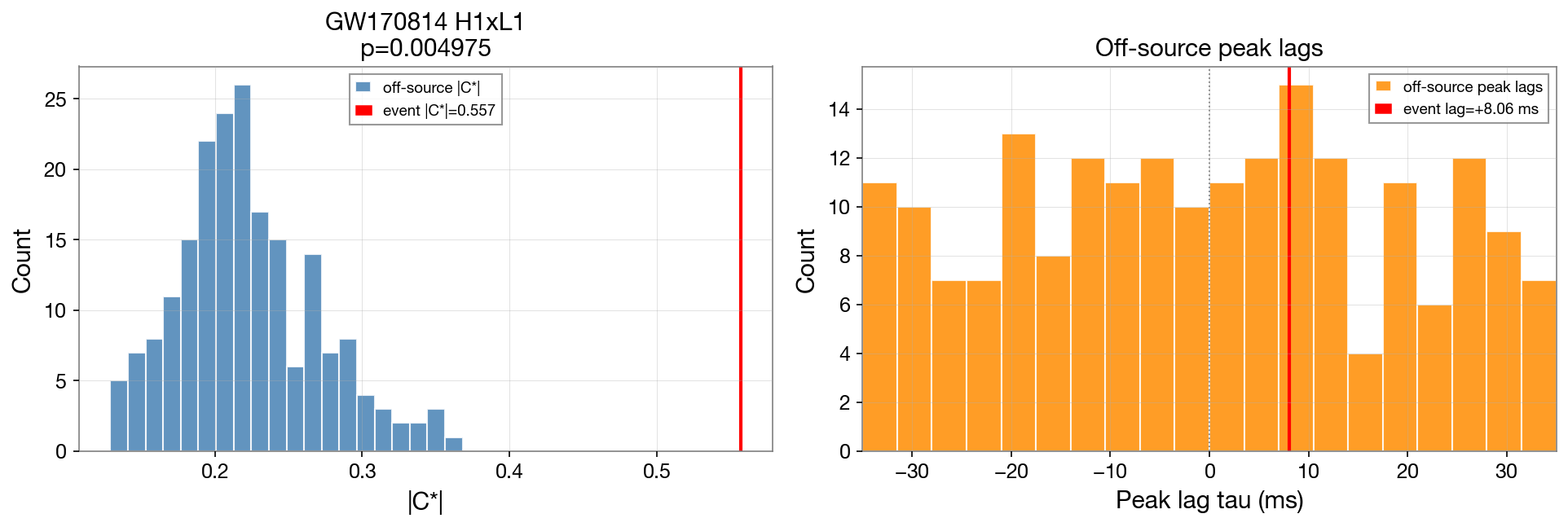}

\includegraphics[width=.94\linewidth,height=.235\textheight,keepaspectratio]{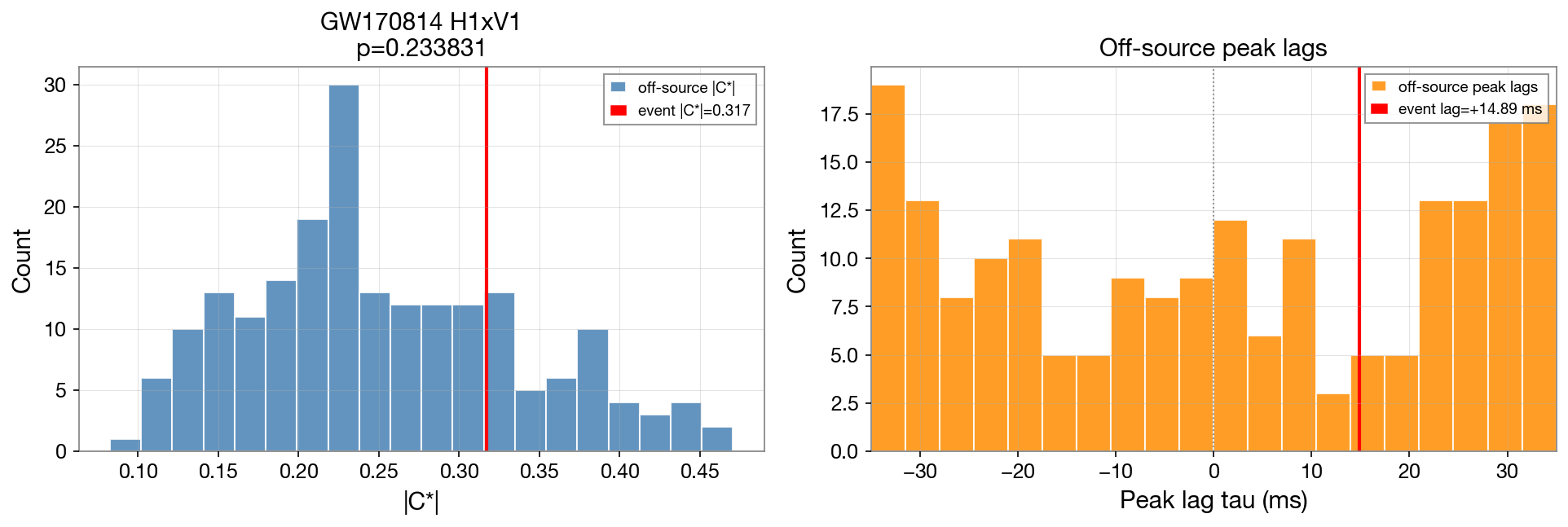}

\includegraphics[width=.94\linewidth,height=.235\textheight,keepaspectratio]{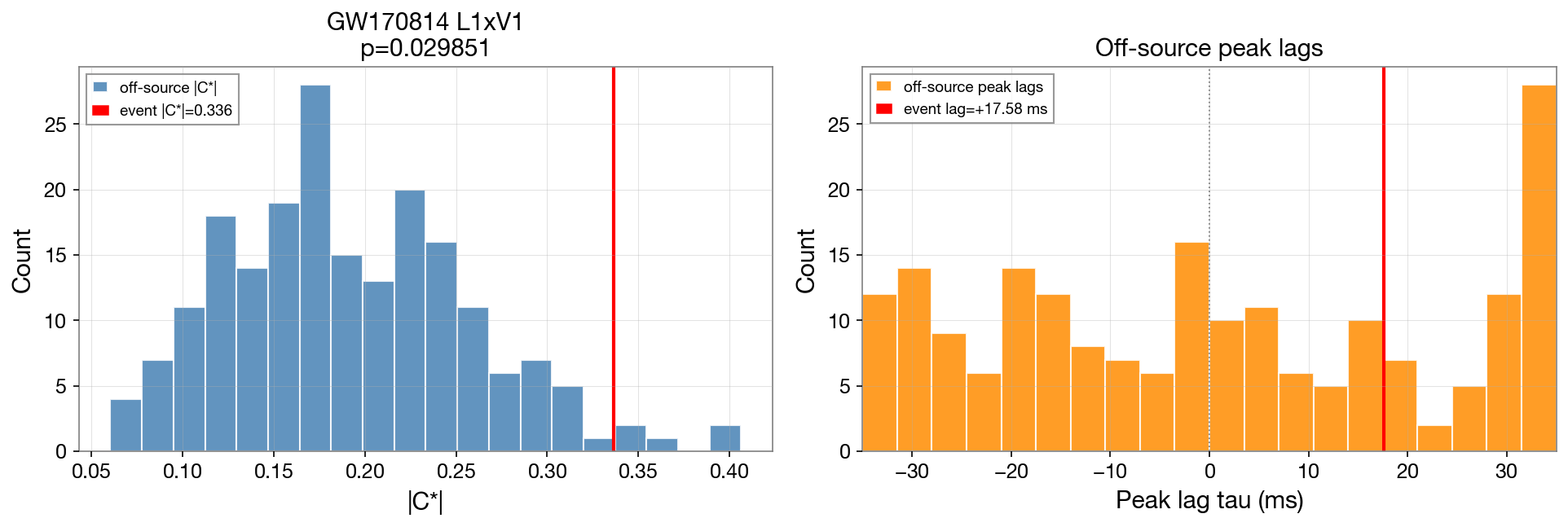}

\captionof{figure}{GW170814: distributions of reference peak magnitudes (left) and maximizing lags (right); the event values are marked in red. All p-values are local.}\label{fig:S2}
\end{minipage}\par
\clearpage
\section*{GW170818 -- Correlation curves}
Figure~\ref{fig:S3} shows the correlation curves.
\par\noindent\begin{minipage}{\linewidth}\centering
\includegraphics[width=.94\linewidth,height=.235\textheight,keepaspectratio]{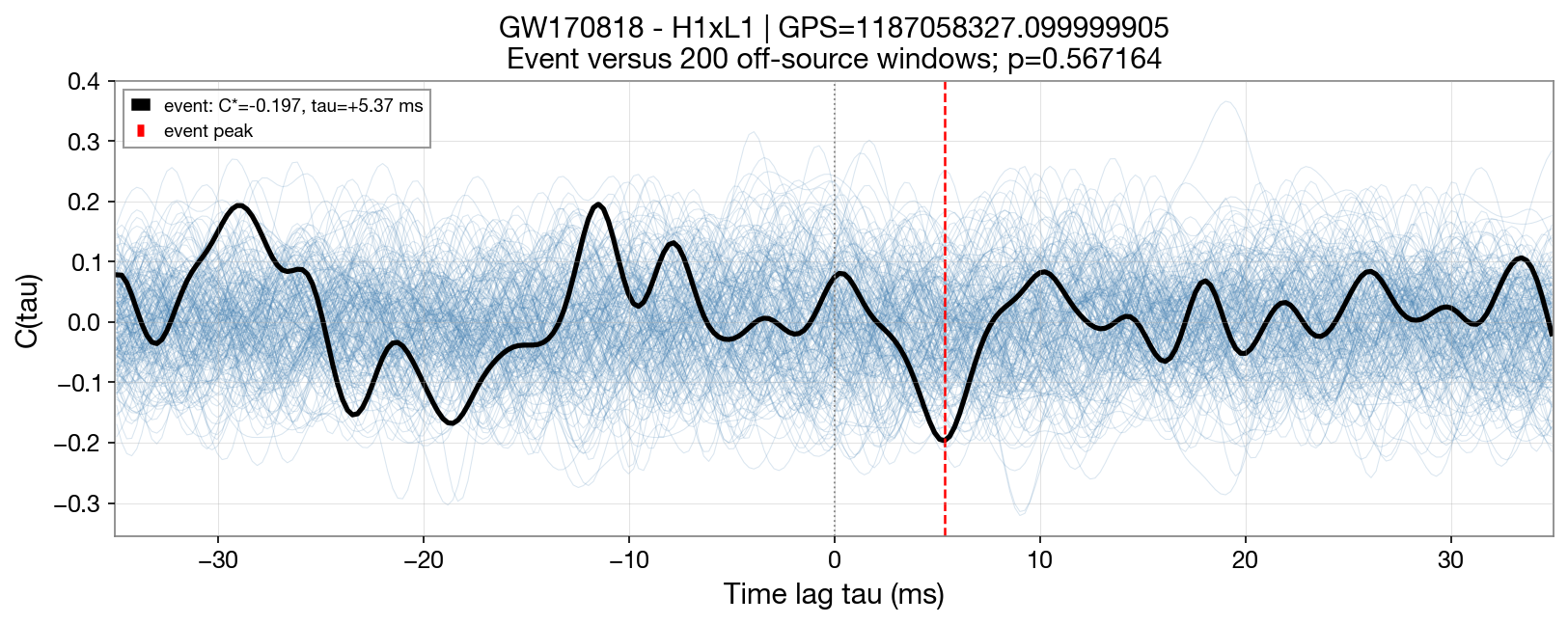}

\includegraphics[width=.94\linewidth,height=.235\textheight,keepaspectratio]{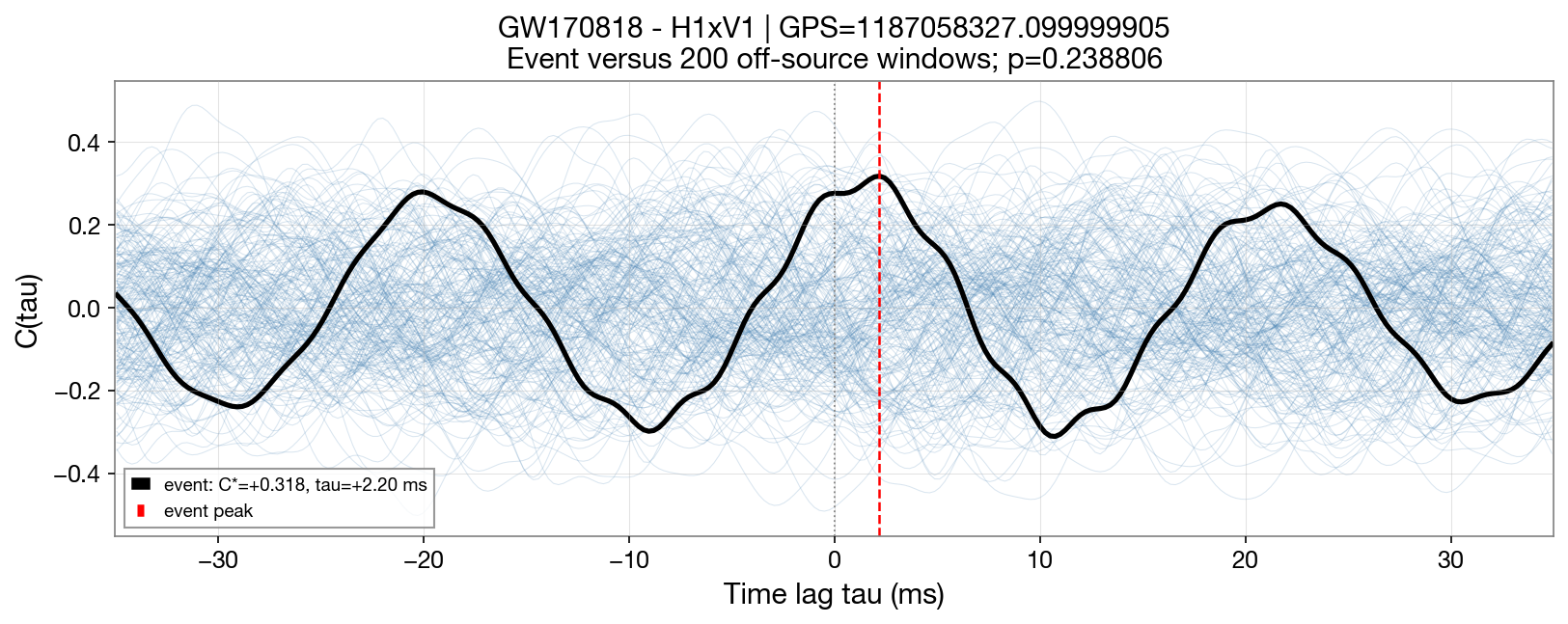}

\includegraphics[width=.94\linewidth,height=.235\textheight,keepaspectratio]{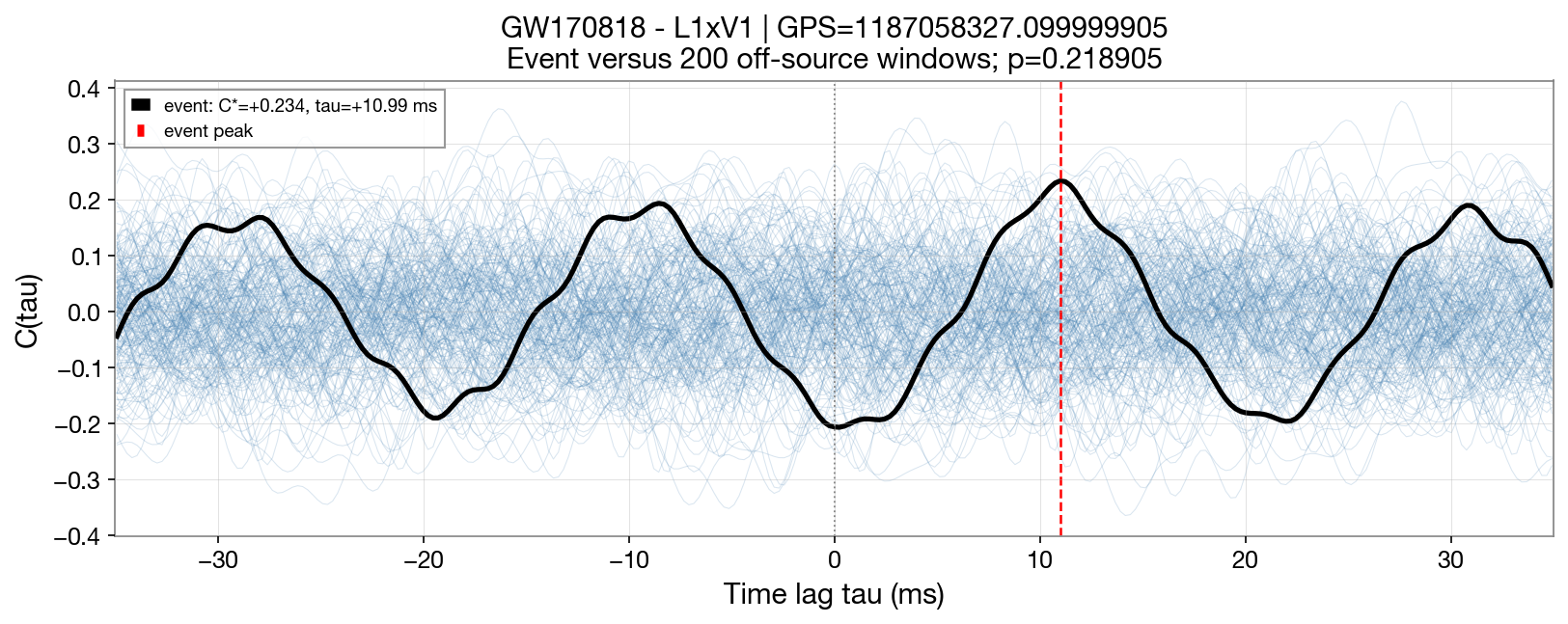}

\captionof{figure}{GW170818: event correlations (black) and 200 reference curves (blue). Positive lag means the first detector is delayed relative to the second. All p-values are local.}\label{fig:S3}
\end{minipage}\par
\clearpage
\section*{GW170818 -- Peak magnitudes and lags}
Figure~\ref{fig:S4} shows the reference distributions.
\par\noindent\begin{minipage}{\linewidth}\centering
\includegraphics[width=.94\linewidth,height=.235\textheight,keepaspectratio]{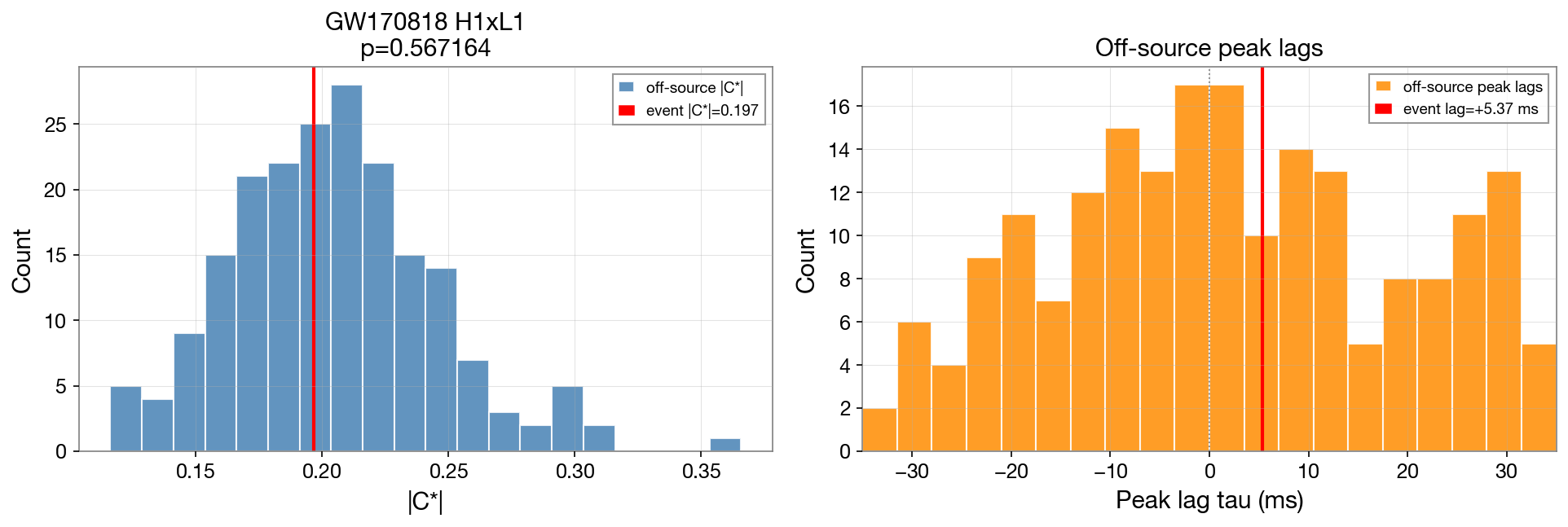}

\includegraphics[width=.94\linewidth,height=.235\textheight,keepaspectratio]{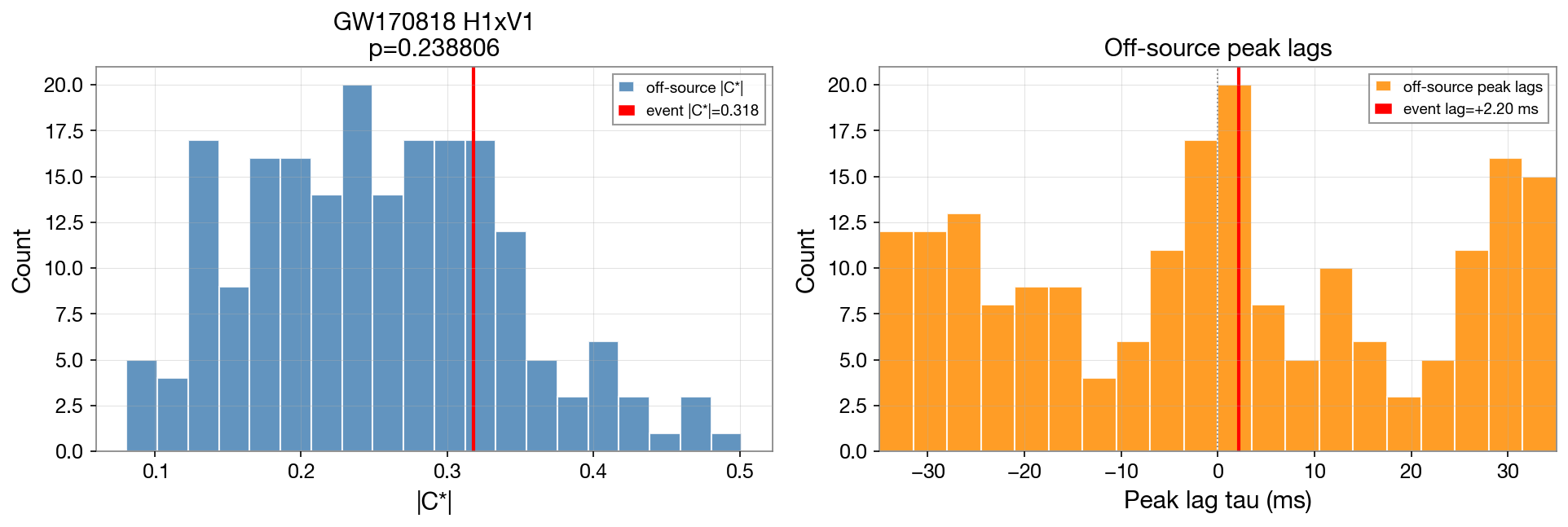}

\includegraphics[width=.94\linewidth,height=.235\textheight,keepaspectratio]{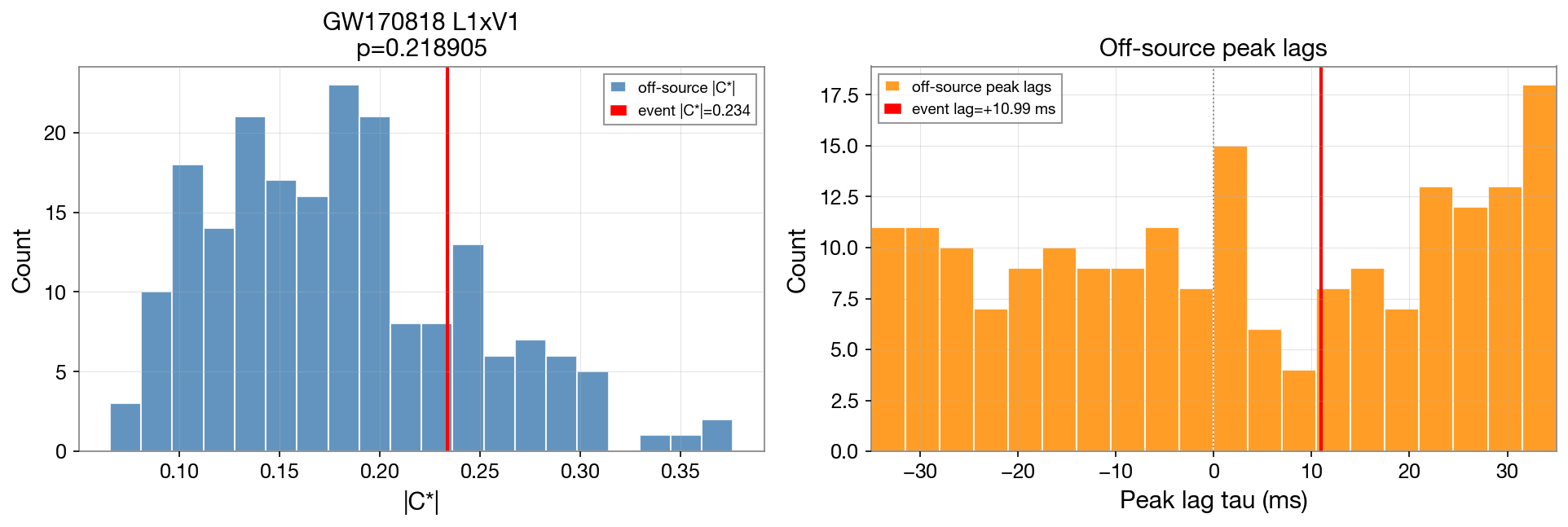}

\captionof{figure}{GW170818: distributions of reference peak magnitudes (left) and maximizing lags (right); the event values are marked in red. All p-values are local.}\label{fig:S4}
\end{minipage}\par
\clearpage
\section*{GW190412 -- Correlation curves}
Figure~\ref{fig:S5} shows the correlation curves.
\par\noindent\begin{minipage}{\linewidth}\centering
\includegraphics[width=.94\linewidth,height=.235\textheight,keepaspectratio]{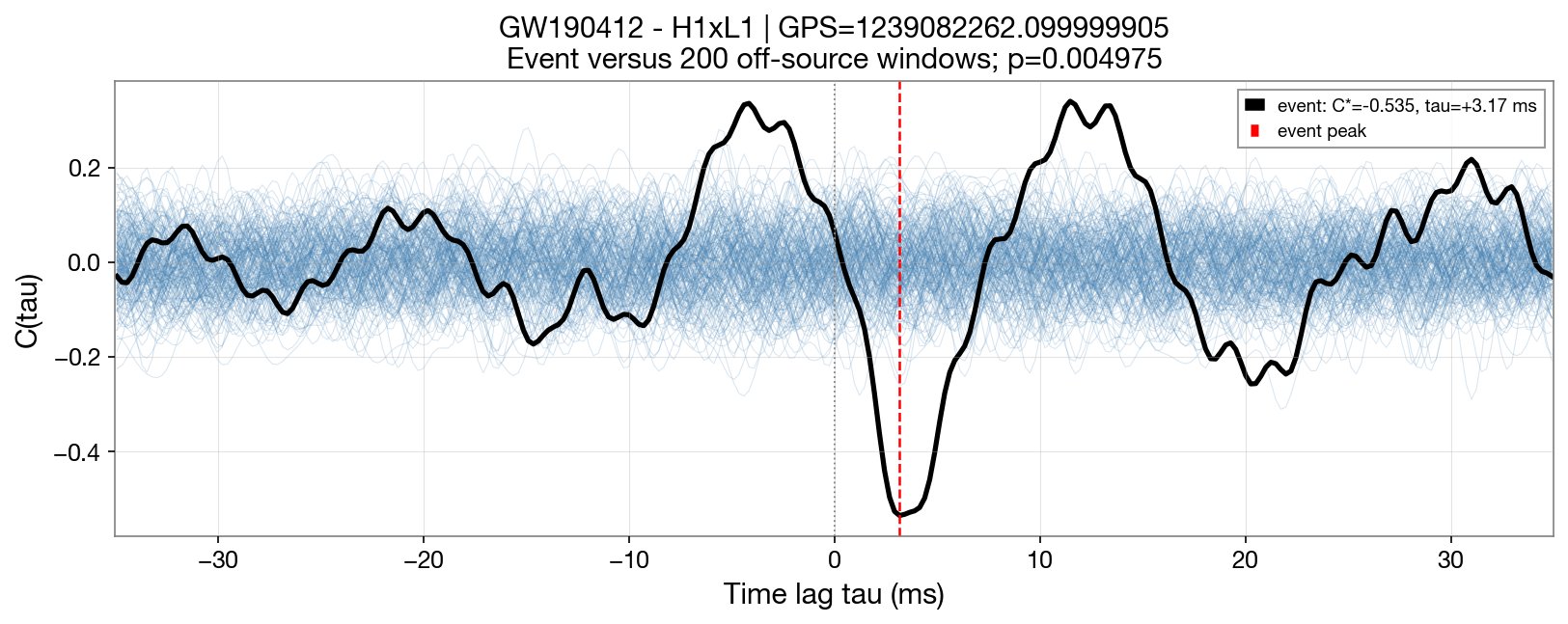}

\includegraphics[width=.94\linewidth,height=.235\textheight,keepaspectratio]{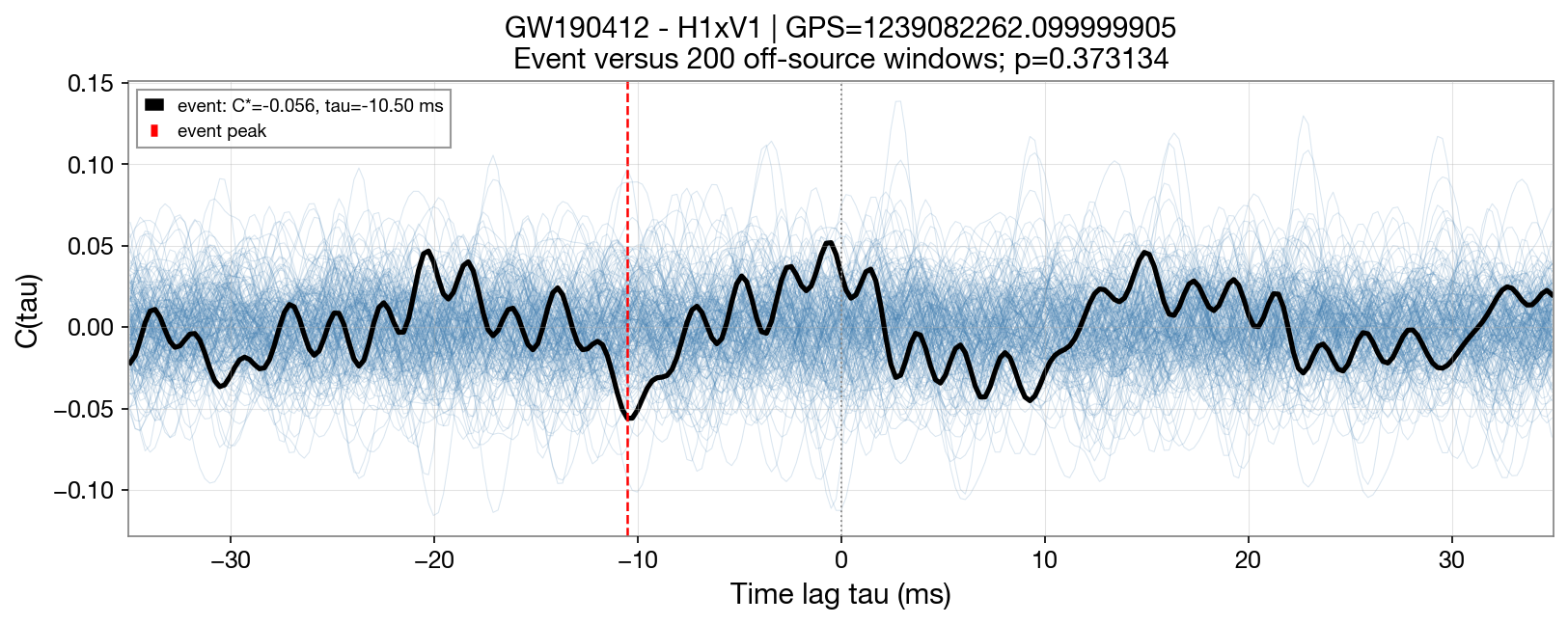}

\includegraphics[width=.94\linewidth,height=.235\textheight,keepaspectratio]{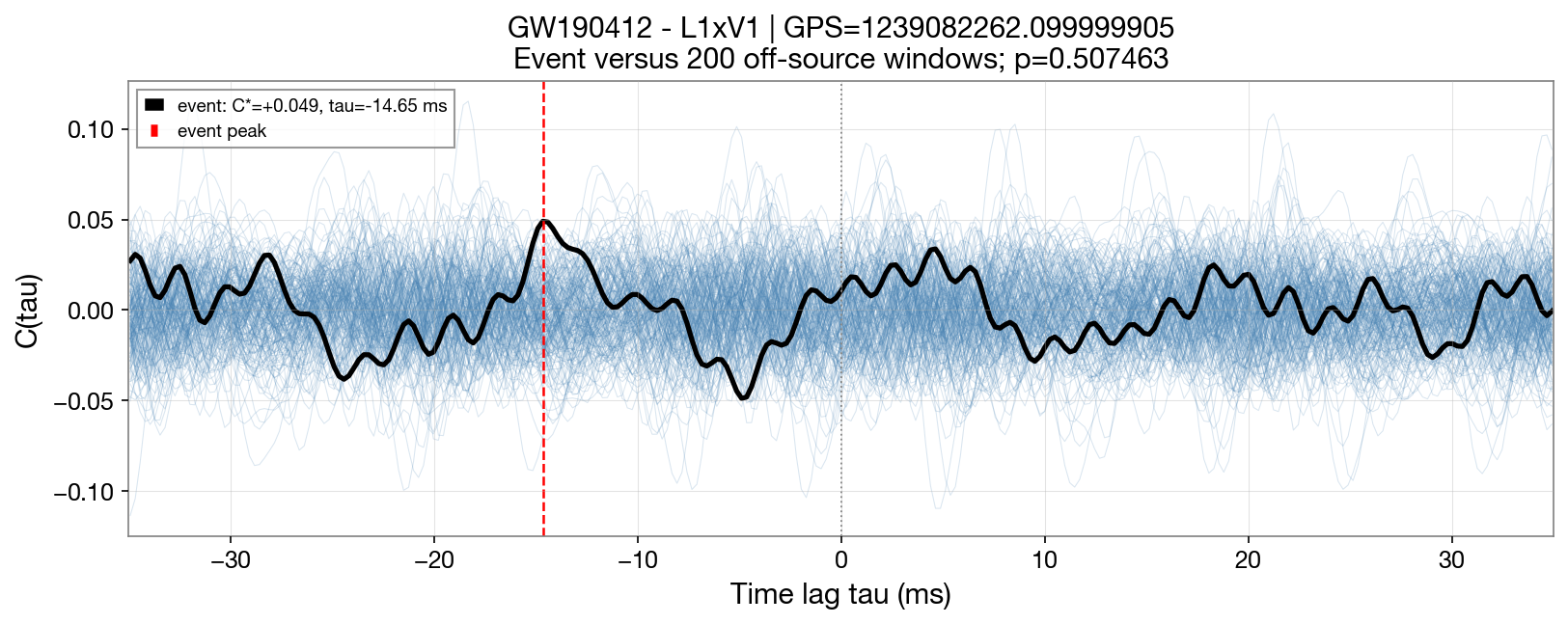}

\captionof{figure}{GW190412: event correlations (black) and 200 reference curves (blue). Positive lag means the first detector is delayed relative to the second. All p-values are local.}\label{fig:S5}
\end{minipage}\par
\clearpage
\section*{GW190412 -- Peak magnitudes and lags}
Figure~\ref{fig:S6} shows the reference distributions.
\par\noindent\begin{minipage}{\linewidth}\centering
\includegraphics[width=.94\linewidth,height=.235\textheight,keepaspectratio]{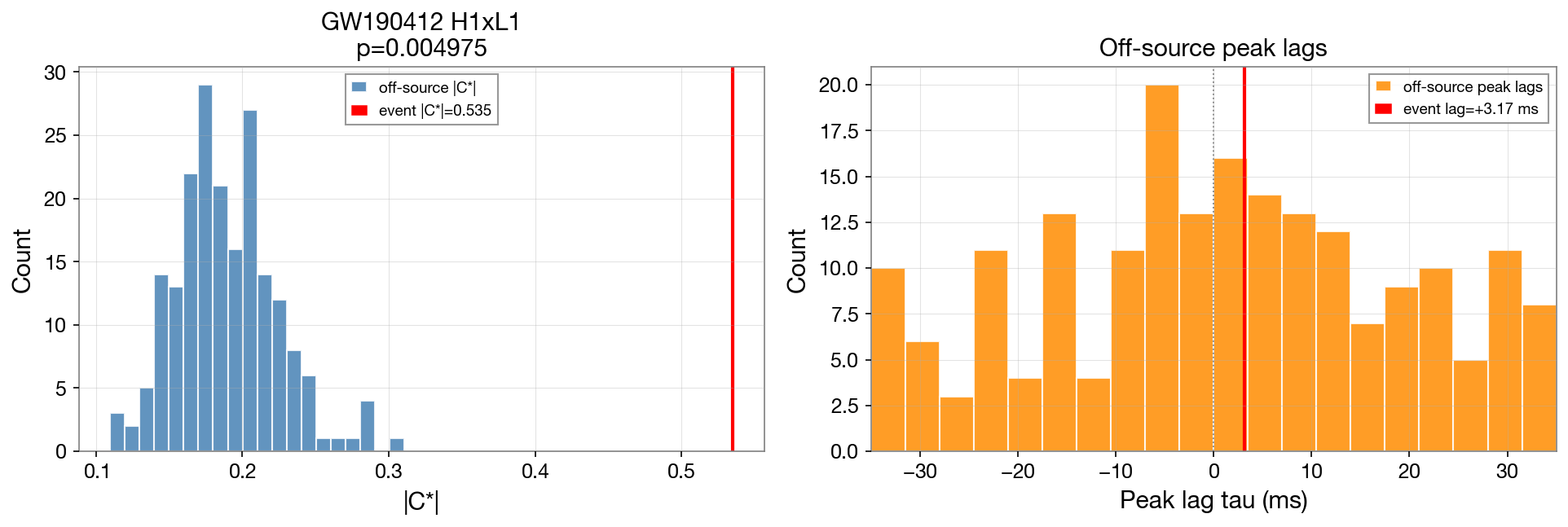}

\includegraphics[width=.94\linewidth,height=.235\textheight,keepaspectratio]{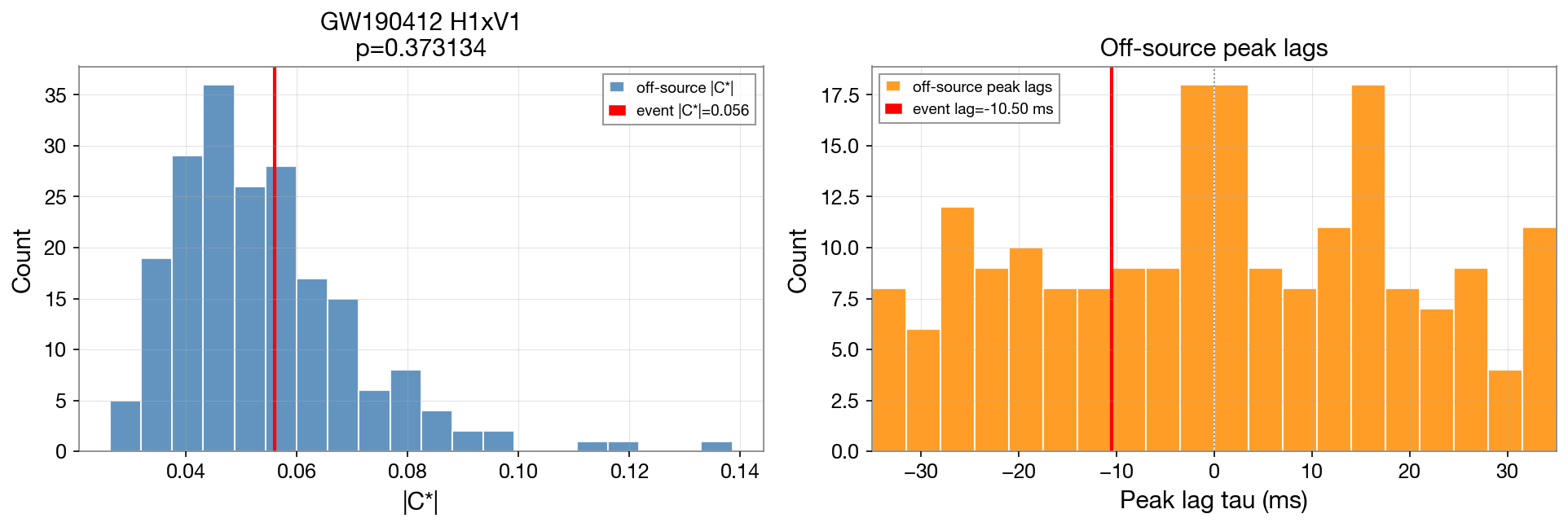}

\includegraphics[width=.94\linewidth,height=.235\textheight,keepaspectratio]{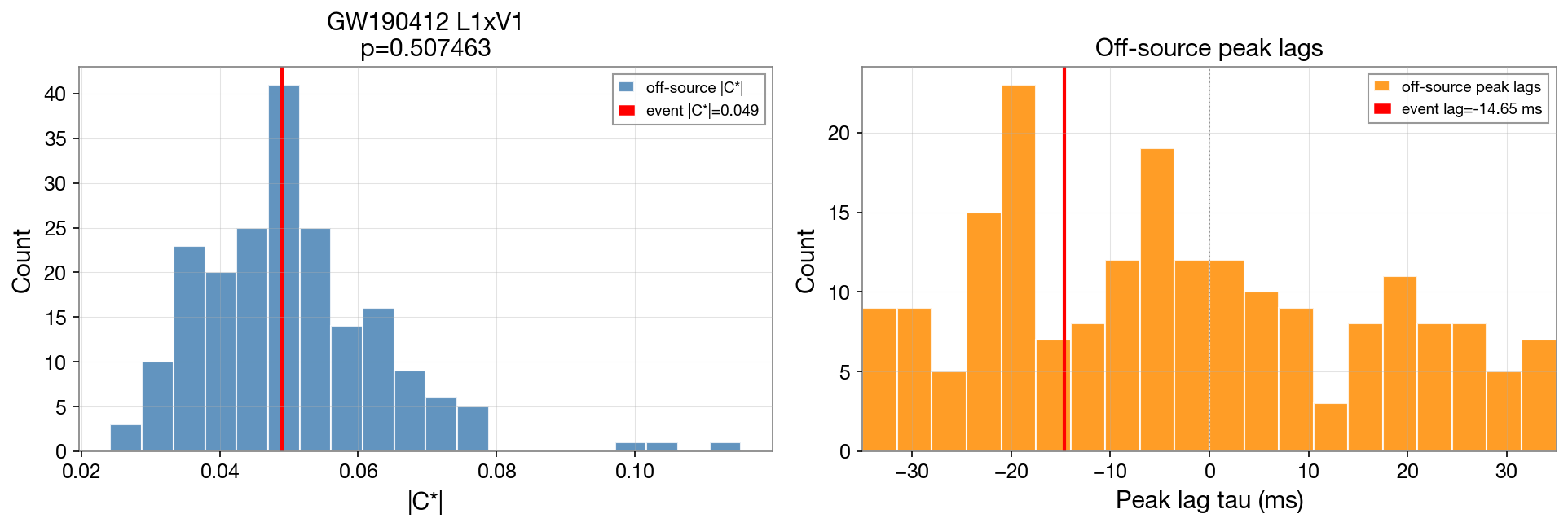}

\captionof{figure}{GW190412: distributions of reference peak magnitudes (left) and maximizing lags (right); the event values are marked in red. All p-values are local.}\label{fig:S6}
\end{minipage}\par
\clearpage
\section*{GW190512\_180714 -- Correlation curves}
Figure~\ref{fig:S7} shows the correlation curves.
\par\noindent\begin{minipage}{\linewidth}\centering
\includegraphics[width=.94\linewidth,height=.235\textheight,keepaspectratio]{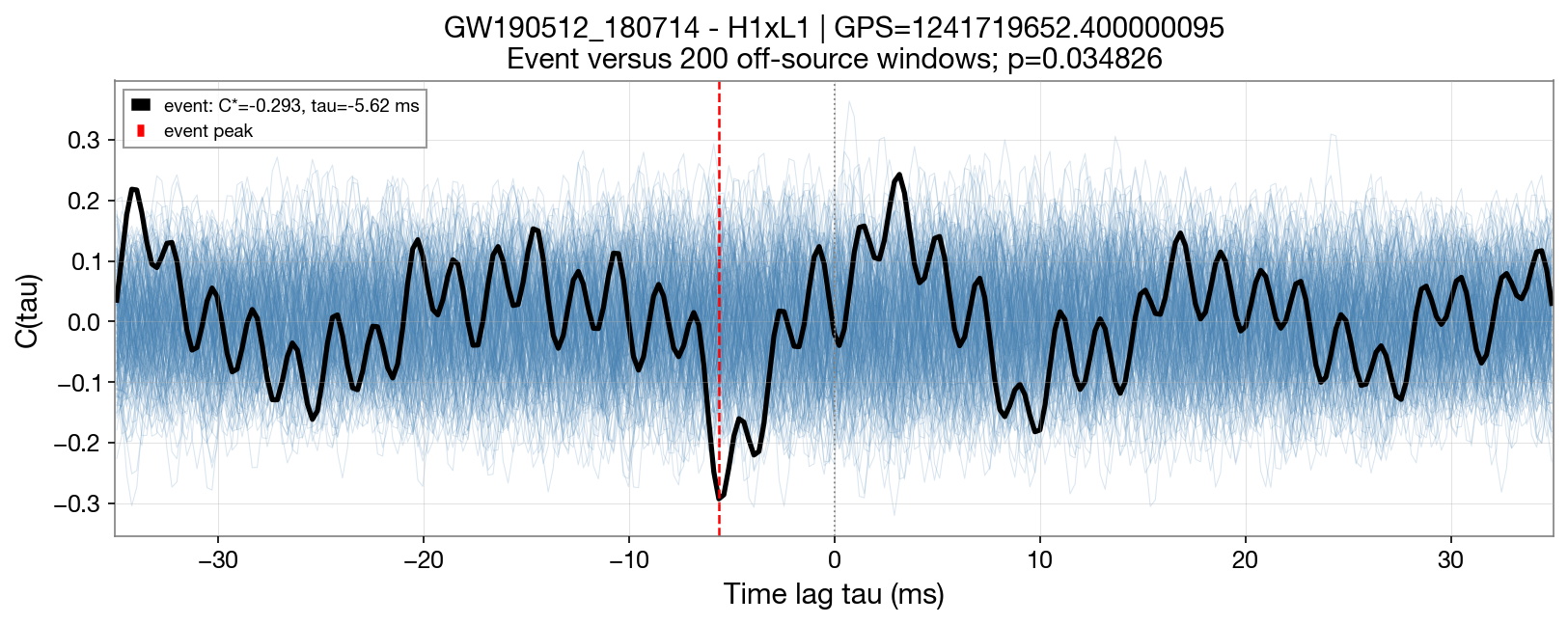}

\includegraphics[width=.94\linewidth,height=.235\textheight,keepaspectratio]{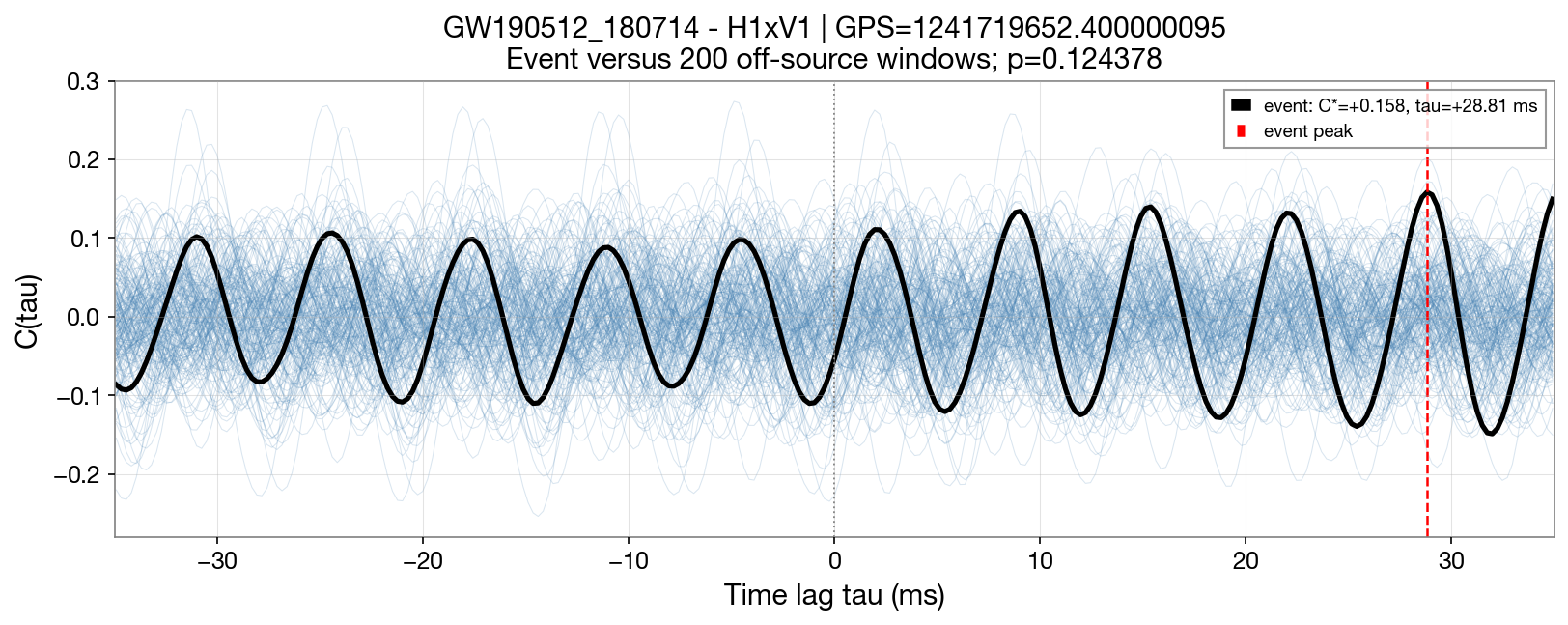}

\includegraphics[width=.94\linewidth,height=.235\textheight,keepaspectratio]{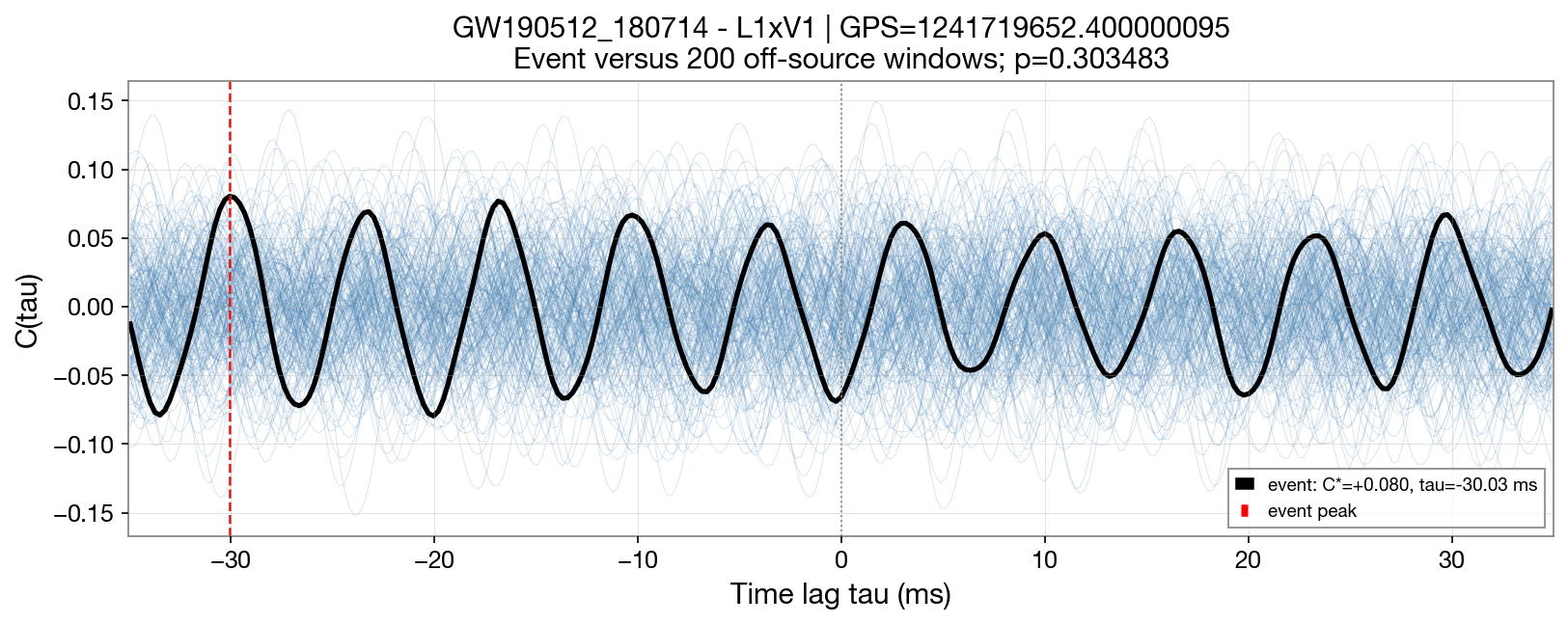}

\captionof{figure}{GW190512\_180714: event correlations (black) and 200 reference curves (blue). Positive lag means the first detector is delayed relative to the second. All p-values are local.}\label{fig:S7}
\end{minipage}\par
\clearpage
\section*{GW190512\_180714 -- Peak magnitudes and lags}
Figure~\ref{fig:S8} shows the reference distributions.
\par\noindent\begin{minipage}{\linewidth}\centering
\includegraphics[width=.94\linewidth,height=.235\textheight,keepaspectratio]{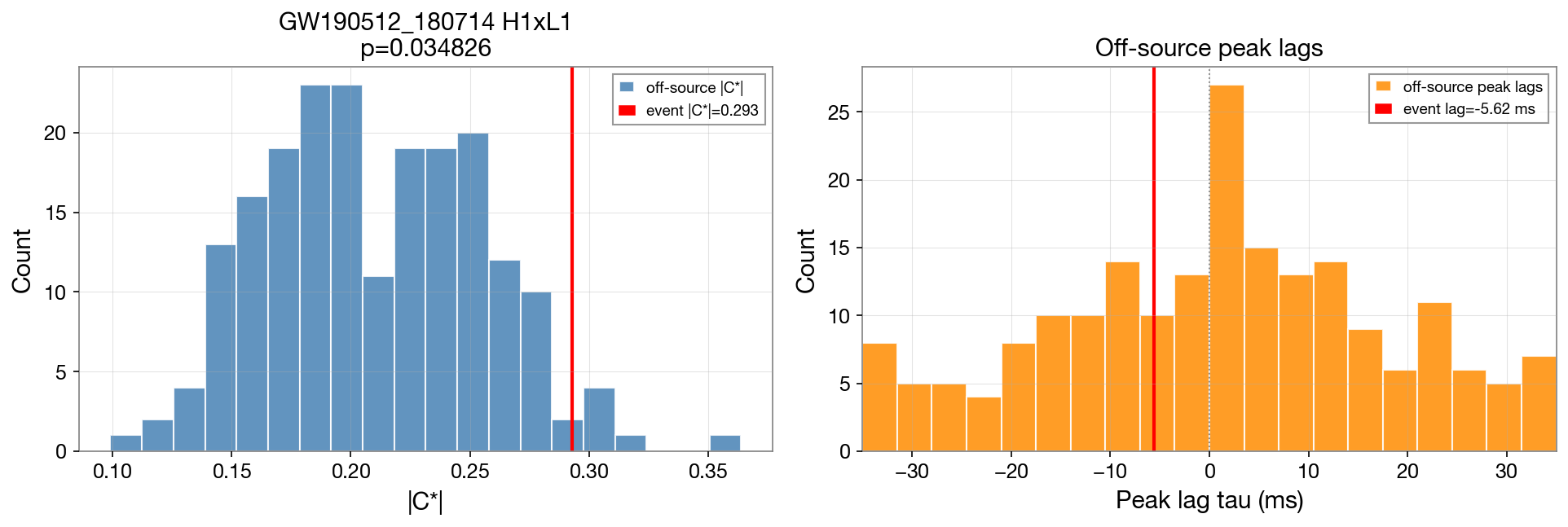}

\includegraphics[width=.94\linewidth,height=.235\textheight,keepaspectratio]{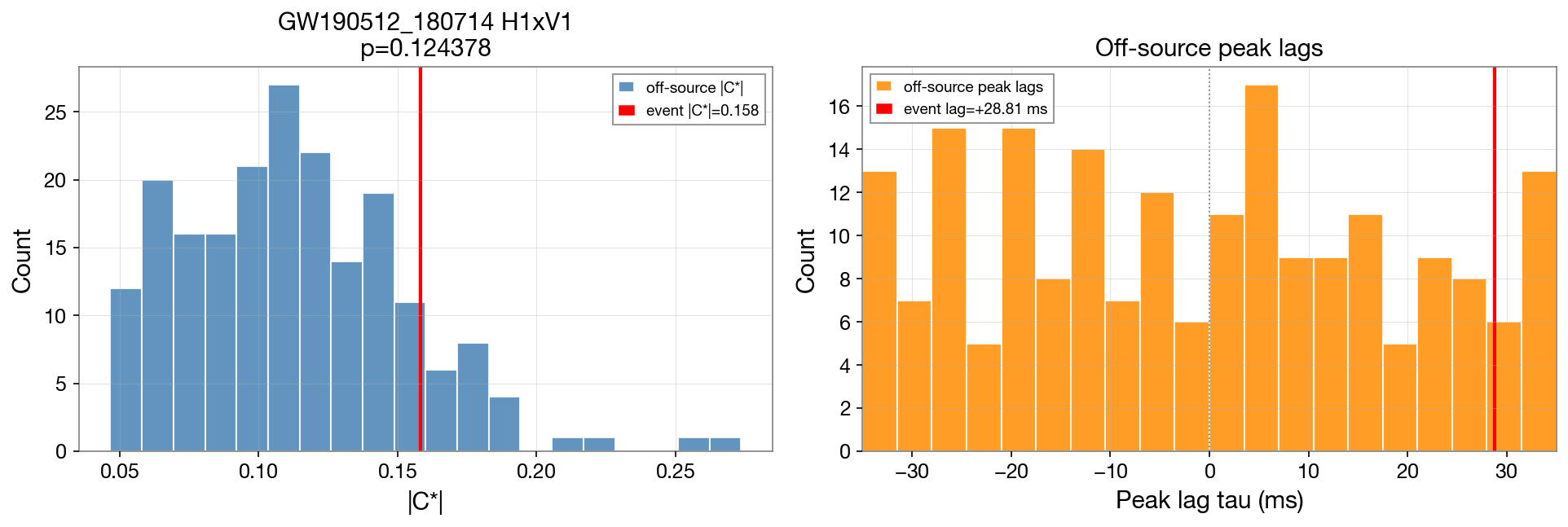}

\includegraphics[width=.94\linewidth,height=.235\textheight,keepaspectratio]{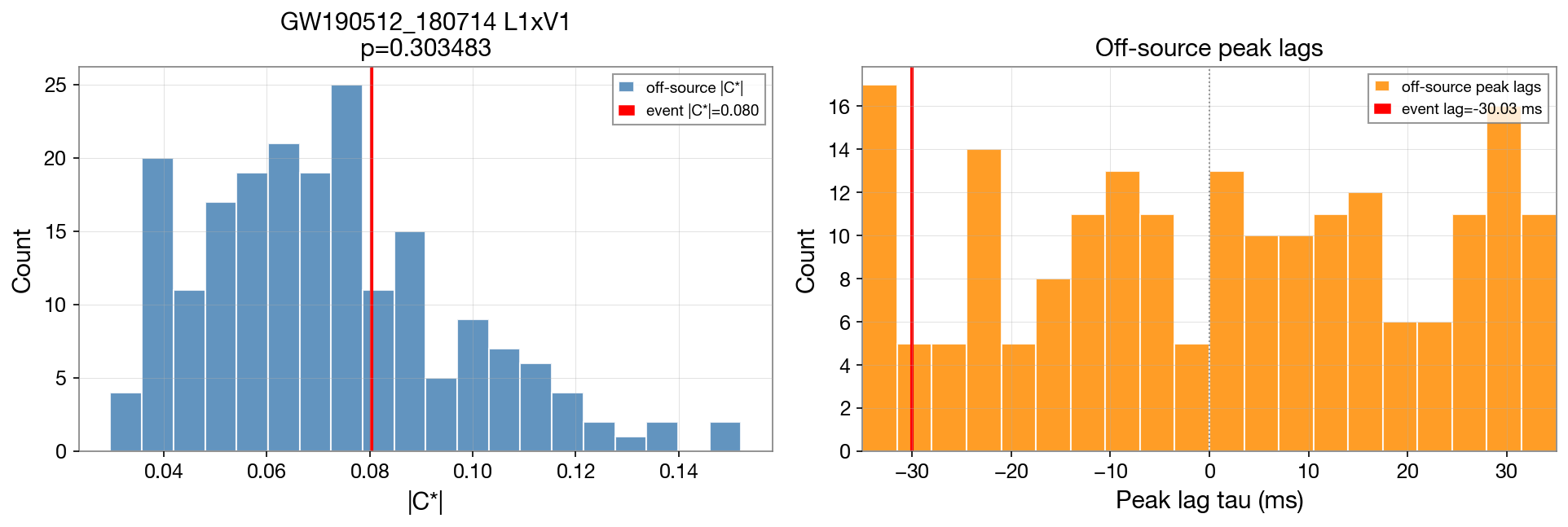}

\captionof{figure}{GW190512\_180714: distributions of reference peak magnitudes (left) and maximizing lags (right); the event values are marked in red. All p-values are local.}\label{fig:S8}
\end{minipage}\par
\clearpage
\section*{GW190521 -- Correlation curves}
Figure~\ref{fig:S9} shows the correlation curves.
\par\noindent\begin{minipage}{\linewidth}\centering
\includegraphics[width=.94\linewidth,height=.235\textheight,keepaspectratio]{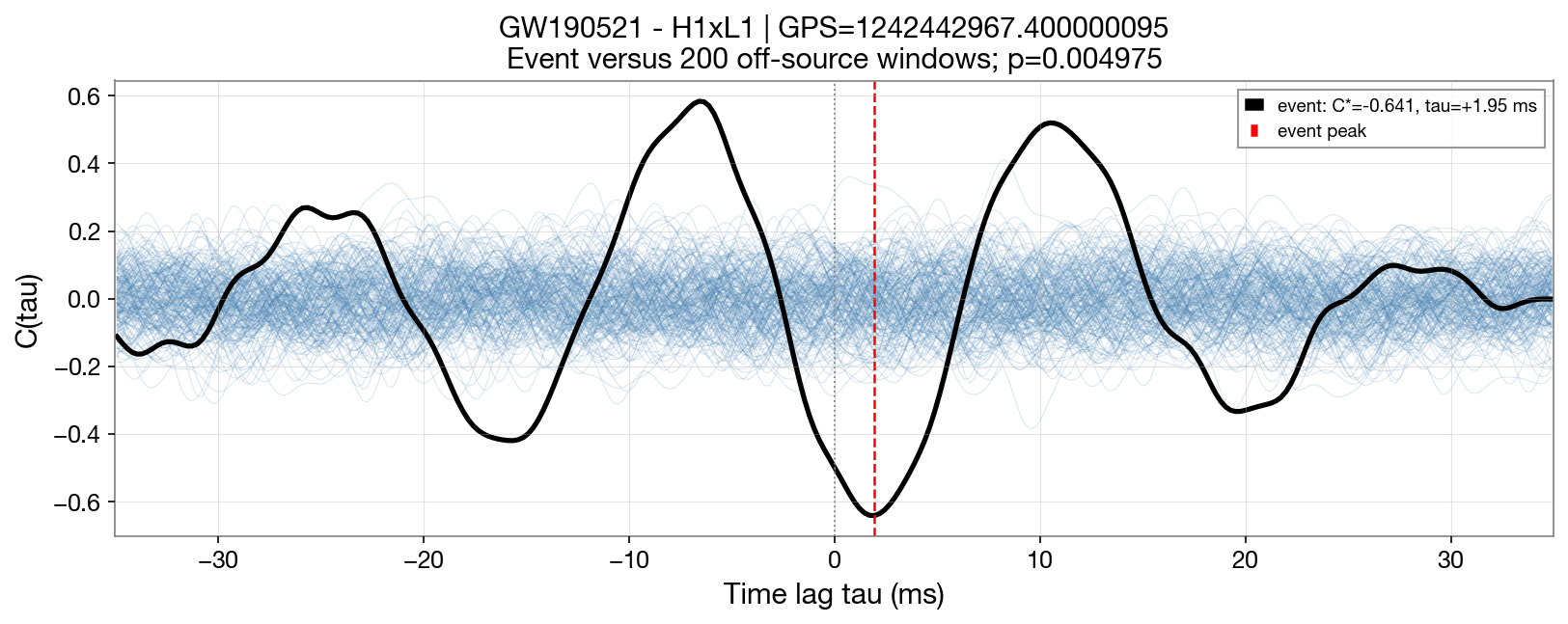}

\includegraphics[width=.94\linewidth,height=.235\textheight,keepaspectratio]{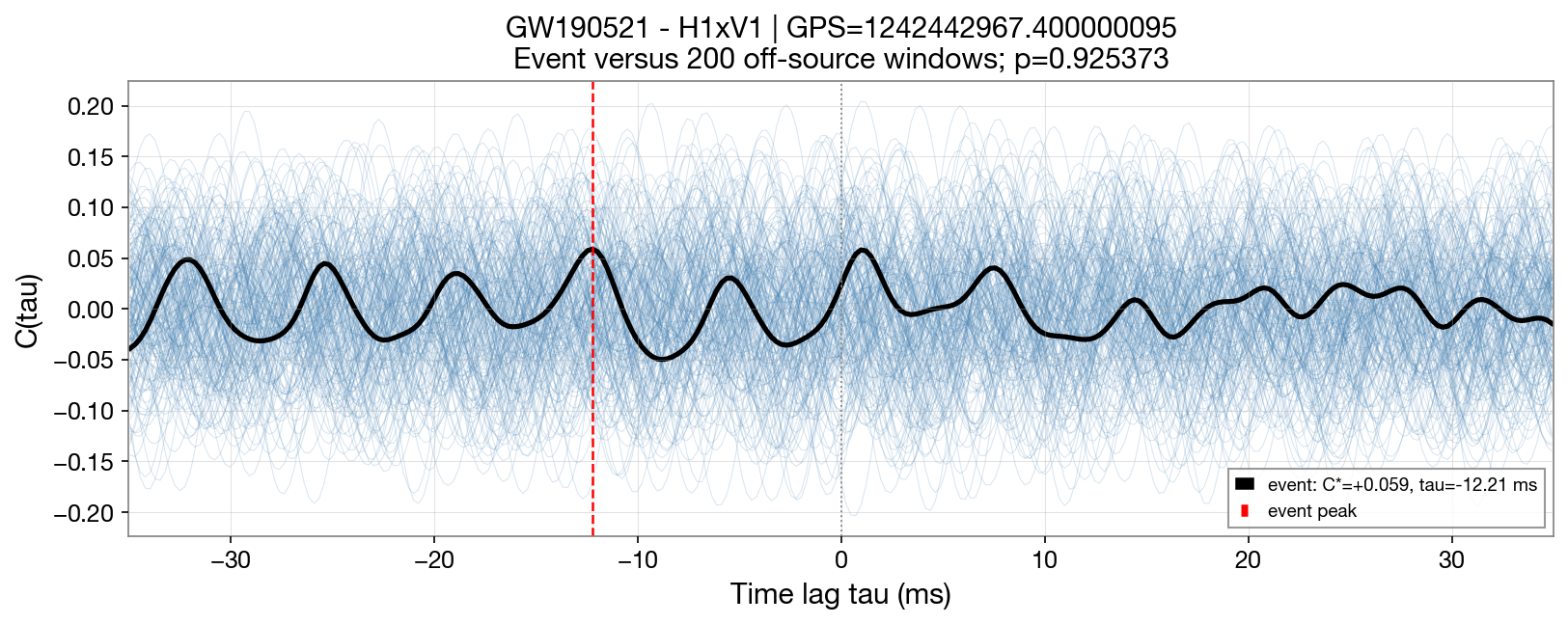}

\includegraphics[width=.94\linewidth,height=.235\textheight,keepaspectratio]{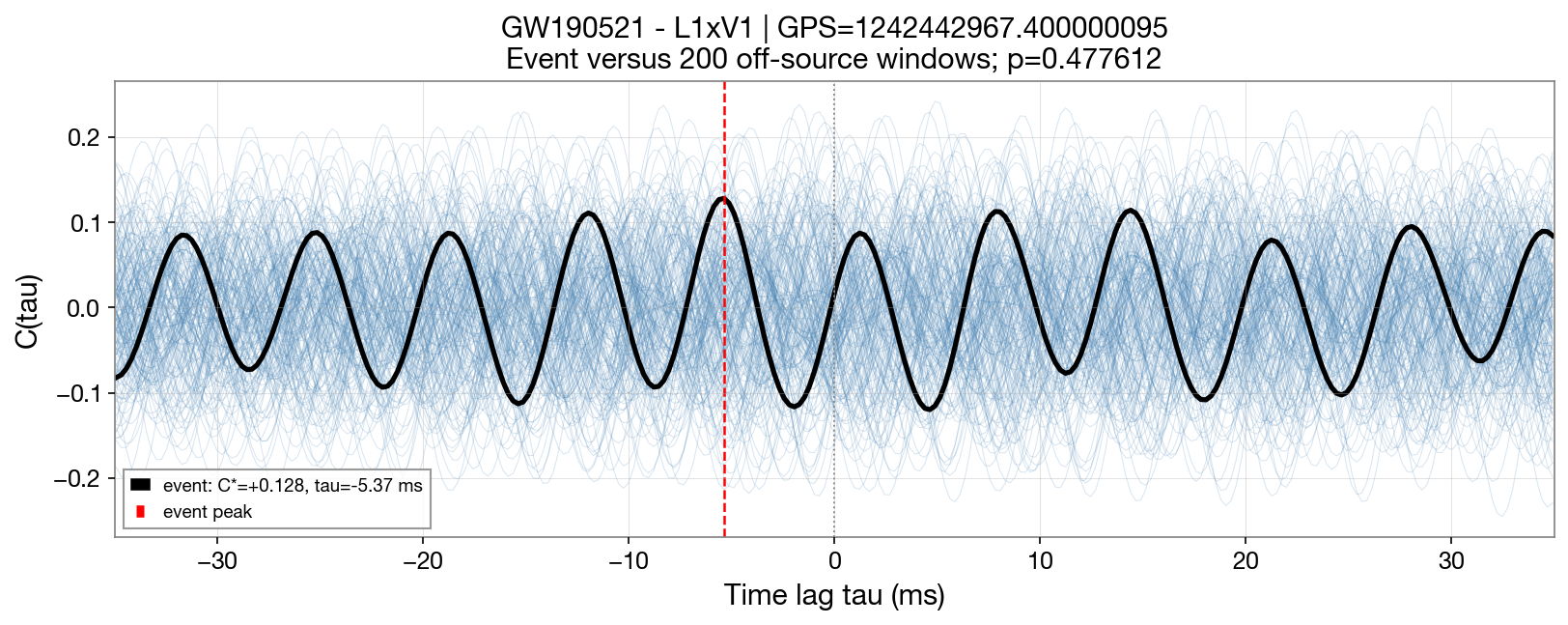}

\captionof{figure}{GW190521: event correlations (black) and 200 reference curves (blue). Positive lag means the first detector is delayed relative to the second. All p-values are local.}\label{fig:S9}
\end{minipage}\par
\clearpage
\section*{GW190521 -- Peak magnitudes and lags}
Figure~\ref{fig:S10} shows the reference distributions.
\par\noindent\begin{minipage}{\linewidth}\centering
\includegraphics[width=.94\linewidth,height=.235\textheight,keepaspectratio]{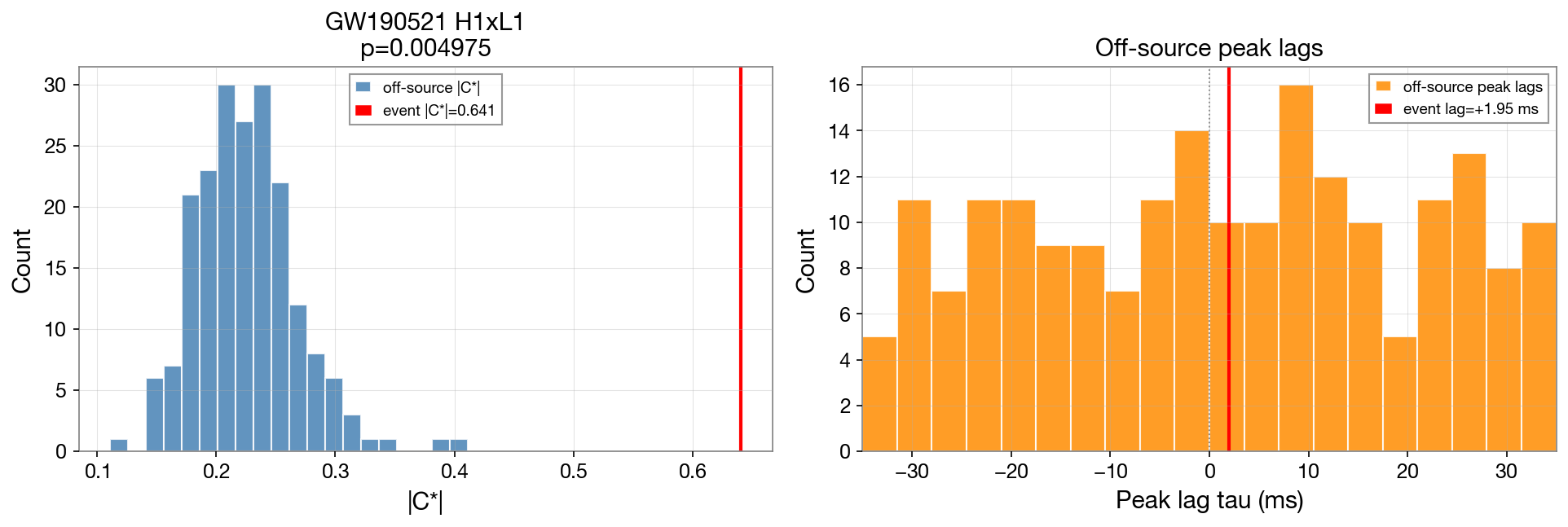}

\includegraphics[width=.94\linewidth,height=.235\textheight,keepaspectratio]{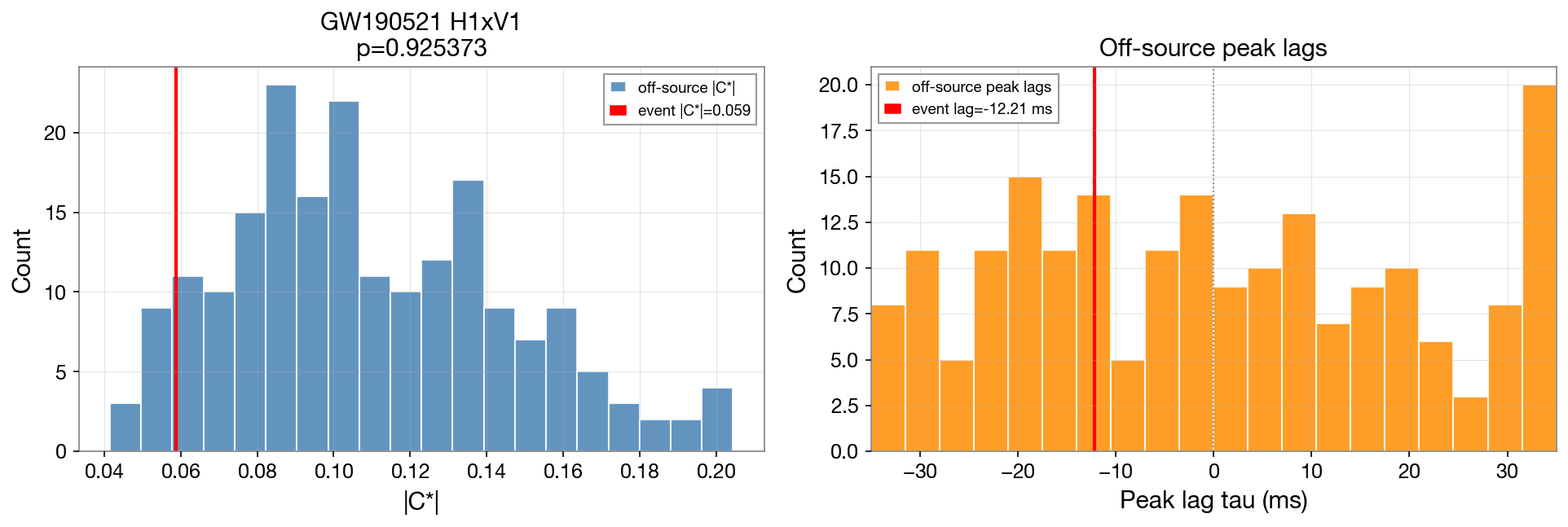}

\includegraphics[width=.94\linewidth,height=.235\textheight,keepaspectratio]{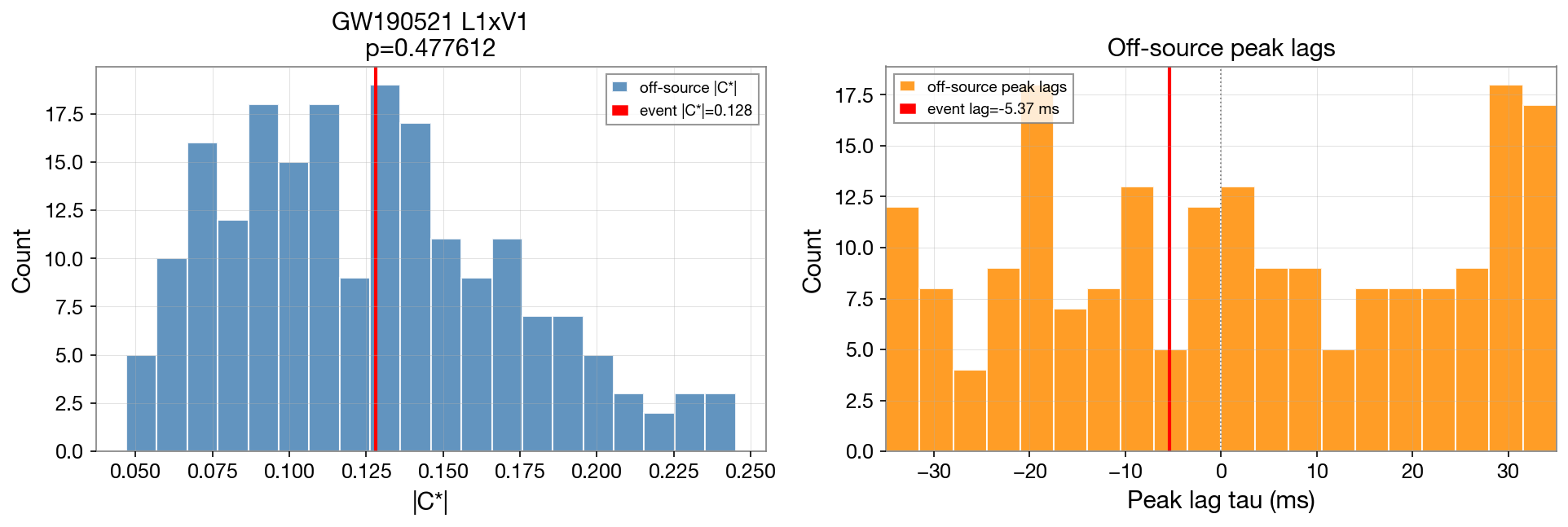}

\captionof{figure}{GW190521: distributions of reference peak magnitudes (left) and maximizing lags (right); the event values are marked in red. All p-values are local.}\label{fig:S10}
\end{minipage}\par
\clearpage
\section*{GW190828\_063405 -- Correlation curves}
Figure~\ref{fig:S11} shows the correlation curves.
\par\noindent\begin{minipage}{\linewidth}\centering
\includegraphics[width=.94\linewidth,height=.235\textheight,keepaspectratio]{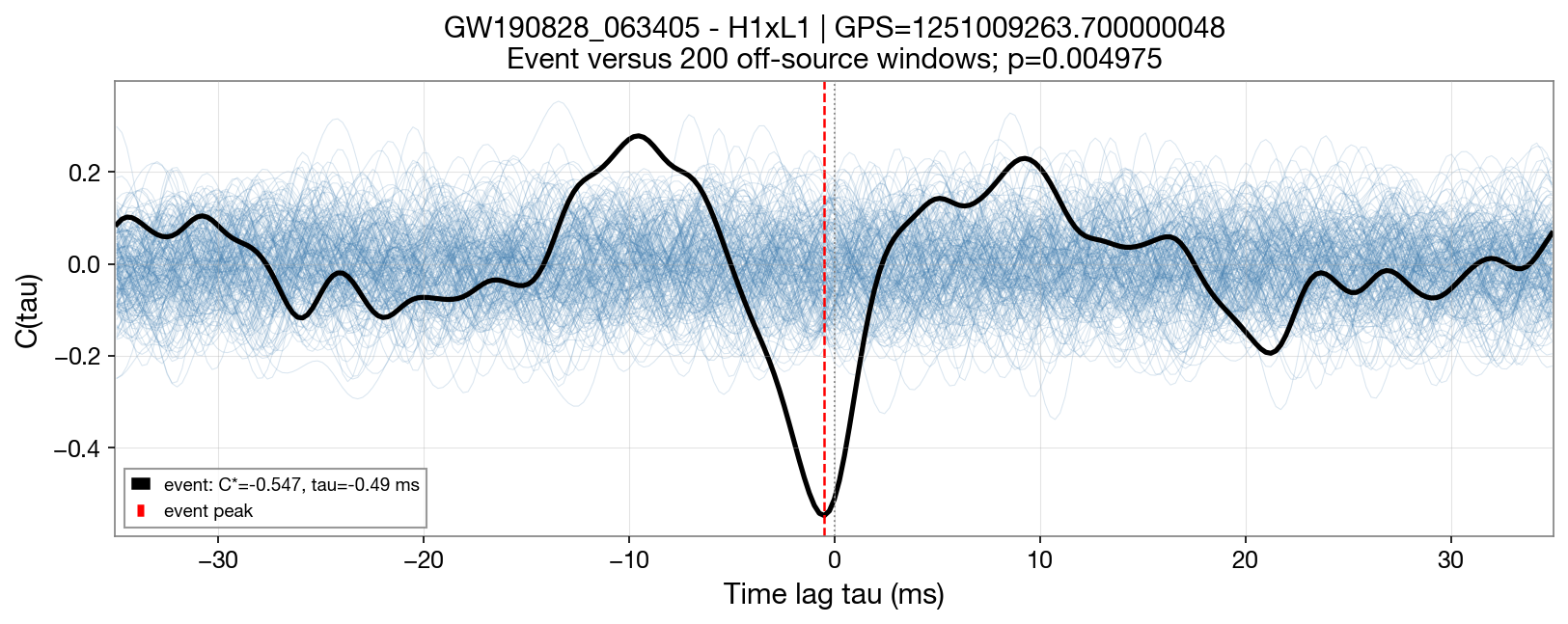}

\includegraphics[width=.94\linewidth,height=.235\textheight,keepaspectratio]{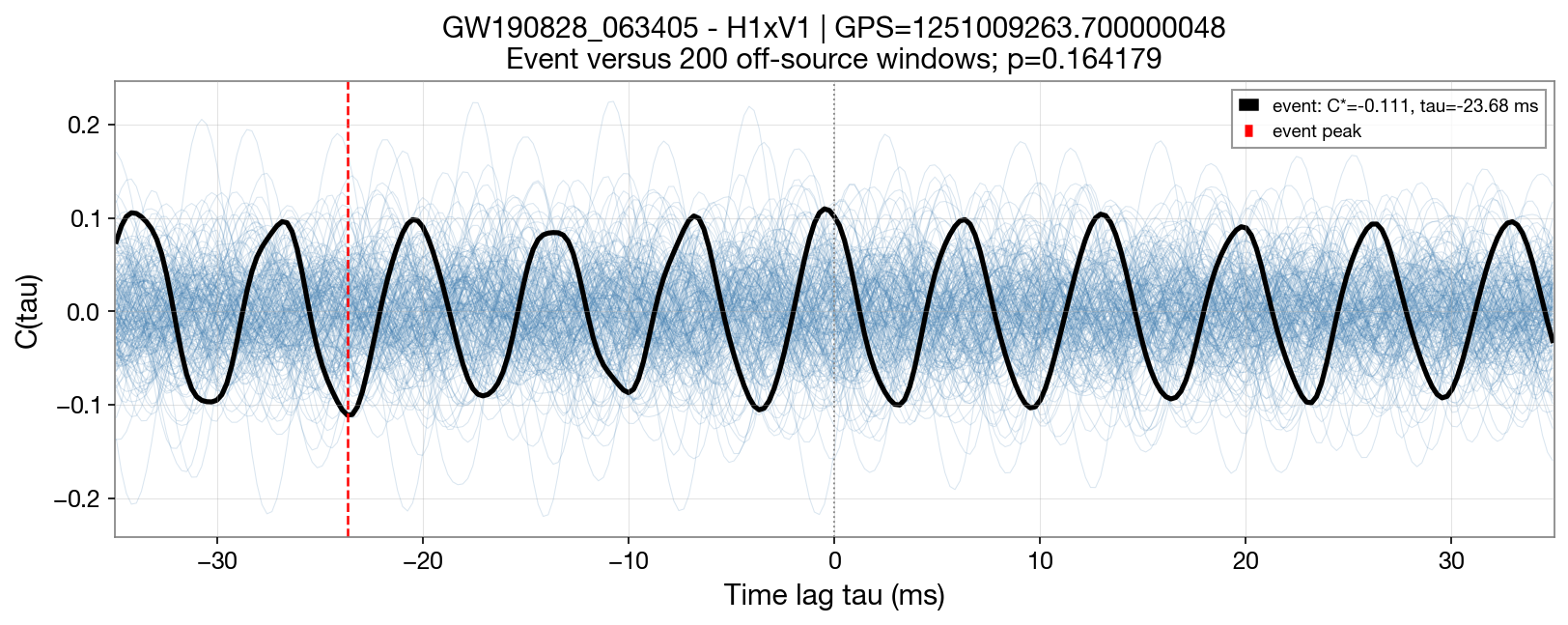}

\includegraphics[width=.94\linewidth,height=.235\textheight,keepaspectratio]{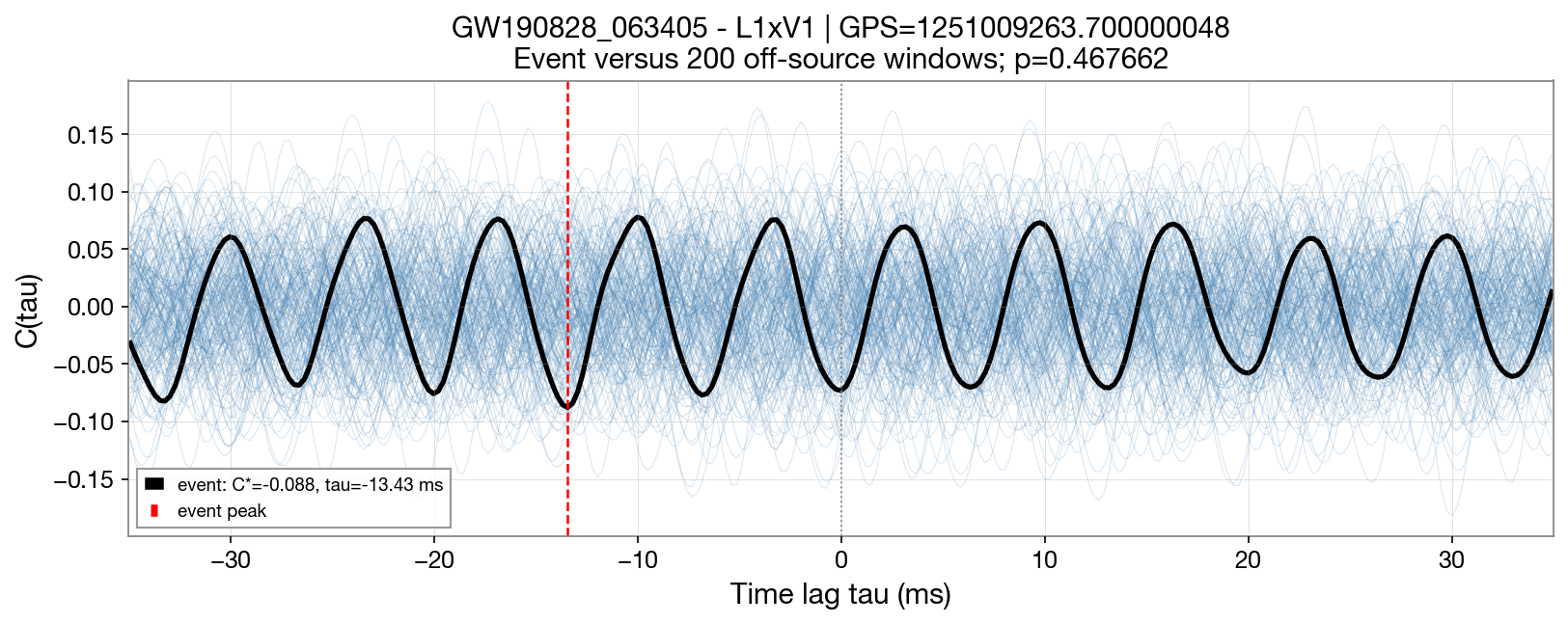}

\captionof{figure}{GW190828\_063405: event correlations (black) and 200 reference curves (blue). Positive lag means the first detector is delayed relative to the second. All p-values are local.}\label{fig:S11}
\end{minipage}\par
\clearpage
\section*{GW190828\_063405 -- Peak magnitudes and lags}
Figure~\ref{fig:S12} shows the reference distributions.
\par\noindent\begin{minipage}{\linewidth}\centering
\includegraphics[width=.94\linewidth,height=.235\textheight,keepaspectratio]{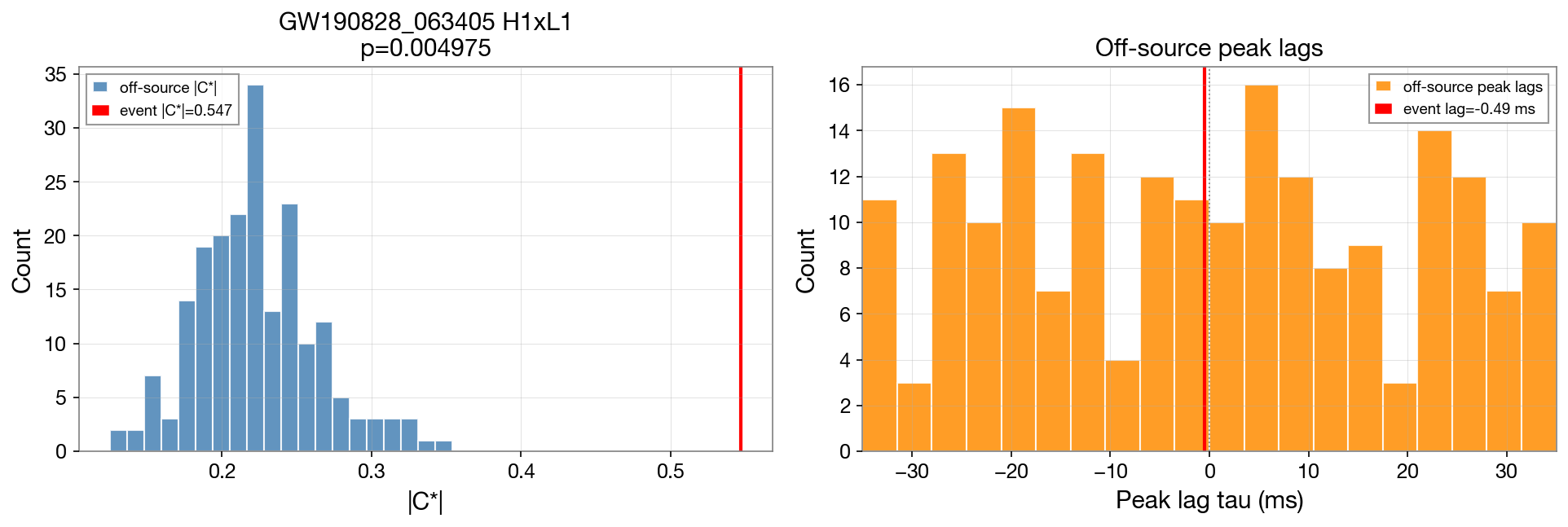}

\includegraphics[width=.94\linewidth,height=.235\textheight,keepaspectratio]{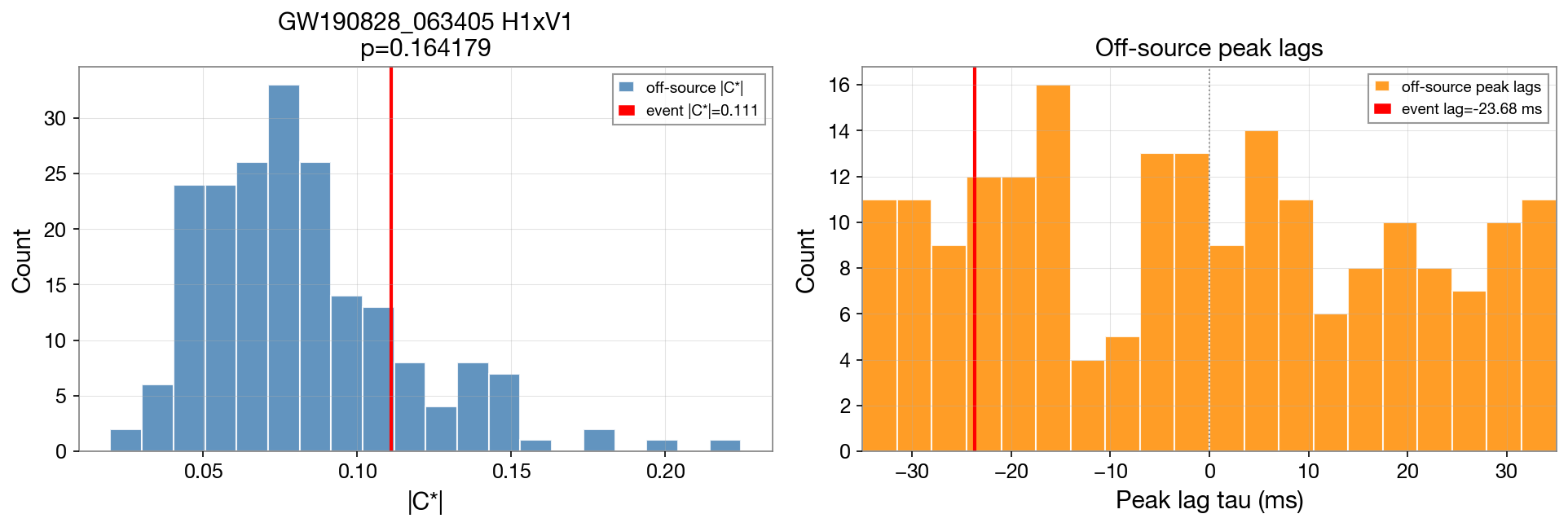}

\includegraphics[width=.94\linewidth,height=.235\textheight,keepaspectratio]{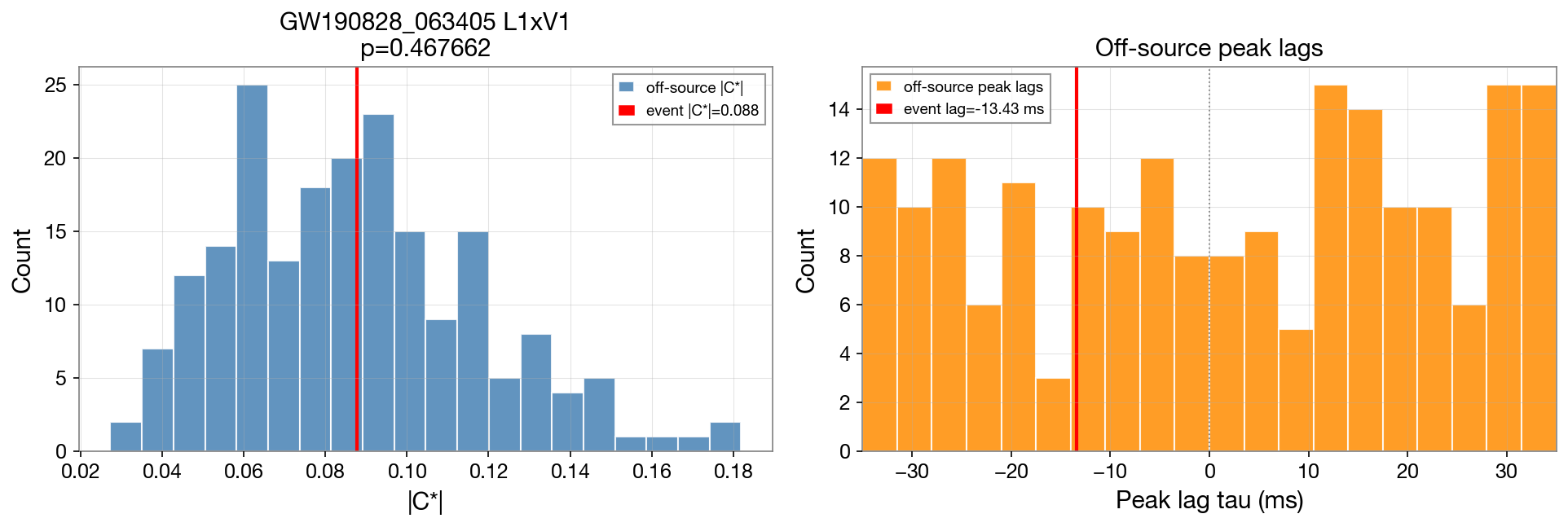}

\captionof{figure}{GW190828\_063405: distributions of reference peak magnitudes (left) and maximizing lags (right); the event values are marked in red. All p-values are local.}\label{fig:S12}
\end{minipage}\par
\clearpage
\section*{GW191127\_050227 -- Correlation curves}
Figure~\ref{fig:S13} shows the correlation curves.
\par\noindent\begin{minipage}{\linewidth}\centering
\includegraphics[width=.94\linewidth,height=.235\textheight,keepaspectratio]{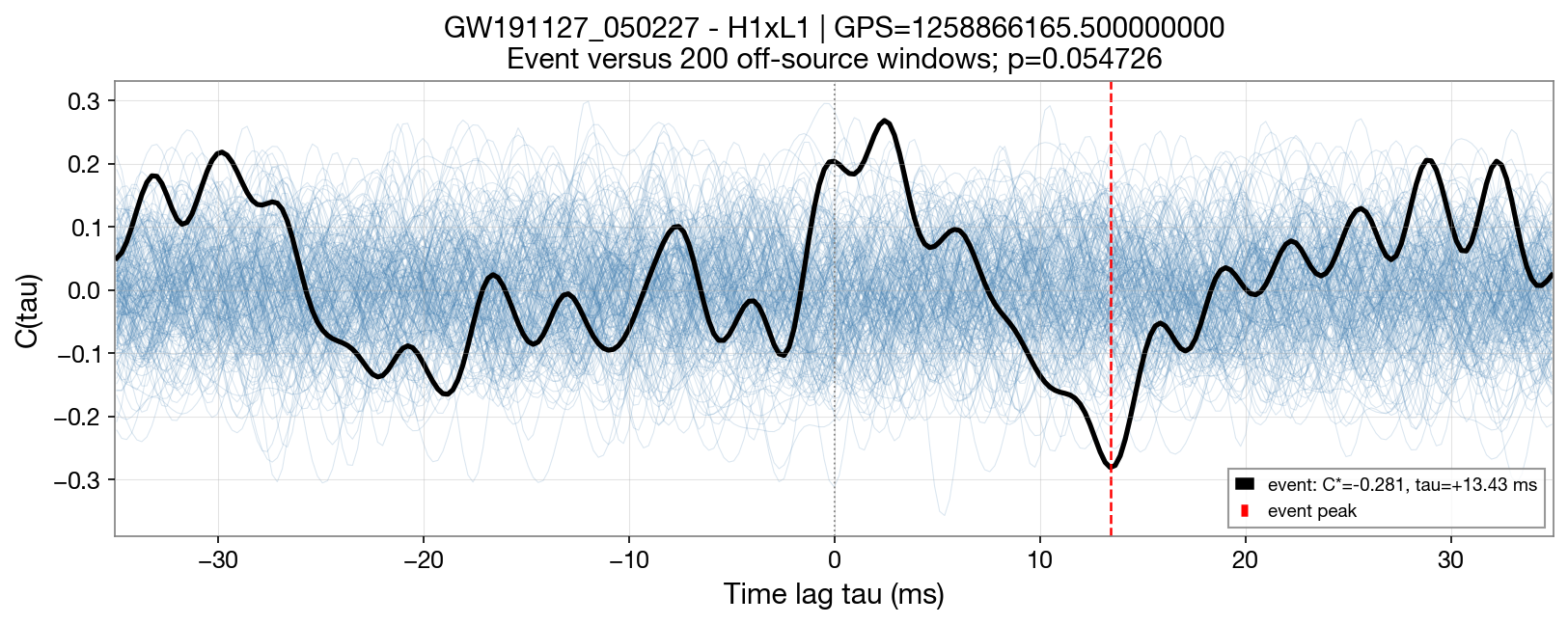}

\includegraphics[width=.94\linewidth,height=.235\textheight,keepaspectratio]{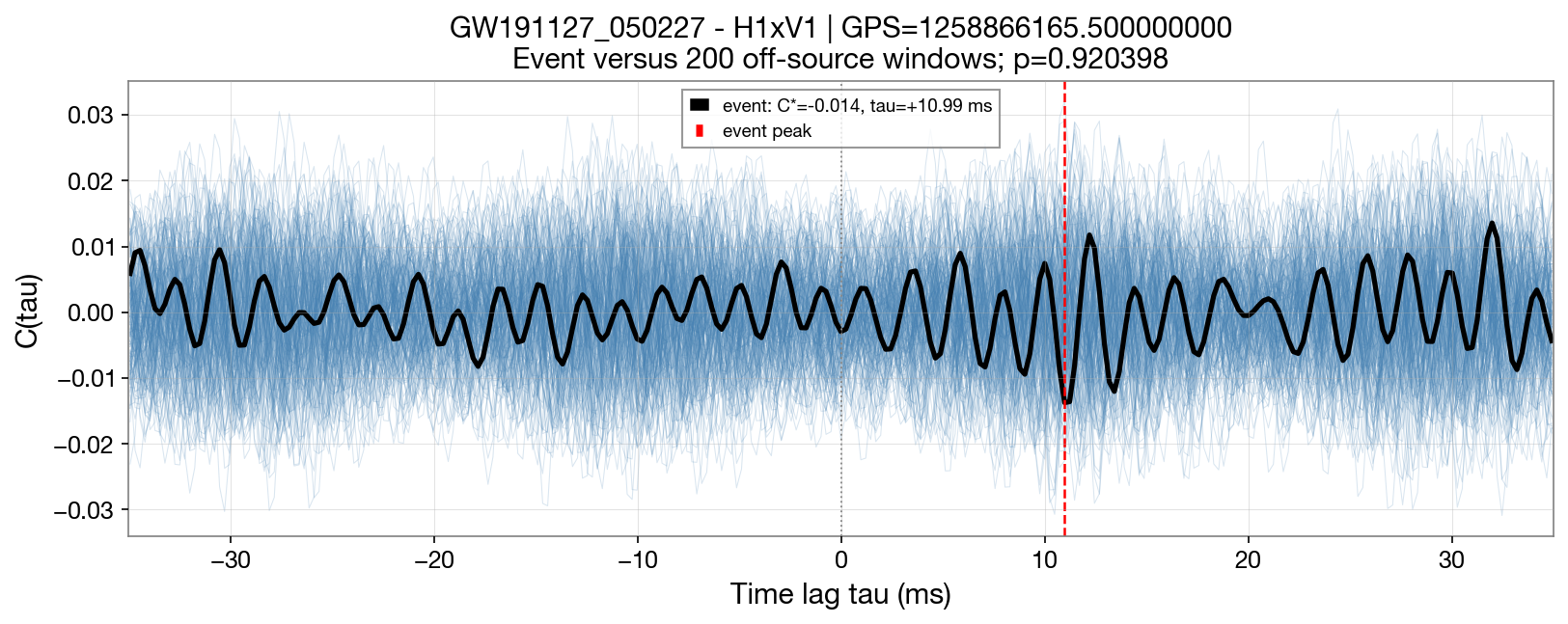}

\includegraphics[width=.94\linewidth,height=.235\textheight,keepaspectratio]{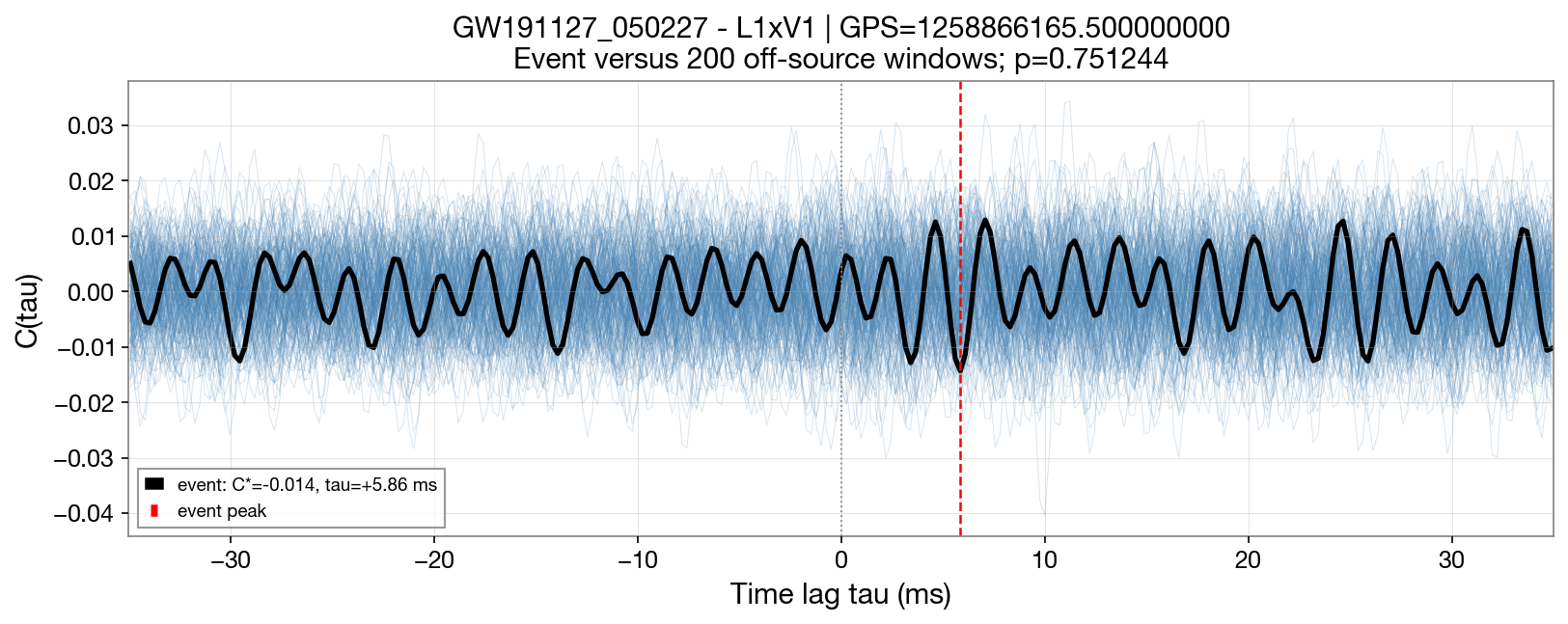}

\captionof{figure}{GW191127\_050227: event correlations (black) and 200 reference curves (blue). Positive lag means the first detector is delayed relative to the second. All p-values are local.}\label{fig:S13}
\end{minipage}\par
\clearpage
\section*{GW191127\_050227 -- Peak magnitudes and lags}
Figure~\ref{fig:S14} shows the reference distributions.
\par\noindent\begin{minipage}{\linewidth}\centering
\includegraphics[width=.94\linewidth,height=.235\textheight,keepaspectratio]{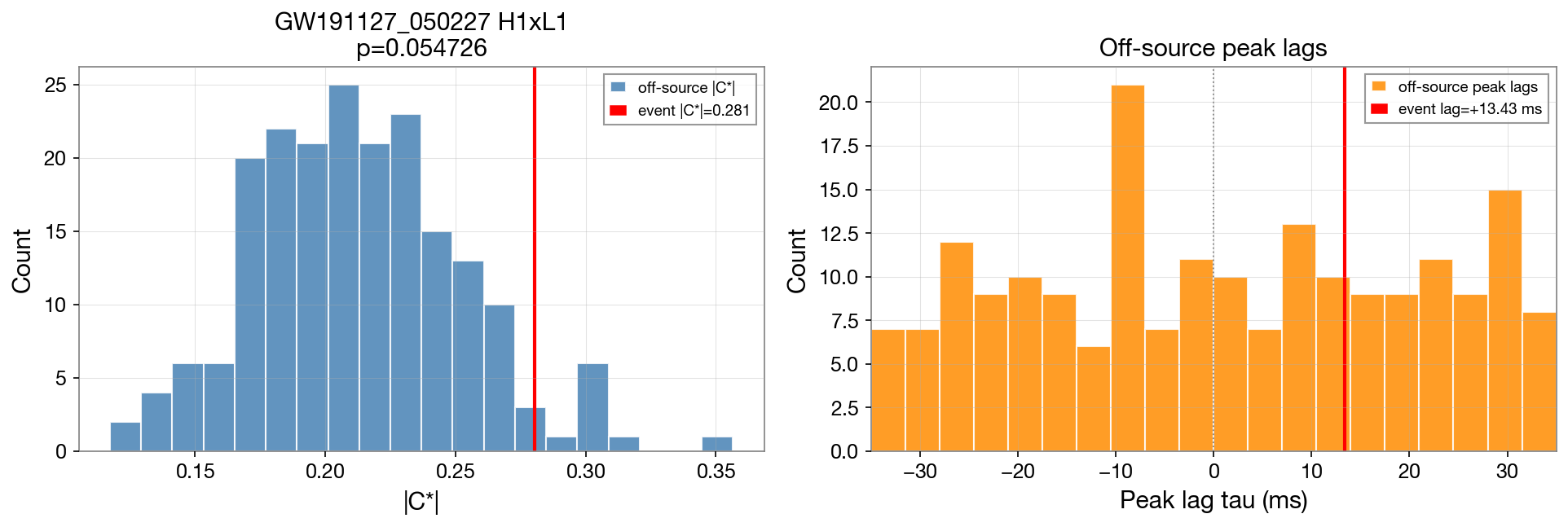}

\includegraphics[width=.94\linewidth,height=.235\textheight,keepaspectratio]{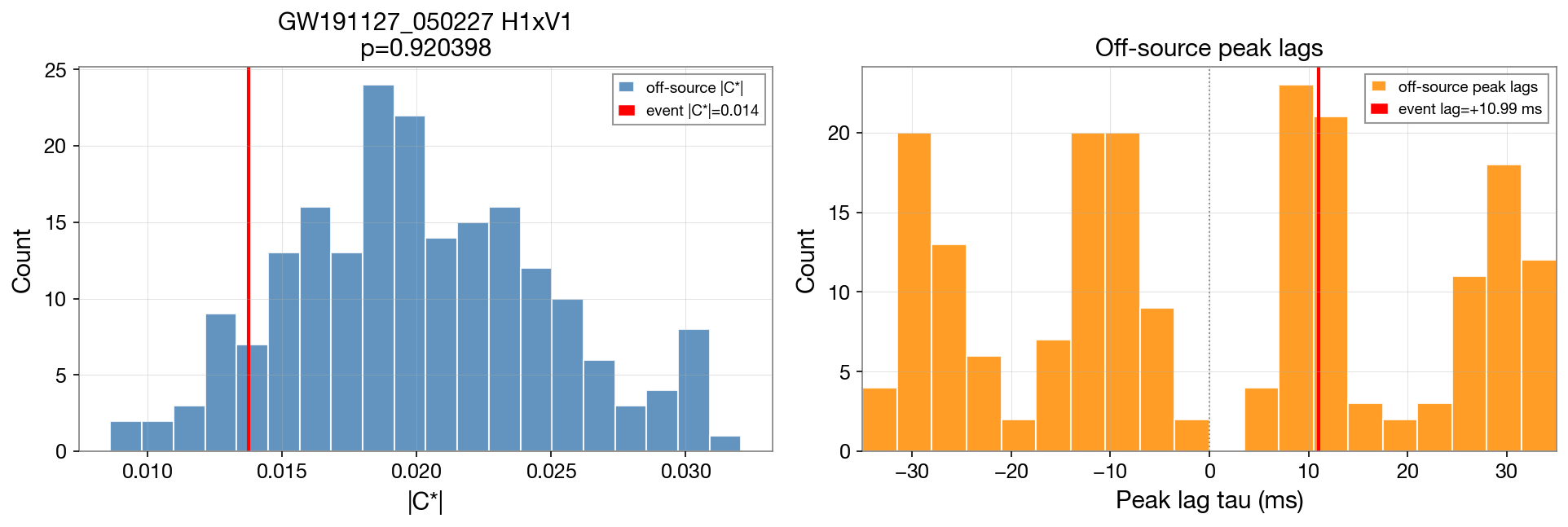}

\includegraphics[width=.94\linewidth,height=.235\textheight,keepaspectratio]{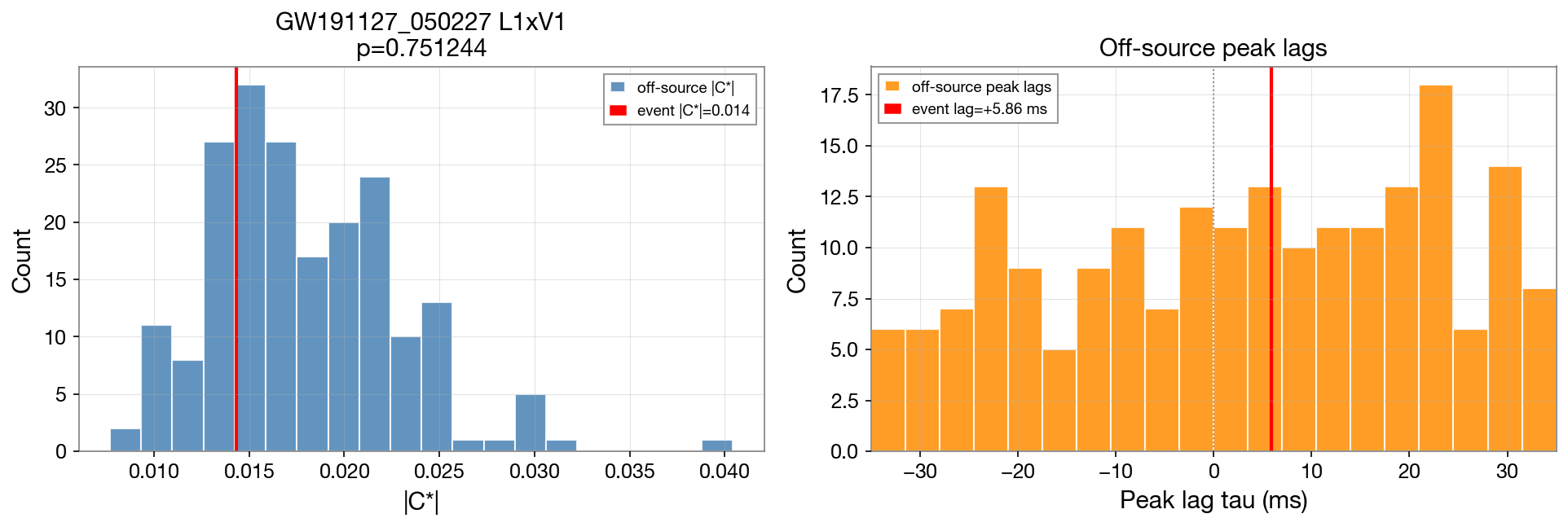}

\captionof{figure}{GW191127\_050227: distributions of reference peak magnitudes (left) and maximizing lags (right); the event values are marked in red. All p-values are local.}\label{fig:S14}
\end{minipage}\par
\clearpage
\section*{GW200129\_065458 -- Correlation curves}
Livingston has category-2/3 quality flags unset at the event epoch; its pairs require a dedicated glitch assessment.\par
Figure~\ref{fig:S15} shows the correlation curves.
\par\noindent\begin{minipage}{\linewidth}\centering
\includegraphics[width=.94\linewidth,height=.235\textheight,keepaspectratio]{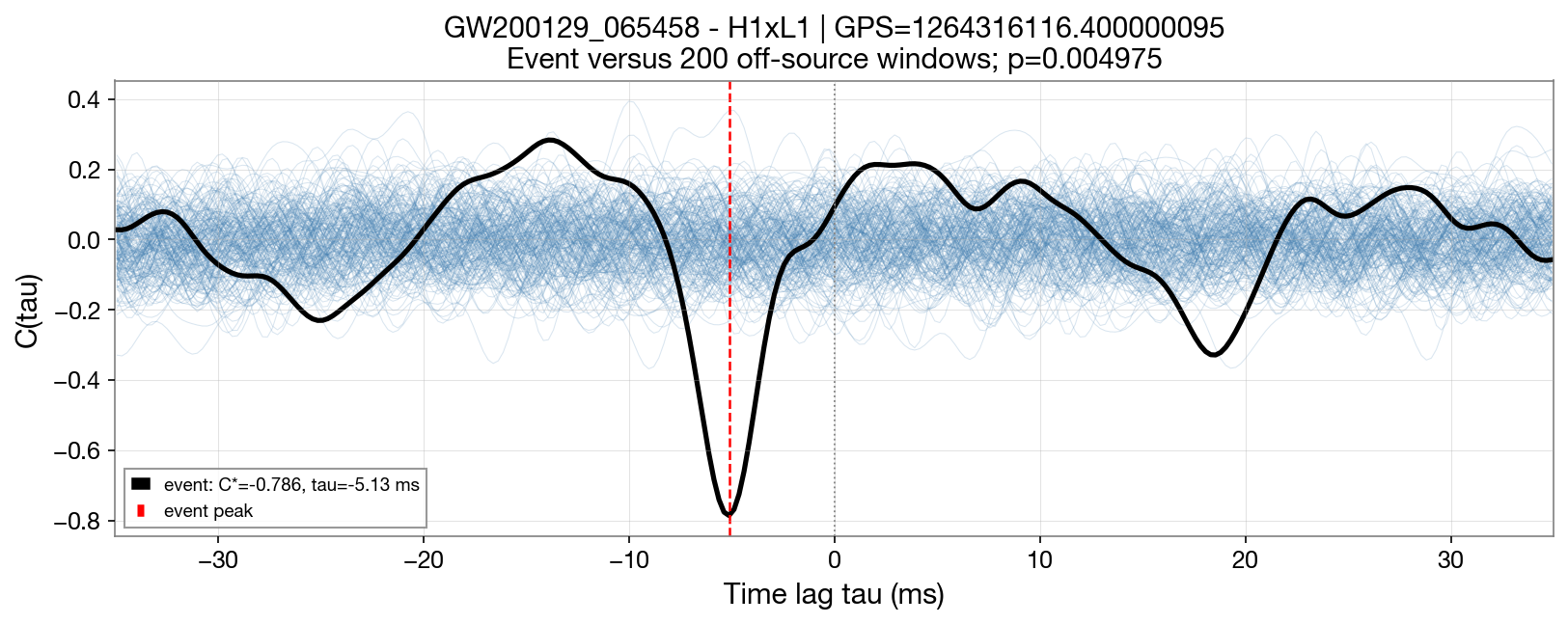}

\includegraphics[width=.94\linewidth,height=.235\textheight,keepaspectratio]{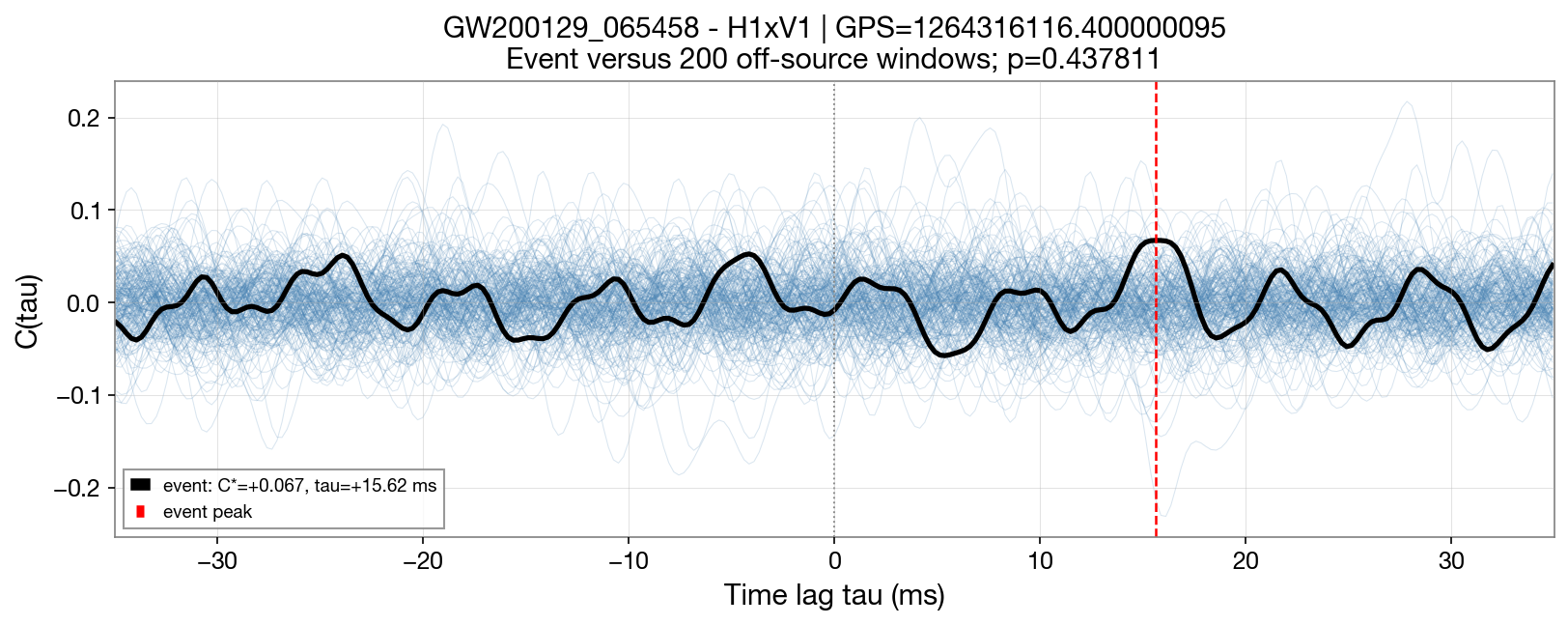}

\includegraphics[width=.94\linewidth,height=.235\textheight,keepaspectratio]{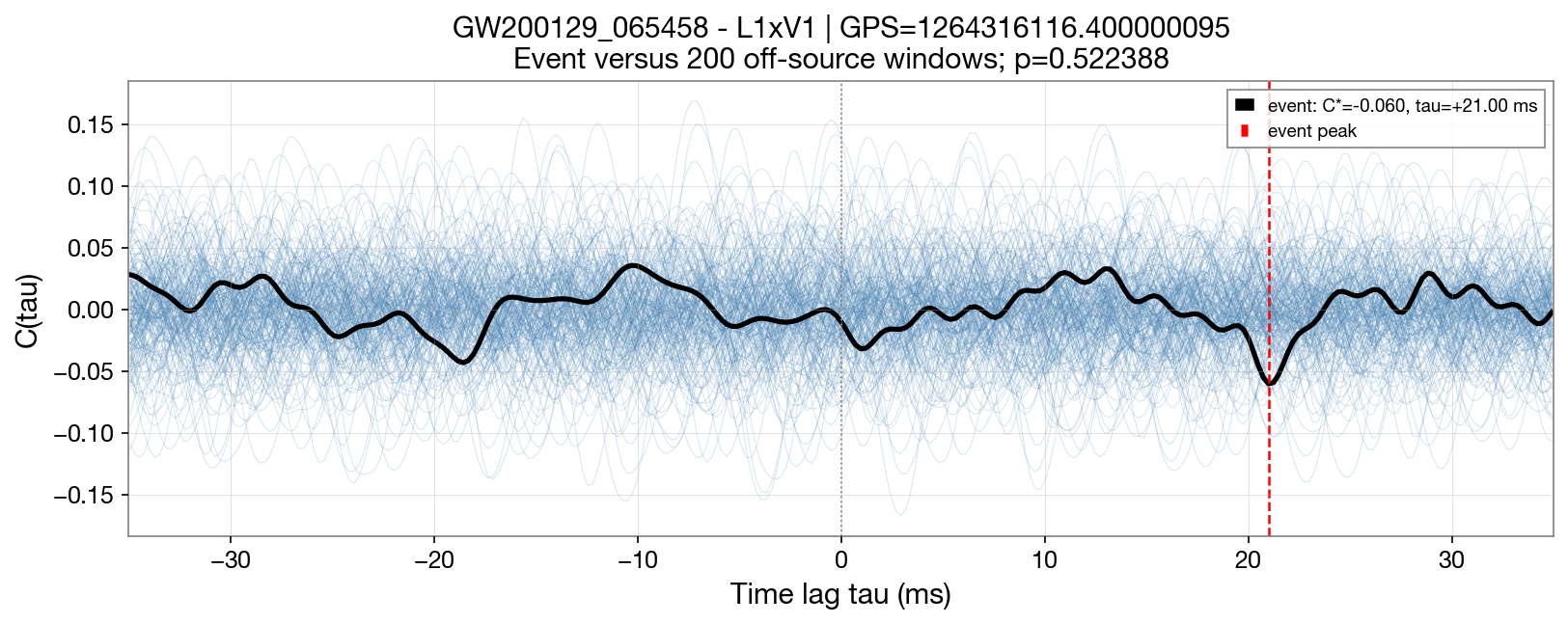}

\captionof{figure}{GW200129\_065458: event correlations (black) and 200 reference curves (blue). Positive lag means the first detector is delayed relative to the second. All p-values are local.}\label{fig:S15}
\end{minipage}\par
\clearpage
\section*{GW200129\_065458 -- Peak magnitudes and lags}
Livingston has category-2/3 quality flags unset at the event epoch; its pairs require a dedicated glitch assessment.\par
Figure~\ref{fig:S16} shows the reference distributions.
\par\noindent\begin{minipage}{\linewidth}\centering
\includegraphics[width=.94\linewidth,height=.235\textheight,keepaspectratio]{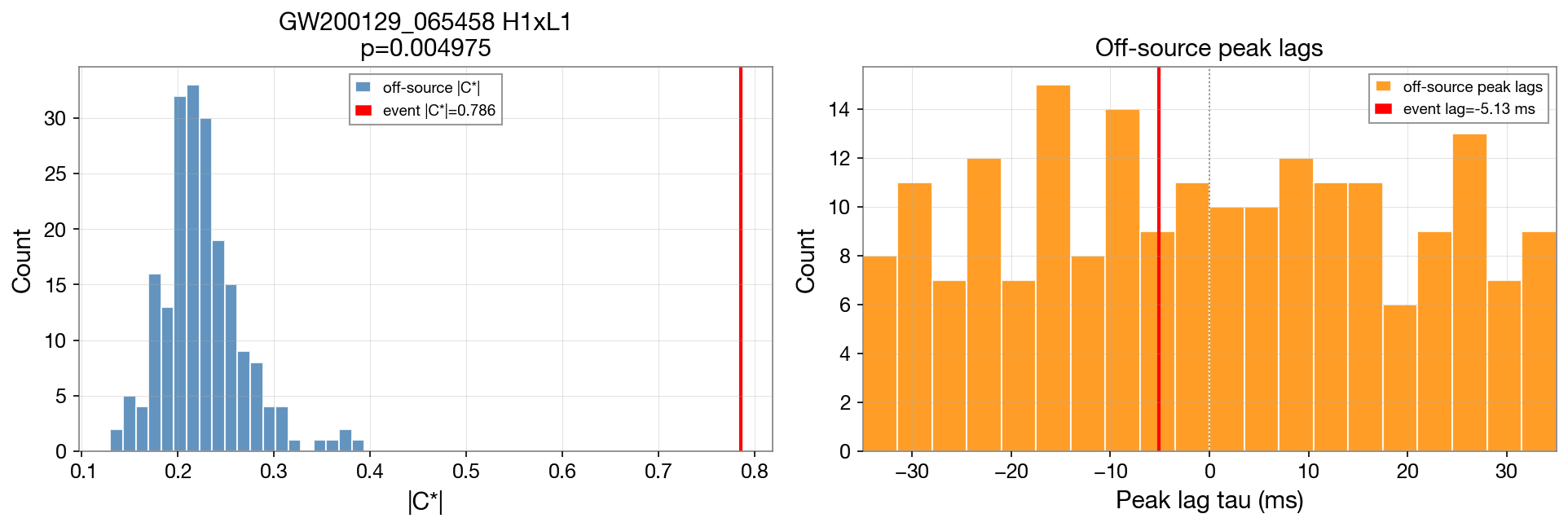}

\includegraphics[width=.94\linewidth,height=.235\textheight,keepaspectratio]{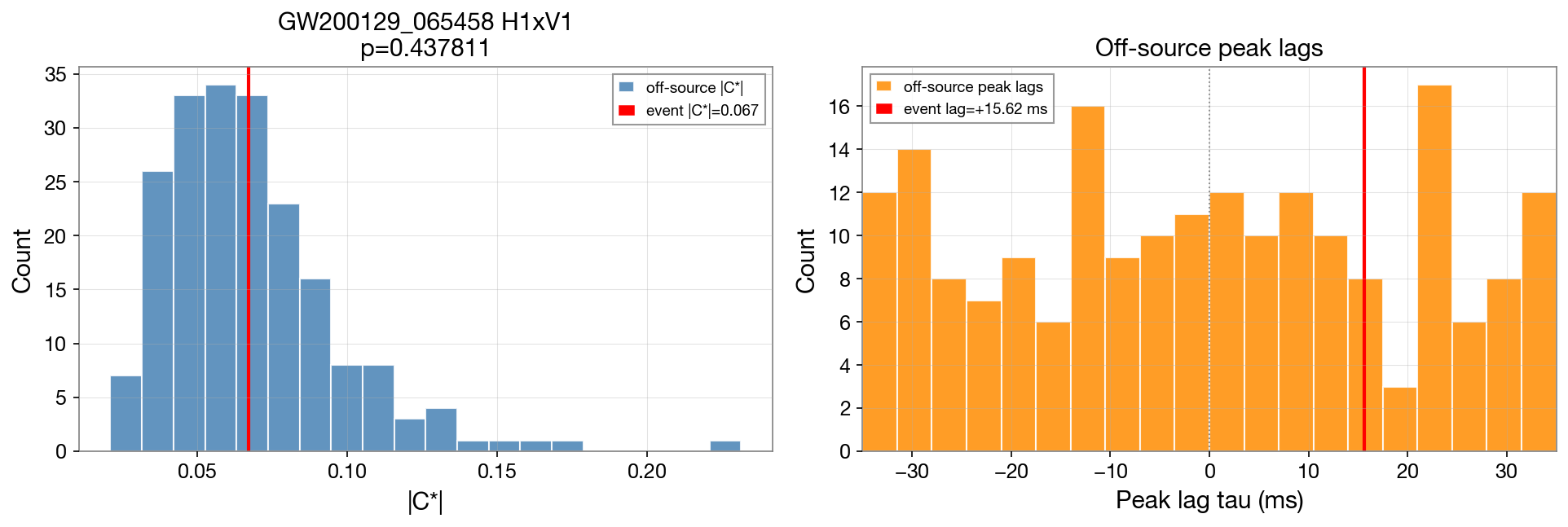}

\includegraphics[width=.94\linewidth,height=.235\textheight,keepaspectratio]{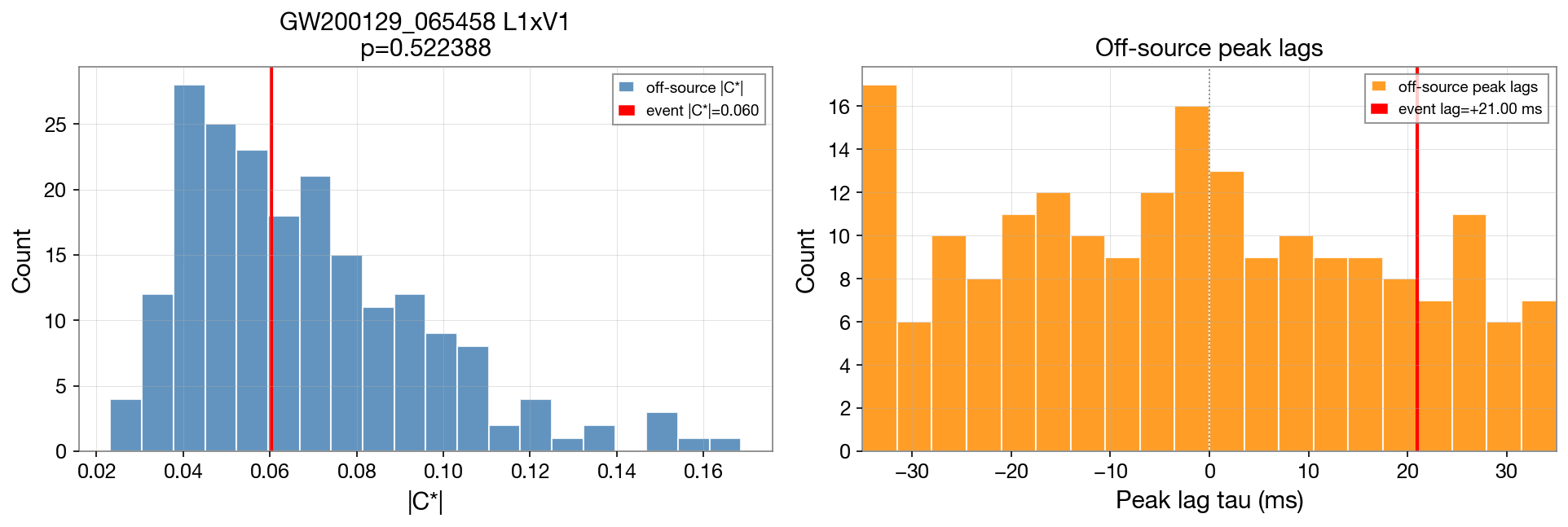}

\captionof{figure}{GW200129\_065458: distributions of reference peak magnitudes (left) and maximizing lags (right); the event values are marked in red. All p-values are local.}\label{fig:S16}
\end{minipage}\par
\clearpage
\section*{GW200219\_094415 -- Correlation curves}
Figure~\ref{fig:S17} shows the correlation curves.
\par\noindent\begin{minipage}{\linewidth}\centering
\includegraphics[width=.94\linewidth,height=.235\textheight,keepaspectratio]{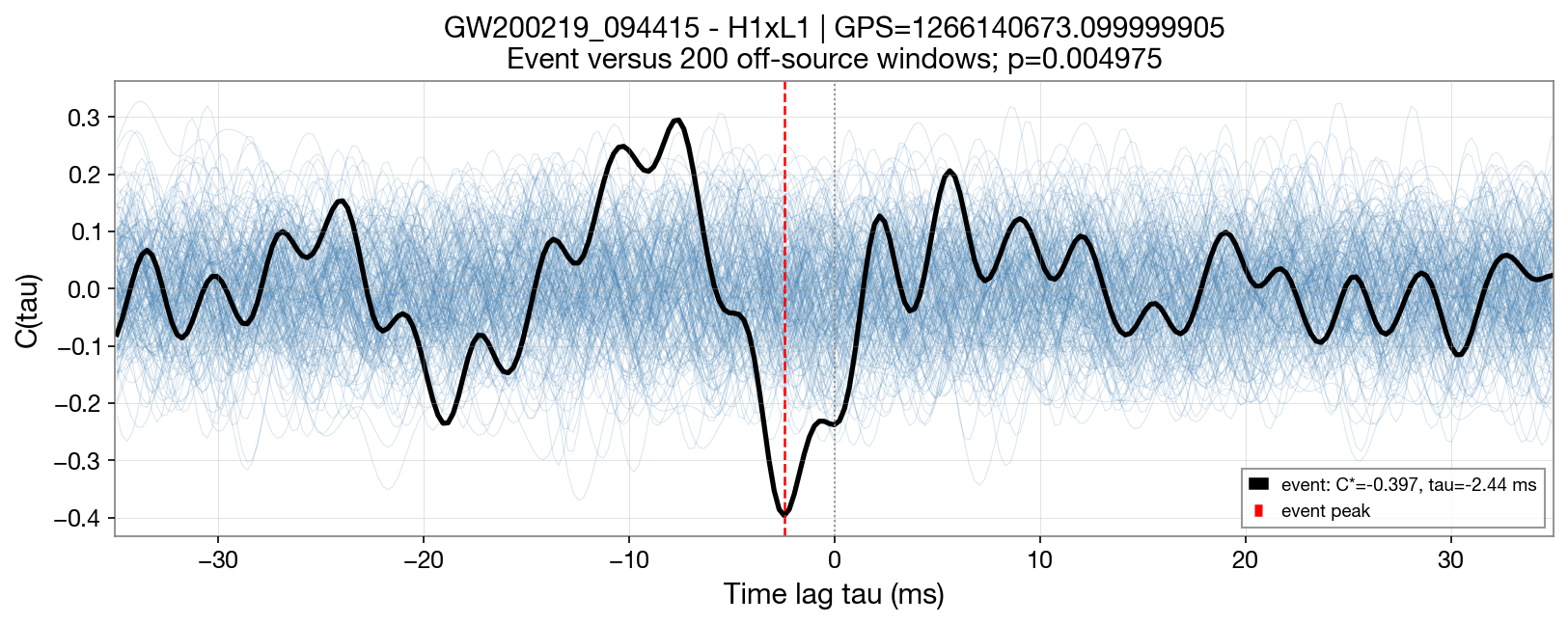}

\includegraphics[width=.94\linewidth,height=.235\textheight,keepaspectratio]{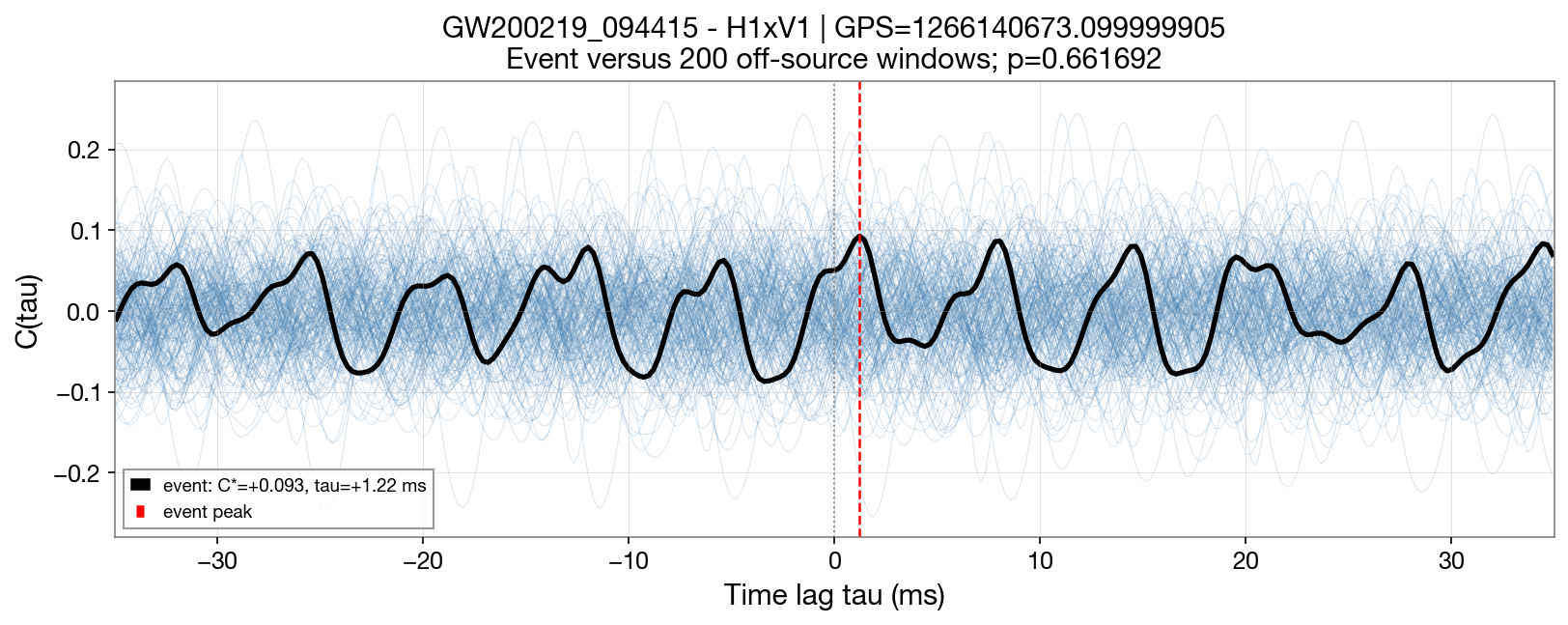}

\includegraphics[width=.94\linewidth,height=.235\textheight,keepaspectratio]{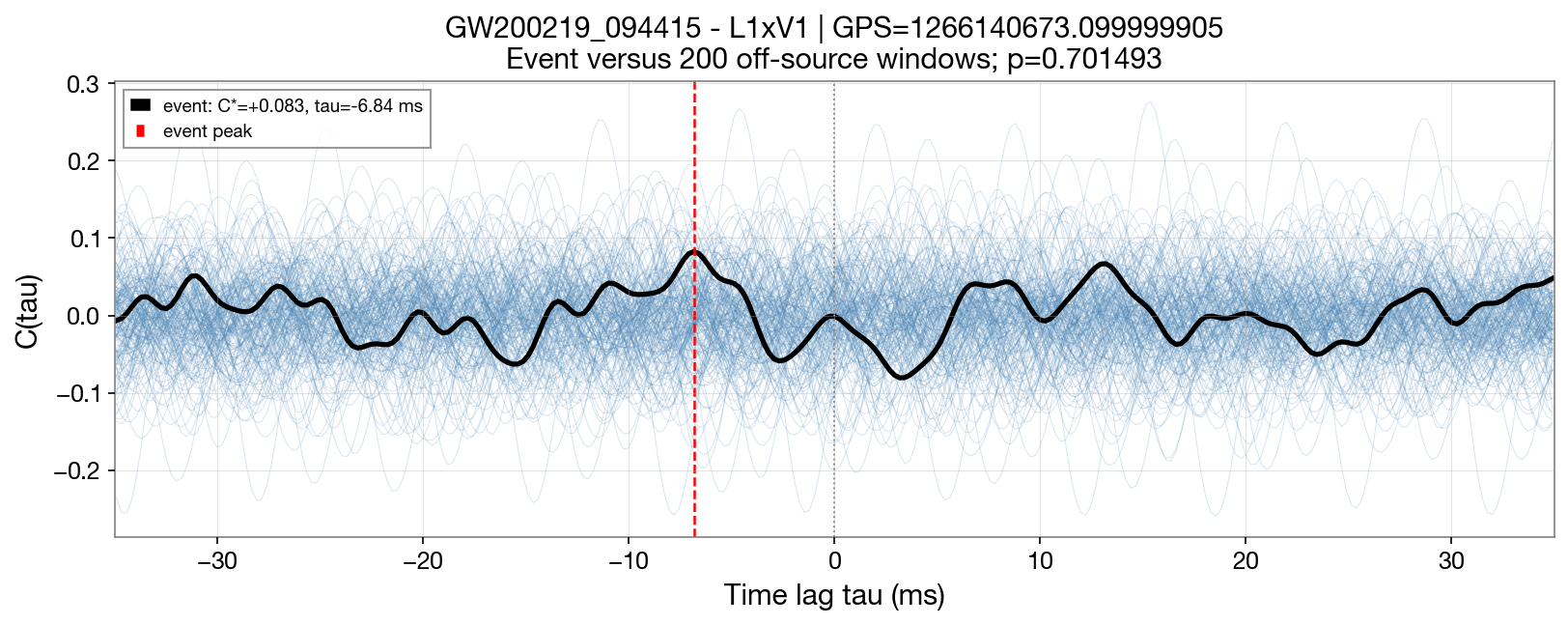}

\captionof{figure}{GW200219\_094415: event correlations (black) and 200 reference curves (blue). Positive lag means the first detector is delayed relative to the second. All p-values are local.}\label{fig:S17}
\end{minipage}\par
\clearpage
\section*{GW200219\_094415 -- Peak magnitudes and lags}
Figure~\ref{fig:S18} shows the reference distributions.
\par\noindent\begin{minipage}{\linewidth}\centering
\includegraphics[width=.94\linewidth,height=.235\textheight,keepaspectratio]{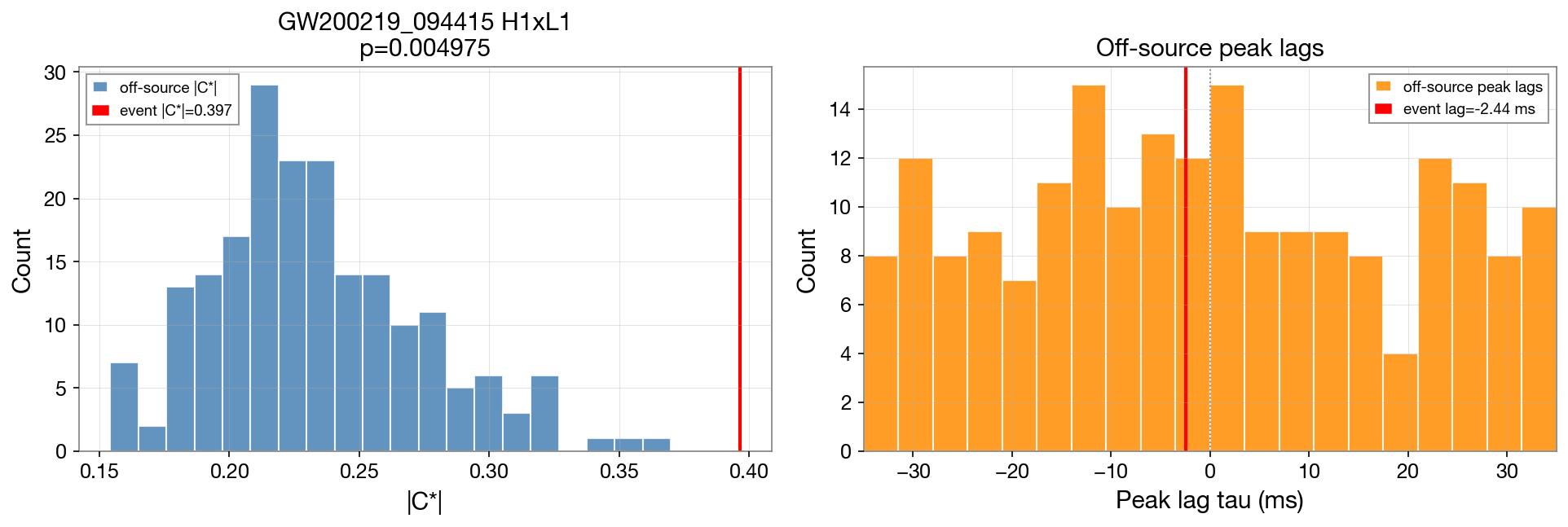}

\includegraphics[width=.94\linewidth,height=.235\textheight,keepaspectratio]{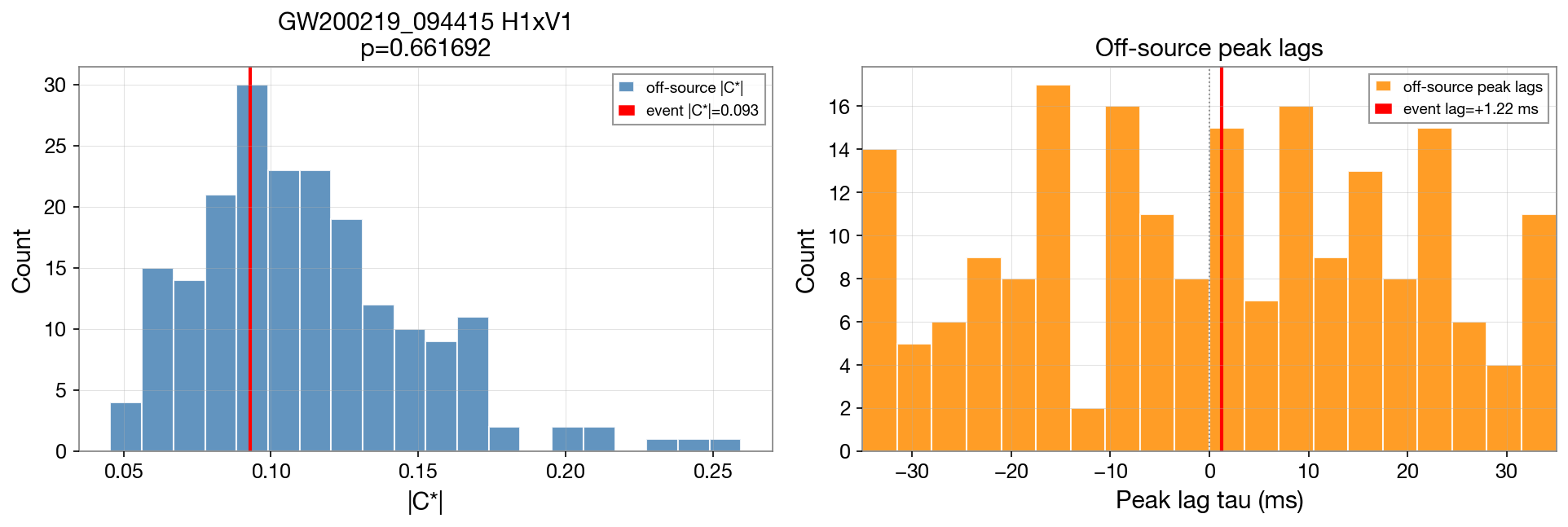}

\includegraphics[width=.94\linewidth,height=.235\textheight,keepaspectratio]{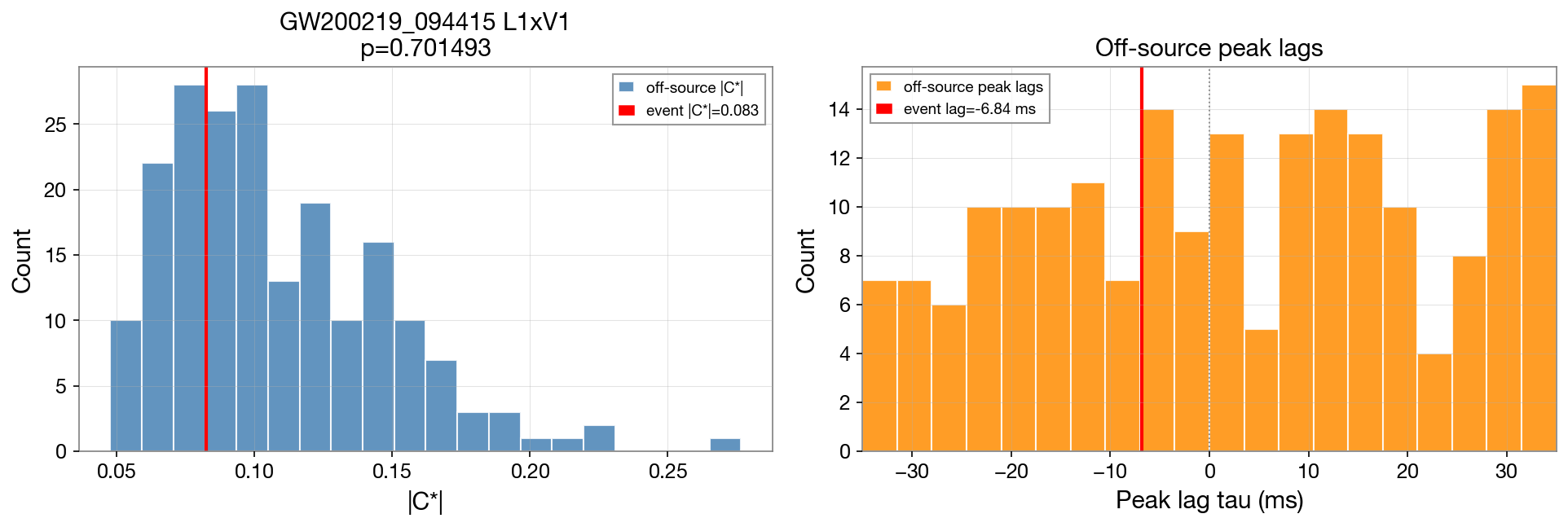}

\captionof{figure}{GW200219\_094415: distributions of reference peak magnitudes (left) and maximizing lags (right); the event values are marked in red. All p-values are local.}\label{fig:S18}
\end{minipage}\par
\clearpage
\section*{GW200224\_222234 -- Correlation curves}
Figure~\ref{fig:S19} shows the correlation curves.
\par\noindent\begin{minipage}{\linewidth}\centering
\includegraphics[width=.94\linewidth,height=.235\textheight,keepaspectratio]{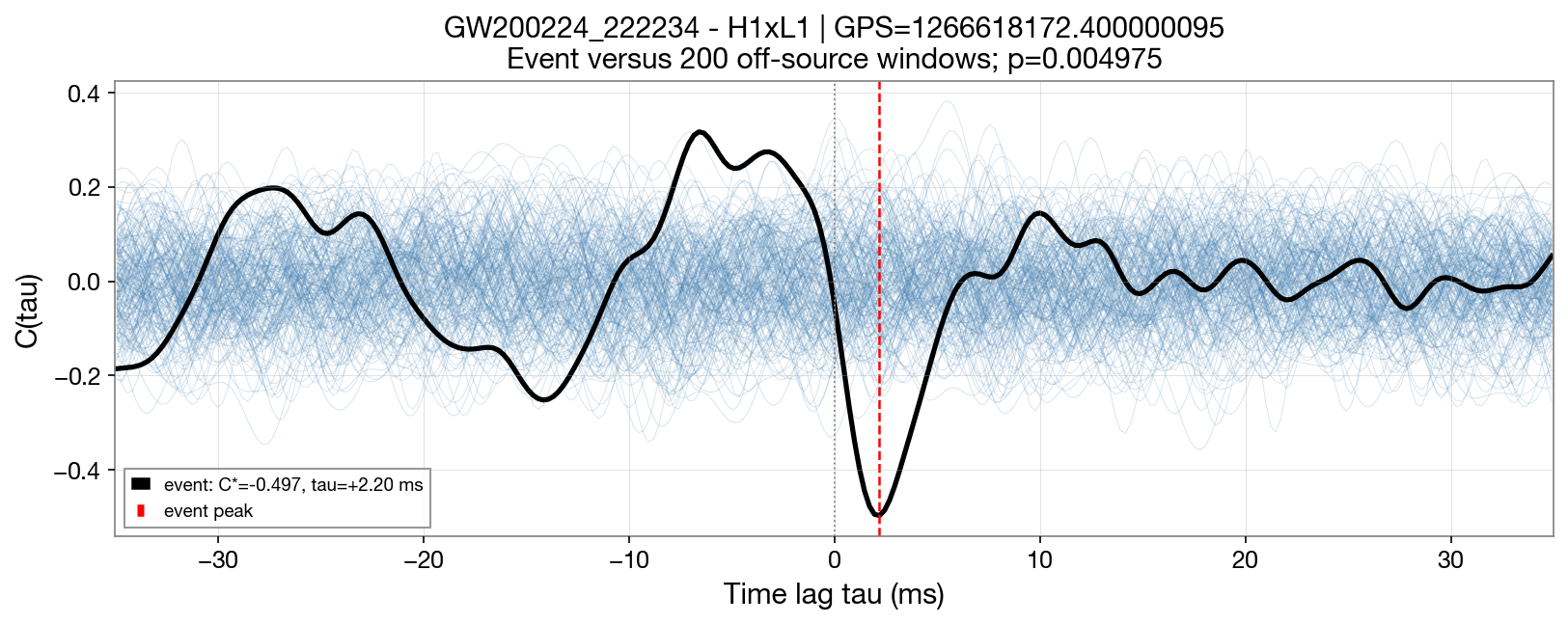}

\includegraphics[width=.94\linewidth,height=.235\textheight,keepaspectratio]{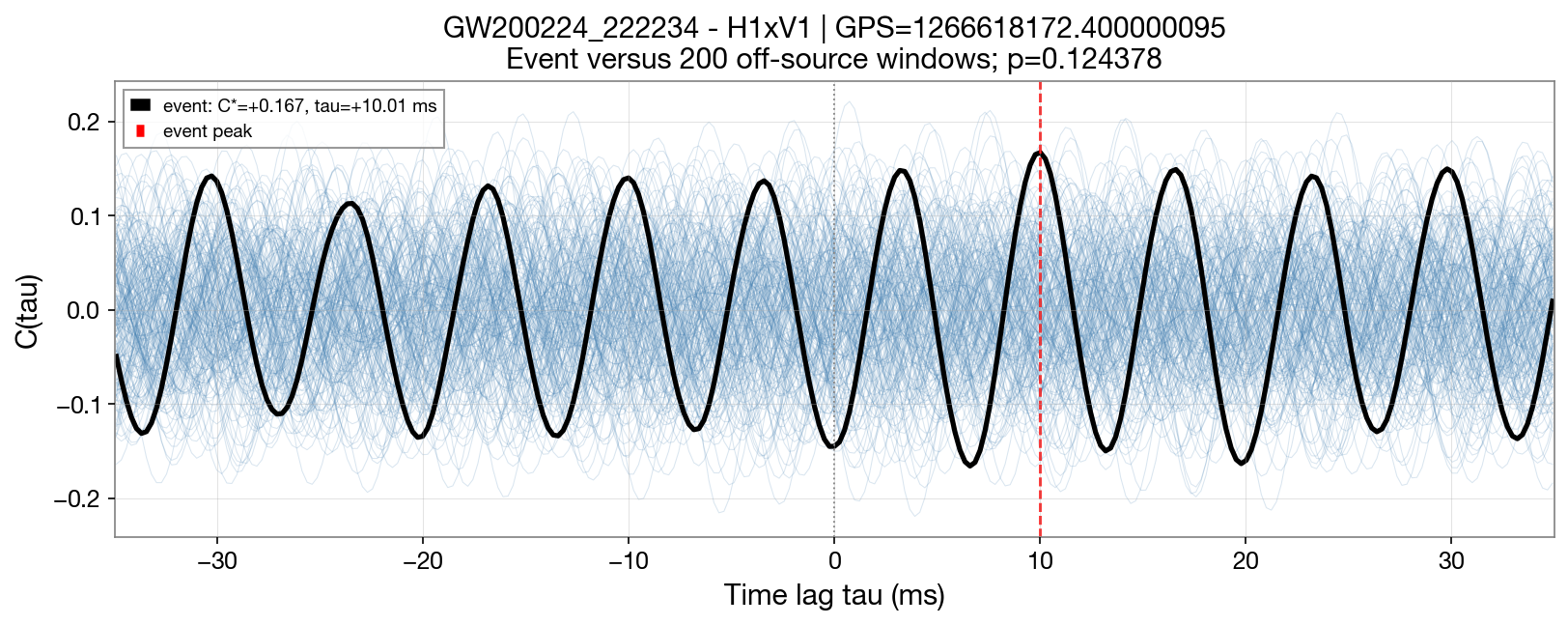}

\includegraphics[width=.94\linewidth,height=.235\textheight,keepaspectratio]{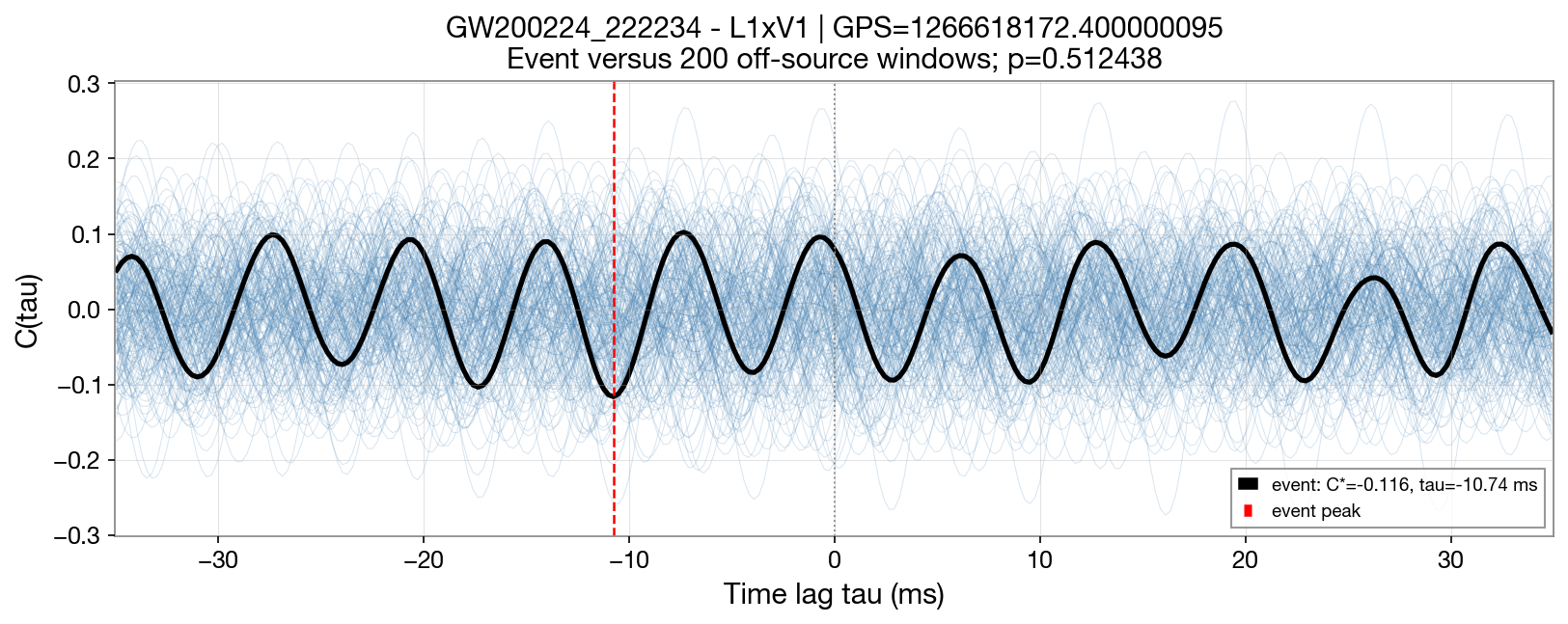}

\captionof{figure}{GW200224\_222234: event correlations (black) and 200 reference curves (blue). Positive lag means the first detector is delayed relative to the second. All p-values are local.}\label{fig:S19}
\end{minipage}\par
\clearpage
\section*{GW200224\_222234 -- Peak magnitudes and lags}
Figure~\ref{fig:S20} shows the reference distributions.
\par\noindent\begin{minipage}{\linewidth}\centering
\includegraphics[width=.94\linewidth,height=.235\textheight,keepaspectratio]{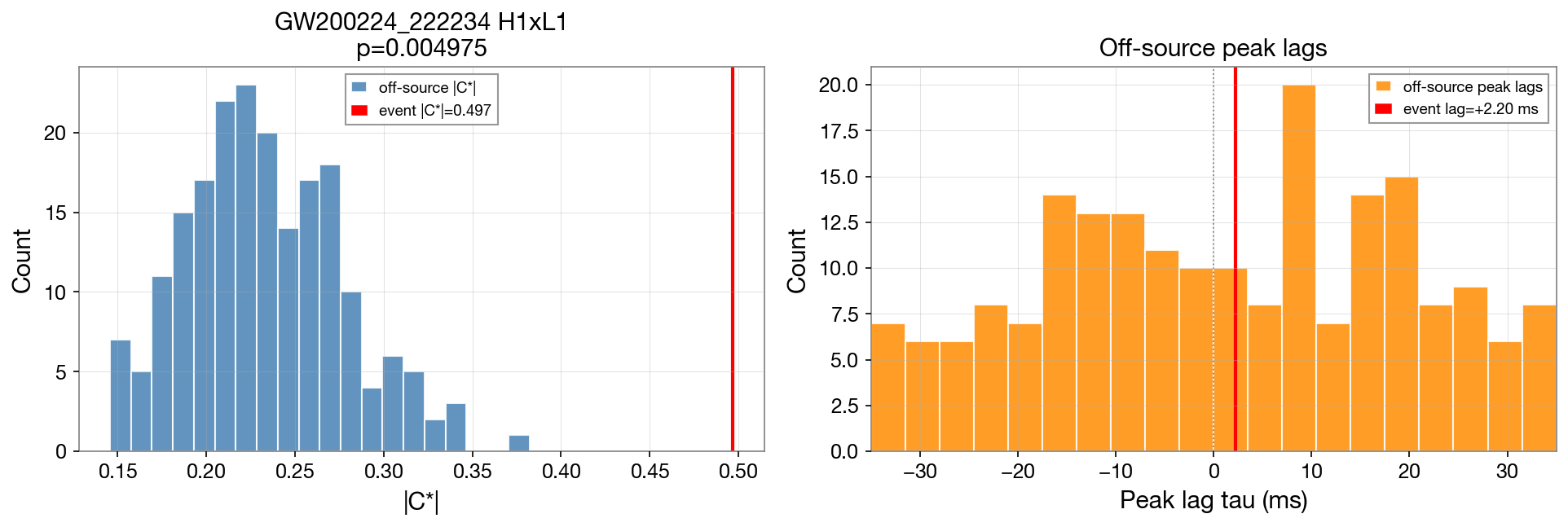}

\includegraphics[width=.94\linewidth,height=.235\textheight,keepaspectratio]{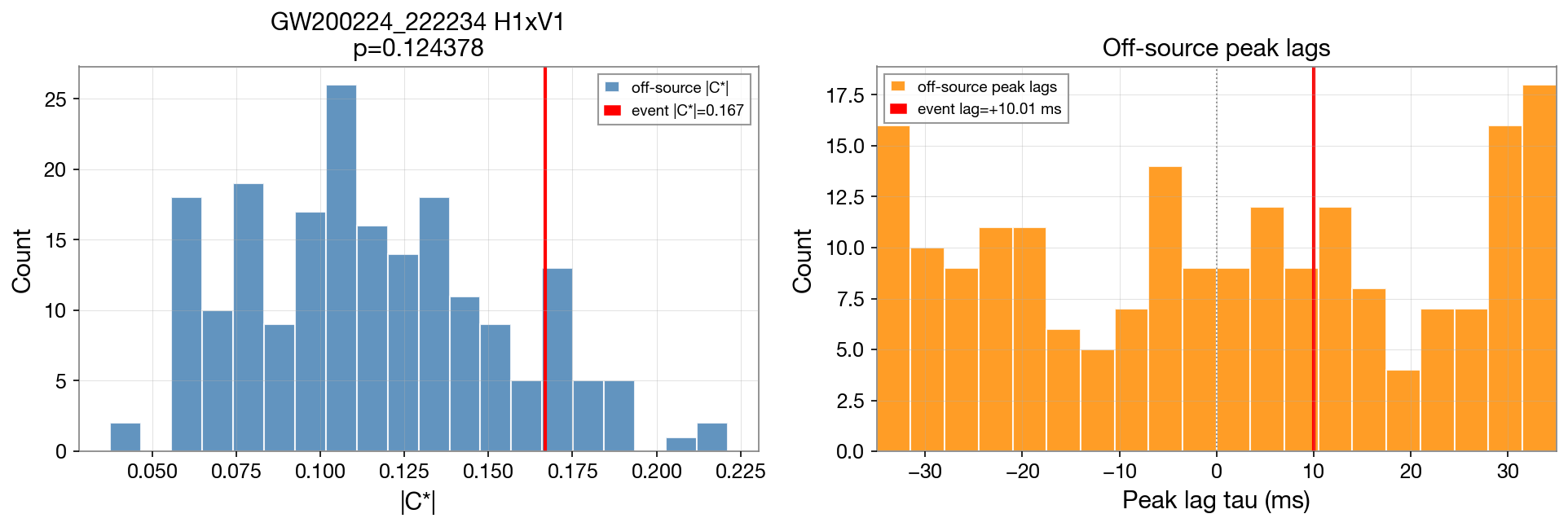}

\includegraphics[width=.94\linewidth,height=.235\textheight,keepaspectratio]{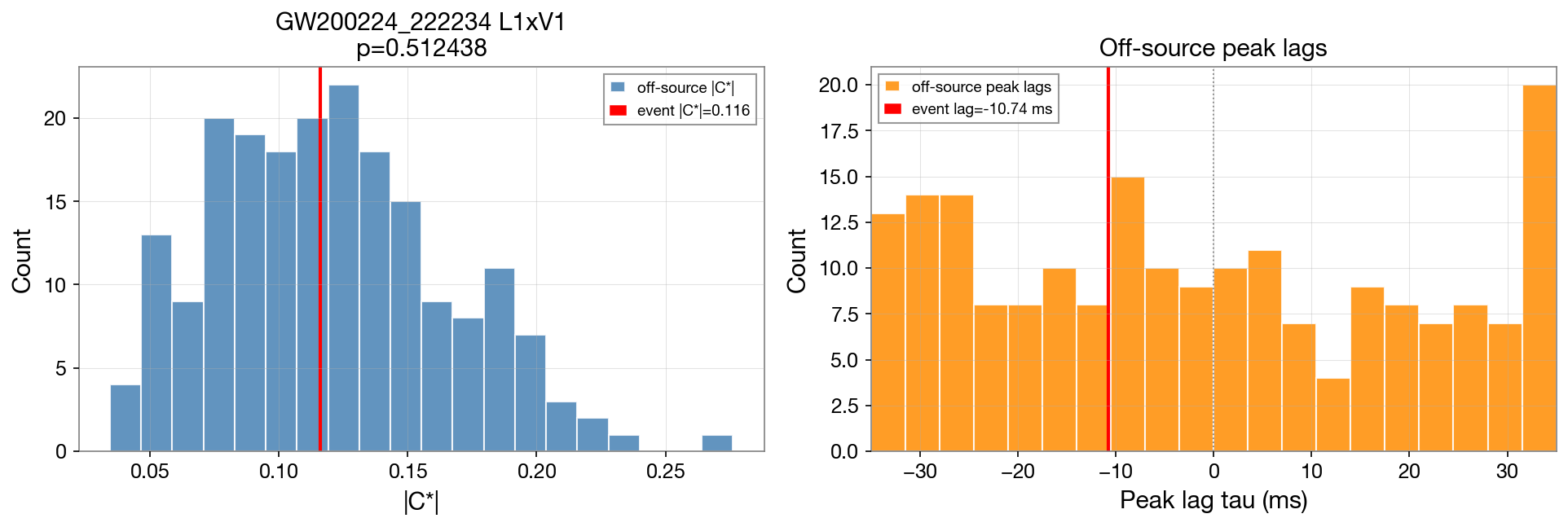}

\captionof{figure}{GW200224\_222234: distributions of reference peak magnitudes (left) and maximizing lags (right); the event values are marked in red. All p-values are local.}\label{fig:S20}
\end{minipage}\par
\clearpage\section*{Additional timing and spectral sensitivity}
Table~\ref{tab:S1} compares the timing, quality and spectral diagnostics.
Fixed 819-sample windows are used for the timing diagnostic; the original table and CW-notch event diagnostic use inclusive masks. Q denotes DATA, CBC/BURST categories 1 and 2 and absence of non-CW injections, with full scan support plus 2 s. The Q-scan p-value compares timing-scan maxima with identical reference scans. CW denotes combined mains and LIGO injection-frequency notches, with original reference centers. These are separate exploratory comparisons; their p-values cannot be selected or pooled to increase significance.
\par\noindent\begin{minipage}{\linewidth}\centering\small
\captionof{table}{Complete additional timing, quality and spectral-sensitivity results.
Q references are selected independently of correlation amplitude. A dash
marks an ineligible event window.}\label{tab:S1}
\begin{tabular}{llrrrrr}\hline Event & Pair & Q refs & Q $p_0$ & Q scan $p$ & CW $p$\\\hline
GW170814 & H1xL1 & 200 & 0.00498 & 0.00498 & 0.00995\\
GW170814 & H1xV1 & 200 & 0.21891 & 0.70647 & 0.50746\\
GW170814 & L1xV1 & 200 & 0.02488 & 0.08955 & 0.96020\\
GW170818 & H1xL1 & 200 & 0.60697 & 0.37811 & 0.63184\\
GW170818 & H1xV1 & 200 & 0.22388 & 0.04478 & 0.32338\\
GW170818 & L1xV1 & 200 & 0.16418 & 0.54726 & 0.94030\\
GW190412 & H1xL1 & 200 & 0.00498 & 0.00498 & 0.00498\\
GW190412 & H1xV1 & 200 & 0.30348 & 0.71144 & 0.32338\\
GW190412 & L1xV1 & 200 & 0.38806 & 0.76119 & 0.50746\\
GW190512\_180714 & H1xL1 & 193 & 0.02577 & 0.02577 & 0.27363\\
GW190512\_180714 & H1xV1 & 200 & 0.12438 & 0.32836 & 0.11443\\
GW190512\_180714 & L1xV1 & 193 & 0.30928 & 0.37113 & 0.37313\\
GW190521 & H1xL1 & 171 & 0.00581 & 0.00581 & 0.00498\\
GW190521 & H1xV1 & 186 & 0.90374 & 0.98396 & 0.08955\\
GW190521 & L1xV1 & 182 & 0.46995 & 0.89071 & 0.03980\\
GW190828\_063405 & H1xL1 & 192 & 0.00518 & 0.00518 & 0.00498\\
GW190828\_063405 & H1xV1 & 200 & 0.12935 & 0.35821 & 0.89552\\
GW190828\_063405 & L1xV1 & 194 & 0.42051 & 0.08718 & 0.93035\\
GW200129\_065458 & H1xL1 & 126 & -- & -- & 0.00498\\
GW200129\_065458 & H1xV1 & 200 & 0.39801 & 0.67662 & 0.32836\\
GW200129\_065458 & L1xV1 & 114 & -- & -- & 0.13930\\
GW200219\_094415 & H1xL1 & 200 & 0.00498 & 0.00498 & 0.00498\\
GW200219\_094415 & H1xV1 & 200 & 0.60697 & 0.59701 & 0.49254\\
GW200219\_094415 & L1xV1 & 200 & 0.78607 & 0.80597 & 0.82090\\
GW200224\_222234 & H1xL1 & 200 & 0.00498 & 0.00498 & 0.00995\\
GW200224\_222234 & H1xV1 & 200 & 0.13433 & 0.21393 & 0.13930\\
GW200224\_222234 & L1xV1 & 200 & 0.52736 & 0.91045 & 0.17413\\
\hline\end{tabular}\end{minipage}\par
Excluded Q event windows are marked --; their retained reference counts do not make the event eligible. Requiring the CW absence flag additionally leaves no eligible primary event pair. CW notches do not guarantee complete injection subtraction. No individual Q comparison passes Holm at 5\% across its 25 eligible pairs.

\end{document}